\documentclass[12pt,aps,prd,preprint,tightenlines,superscriptaddress,
amsmath,amssymb,nofootinbib]{revtex4-1} 

\RequirePackage[colorlinks=true
,urlcolor=blue
,anchorcolor=blue
,citecolor=blue
,filecolor=blue
,linkcolor=blue
,menucolor=blue
,pagecolor=blue
,linktocpage=true
,pdfproducer=medialab
,pdfa=true
]{hyperref}

\allowdisplaybreaks

\usepackage{amsmath,amssymb,amsthm,amsfonts}
\usepackage{graphicx}
\usepackage{subfigure}
\usepackage{dcolumn}	
\usepackage{hyperref}      	
\usepackage{bm}		
\usepackage{epsfig}
\usepackage{epstopdf}
\usepackage{setspace}
\usepackage[usenames, dvipsnames]{color}
\usepackage{slashed}
\usepackage{comment}
\usepackage{enumitem}
\usepackage{siunitx}
\usepackage{multirow}
\usepackage{datetime}
\usepackage{fancyvrb}
\usepackage{enumitem}
\setlist[description]{leftmargin=0.4cm}

\newcommand{\PRE}[1]{{#1}} 

\newcommand{\be}{\begin{equation}\begin{aligned}}
\newcommand{\ee}{\end{aligned}\end{equation}}
\newcommand{\beq}{\begin{equation}}
\newcommand{\eeq}{\end{equation}}
\newcommand{\beqa}{\begin{eqnarray}}
\newcommand{\eeqa}{\end{eqnarray}}

\newcommand{\ifb}{\text{fb}^{-1}}
\newcommand{\iab}{\text{ab}^{-1}}

\newcommand{\ev}{\text{eV}}

\newcommand{\mev}{\text{MeV}}
\newcommand{\gev}{\text{GeV}}
\newcommand{\tev}{\text{TeV}}

\newcommand{\pb}{\text{pb}}

\newcommand{\micm}{\mu\text{m}}
\newcommand{\mm}{\text{mm}}
\newcommand{\cm}{\text{cm}}
\newcommand{\m}{\text{m}}

\newcommand{\mrad}{\text{mrad}}
\newcommand{\murad}{\mu\text{rad}}

\newcommand{\kg}{\text{kg}}

\renewcommand{\eqref}[1]{Eq.~(\ref{eq:#1})}

\newcommand{\secref}[1]{Sec.~\ref{sec:#1}}
\newcommand{\secsref}[2]{Secs.~\ref{sec:#1} and \ref{sec:#2}}
\newcommand{\Secref}[1]{Section~\ref{sec:#1}}
\newcommand{\figref}[1]{Fig.~\ref{fig:#1}}
\newcommand{\figsref}[2]{Figs.~\ref{fig:#1} and \ref{fig:#2}}
\newcommand{\figssref}[3]{Figs.~\ref{fig:#1}, \ref{fig:#2}, and \ref{fig:#3}}

\newcommand{\tableref}[1]{Table~\ref{table:#1}}

\newcommand{\slas}[1]{\! \not{\! \! #1}}
\newcommand{\slass}[1]{\! \not{\! #1}}

\newcommand{\llp}{\text{LLP}}

\newcommand{\mphi}{m_{\phi}}

\newcommand{\red}[1]{\textcolor{red}{#1}}

\RequirePackage[normalem]{ulem}

\begin{document}
 
\count\footins = 1000
 
\preprint{UCI-TR-2018-19, KYUSHU-RCAPP-2018-06}

\title{
{\Large FASER's Physics Reach for Long-Lived Particles}
\PRE{\vspace*{0.5in} \\
FASER Collaboration}
}

\begin{figure*}[h]
\centering
\includegraphics[width=0.6\textwidth]{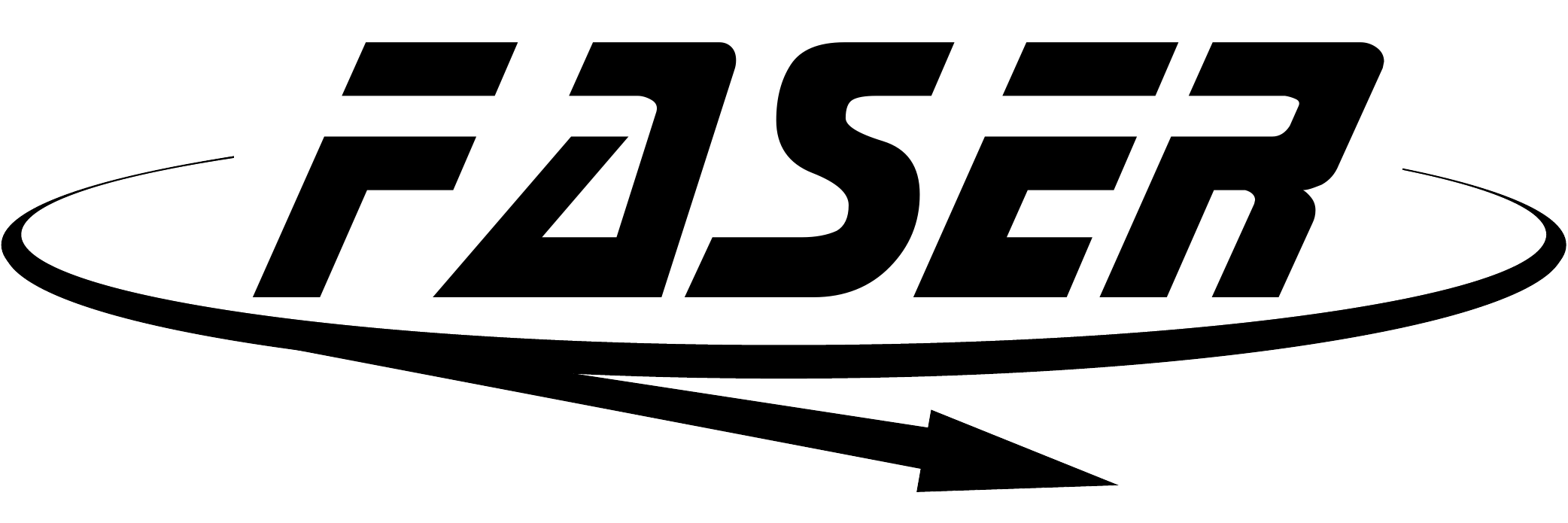}
\end{figure*}

\author{Akitaka Ariga}
\affiliation{Universit\"at Bern, Sidlerstrasse 5, CH-3012 Bern, Switzerland}

\author{Tomoko Ariga}
\affiliation{Universit\"at Bern, Sidlerstrasse 5, CH-3012 Bern, Switzerland}
\affiliation{Kyushu University, Nishi-ku, 819-0395 Fukuoka, Japan}

\author{Jamie Boyd}
\affiliation{CERN, CH-1211 Geneva 23, Switzerland}

\author{Franck Cadoux}
\affiliation{D\'epartement de Physique Nucl\'eaire et Corpusculaire, 
University of Geneva, CH-1211 Geneva 4, Switzerland}

\author{David~W.~Casper}
\affiliation{Department of Physics and Astronomy, 
University of California, Irvine, CA 92697-4575, USA}

\author{Yannick Favre}
\affiliation{D\'epartement de Physique Nucl\'eaire et Corpusculaire, 
University of Geneva, CH-1211 Geneva 4, Switzerland}

\author{Jonathan~L.~Feng}
\affiliation{Department of Physics and Astronomy, 
University of California, Irvine, CA 92697-4575, USA}

\author{Didier Ferrere}
\affiliation{D\'epartement de Physique Nucl\'eaire et Corpusculaire, 
University of Geneva, CH-1211 Geneva 4, Switzerland}

\author{Iftah Galon}
\affiliation{New High Energy Theory Center, Rutgers, The State University of New Jersey, Piscataway, New Jersey 08854-8019, USA}

\author{Sergio Gonzalez-Sevilla}
\affiliation{D\'epartement de Physique Nucl\'eaire et Corpusculaire, 
University of Geneva, CH-1211 Geneva 4, Switzerland}

\author{Shih-Chieh Hsu}
\affiliation{Department of Physics, University of Washington, PO Box 351560, Seattle, WA 98195-1560, USA}

\author{Giuseppe Iacobucci}
\affiliation{D\'epartement de Physique Nucl\'eaire et Corpusculaire, 
University of Geneva, CH-1211 Geneva 4, Switzerland}

\author{Enrique Kajomovitz}
\affiliation{Technion -- Israel Institute of Technology, Haifa 32000, Israel}

\author{Felix Kling}
\affiliation{Department of Physics and Astronomy, 
University of California, Irvine, CA 92697-4575, USA}

\author{Susanne Kuehn}
\affiliation{CERN, CH-1211 Geneva 23, Switzerland}

\author{Lorne Levinson}
\affiliation{Weizmann Institute of Science, Rehovot 761001, Israel}

\author{Hidetoshi Otono}
\affiliation{Kyushu University, Nishi-ku, 819-0395 Fukuoka, Japan}

\author{Brian Petersen}
\affiliation{CERN, CH-1211 Geneva 23, Switzerland}

\author{Osamu Sato}
\affiliation{Nagoya University, Furo-cho, Chikusa-ku, Nagoya-shi 464-8602, Japan}

\author{Matthias Schott}
\affiliation{Institut f\"ur Physik, Universit\"at Mainz, Mainz, Germany}

\author{Anna Sfyrla}
\affiliation{D\'epartement de Physique Nucl\'eaire et Corpusculaire, 
University of Geneva, CH-1211 Geneva 4, Switzerland}

\author{Jordan Smolinsky}
\affiliation{Department of Physics and Astronomy, 
University of California, Irvine, CA 92697-4575, USA}

\author{Aaron~M.~Soffa}
\affiliation{Department of Physics and Astronomy, 
University of California, Irvine, CA 92697-4575, USA}

\author{Yosuke Takubo}
\affiliation{Institute of Particle and Nuclear Study, 
KEK, Oho 1-1, Tsukuba, Ibaraki 305-0801, Japan}

\author{Eric Torrence}
\affiliation{University of Oregon, Eugene, OR 97403, USA}

\author{Sebastian Trojanowski}
\affiliation{\mbox{National Centre for Nuclear Research, Ho{\. z}a 69, 00-681 Warsaw, Poland}}
\affiliation{Consortium for Fundamental Physics, School of Mathematics and  Statistics, University of Sheffield, Hounsfield Road, Sheffield, S3 7RH, UK}

\author{Gang Zhang\PRE{\vspace*{.2in}}}
\affiliation{Tsinghua University, Beijing, China
}


\begin{abstract}
\PRE{\vspace*{0.2in}}
FASER, the ForwArd Search ExpeRiment, is an approved experiment dedicated to searching for light, extremely weakly-interacting particles at the LHC.  Such particles may be produced in the LHC's high-energy collisions and travel long distances through concrete and rock without interacting.  They may then decay to visible particles in FASER, which is placed 480 m downstream of the ATLAS interaction point.  In this work we briefly describe the FASER detector layout and the status of potential backgrounds. We then present the sensitivity reach for FASER for a large number of long-lived particle models, updating previous results to a uniform set of detector assumptions, and analyzing new models.  In particular, we consider all of the renormalizable portal interactions, leading to dark photons, dark Higgs bosons, and heavy neutral leptons (HNLs); light $B-L$ and $L_i - L_j$ gauge bosons; axion-like particles (ALPs) that are coupled dominantly to photons, fermions, and gluons through non-renormalizable operators; and pseudoscalars with Yukawa-like couplings.  We find that FASER and its follow-up, FASER 2, have a full physics program, with discovery sensitivity in all of these models and potentially far-reaching implications for particle physics and cosmology.
\end{abstract}


\maketitle

\renewcommand{\baselinestretch}{0.95}\normalsize
\tableofcontents
\renewcommand{\baselinestretch}{1.0}\normalsize


\clearpage
\section{Introduction}
\label{sec:introduction}

For decades, a focus of energy-frontier particle colliders, such as the LHC, has been searches for new particles with TeV-scale masses and ${\cal O}(1)$ couplings.  The common lore was to target large transverse momentum ($p_T$) signatures that emerge in the roughly isotropic decays of such particles.  There is, however, a complementary class of viable new particles that are much lighter, with masses in the MeV to GeV range, and much more weakly coupled to the standard model (SM)~\cite{Battaglieri:2017aum}. In recent years, these particles have attracted growing interest, in part because they can yield dark matter with the correct relic density~\cite{Boehm:2003hm,Feng:2008ya} and may resolve discrepancies between low-energy experiments and theoretical predictions~\cite{Bennett:2006fi, Pohl:2010zza, Krasznahorkay:2015iga}. Perhaps most importantly, they can be discovered at a wide variety of experiments, reinvigorating efforts to find creative ways to search for new particles.

Such weakly coupled particles are typically long-lived and travel macroscopic distances without interacting before decaying to SM particles.  At the LHC, searching for such particles in the high-$p_T$ region is ineffective, because the high-$p_T$ SM cross sections are insufficient to produce such weakly coupled particles in large enough numbers. The situation is very different at low $p_T$, however, since the inelastic cross section is many orders of magnitude larger. The LHC's discovery potential can, therefore, be augmented tremendously if a detector is placed in the far-forward region of an existing interaction point (IP) after the beam has curved.  FASER~\cite{Feng:2017uoz}, the ForwArd Search ExpeRiment, is a small and inexpensive experiment dedicated to exploiting this opportunity to discover new physics.  It was approved by the CERN Research Board in March 2019.

To be slightly more quantitative, the total inelastic scattering cross section at the 14 TeV LHC is similar to the one measured at 13 TeV: $\sigma_{\text{inel}} \sim 75~\text{mb}$~\cite{Aaboud:2016mmw, VanHaevermaet:2016gnh}. For LHC Run 3, which is expected to gather an integrated luminosity of $150~\ifb$ in the years 2021-23, we therefore expect
\be
N_{\rm{inel}} = 1.1 \times 10^{16}
\ee
inelastic $pp$ scattering events.  This, in turn, implies extraordinary meson production rates of
\be
N_{\pi^0} \approx 2.3 \times 10^{17}, \ \ N_{\eta} \approx 2.5 \times 10^{16} , 
\ \ N_D \approx 1.1 \times 10^{15} , \ \ \text{and} \ \ N_B \approx 7.1 \times 10^{13} \ 
\ee
in each hemisphere.  A further, 20-fold increase can be expected in the high luminosity LHC (HL-LHC) era. These particles are highly concentrated in the very forward direction; for example, as will be discussed in detail below, approximately 0.6\% (10\%) of all neutral pions are produced within 0.2 mrad (2 mrad) of the beam collision axis, which is the angular acceptance for FASER (FASER 2). If one focuses on high-energy pions, the fraction in the very forward direction is even larger.  This can be compared to the tiny angular size as seen from the IP of FASER (FASER 2), which covers only $2\times 10^{-8}$ ($2\times 10^{-6}$) of the solid angle of the forward hemisphere. Moreover, light new particles produced in meson decays are highly collimated, with characteristic angles relative to the parent meson's direction of $\theta \sim \Lambda_{\text{QCD}}/E$, $m_D/E$, and $m_B /E$ for particles produced in pion, $D$, and $B$ decays, respectively, where $E$ is the energy of the particle. For $E\sim \tev$, even hundreds of meters downstream from the IP, the transverse spread is only $\sim 10~\rm{cm}-1~\rm{m}$. 

In addition, the high LHC beam energies give rise to large boost factors that can increase the probability of long-lived particles (LLPs) decaying in a faraway detector in some of the most interesting cases.  Finally, the shielding between the IP and a distant detector, including rock, magnets, absorbers, and concrete walls, eliminates most of the potential backgrounds. A small detector placed hundreds of meters from the IP may therefore harness the extraordinary, previously wasted, SM events rates in the forward region in an extremely low-background environment.

The side tunnels TI12 and TI18 are nearly ideal locations for FASER~\cite{Feng:2017uoz}. These side tunnels were formerly used to connect the SPS to the LEP (now LHC) tunnel, but they are currently unused. The LHC beam collision axis intersects TI12 and TI18 at a distance of 480 m to the west and east of the ATLAS IP, respectively.  Estimates based on detailed simulations using FLUKA~\cite{Ferrari:2005zk,Bohlen:2014buj} by CERN's Sources, Targets, and Interaction (STI) group~\cite{FLUKAstudy}, combined with {\it in situ} measurements using emulsion detectors, have now confirmed a low rate of high-energy SM particles in these locations. Additionally, the FLUKA results combined with radiation monitor measurements have confirmed low radiation levels in these tunnels.  These locations, then, provide extremely low background environments for FASER to search for LLPs that are produced at or close to the IP, propagate in the forward direction close to the beam collision axis, and decay visibly within FASER's decay volume.  

Although TI12 and TI18 are roughly symmetric, it now appears that TI12 provides slightly more space for FASER along the beam collision axis. The proposed timeline, then, is to install FASER in TI12 during Long Shutdown 2 (LS2) from 2019-20 in time to collect data in Run 3 from 2021-23.  In the following LS3 from 2024-26, a larger detector, FASER 2, could be constructed to collect data in the HL-LHC era. The size and layout of these detectors is discussed further in  \secref{FASERdetails}. 

FASER's potential for discovering new light and weakly-interacting particles is based on the general considerations given above.  However, it is also important to quantify FASER's reach relative to existing constraints, as well as to compare FASER to the many other complementary experiments with similar physics targets, including HPS~\cite{Moreno:2013mja}, Belle-II~\cite{Dolan:2017osp}, LHCb~\cite{Ilten:2015hya, Ilten:2016tkc}, NA62~\cite{Dobrich:2018ezn}, NA64~\cite{Gninenko:2018tlp}, SeaQuest~\cite{Berlin:2018pwi}, SHiP~\cite{Alekhin:2015byh}, MATHUSLA~\cite{Evans:2017lvd,Curtin:2018mvb}, CODEX-b~\cite{Gligorov:2017nwh}, AL3X~\cite{Gligorov:2018vkc}, LDMX~\cite{Berlin:2018bsc}, and others mentioned below. For this, it is necessary to evaluate FASER's sensitivity in specific models~\cite{Feng:2017uoz, Feng:2017vli, Batell:2017kty, Kling:2018wct, Helo:2018qej, Bauer:2018onh, Cheng:2018vaj, Feng:2018noy, Hochberg:2018rjs, Berlin:2018jbm, Dercks:2018eua}.  In this study, we determine the sensitivity reach of both FASER and FASER 2 for a wide variety of proposed particles, updating previous results to a uniform set of detector assumptions, and analyzing new models.  In particular, we consider all of the renormalizable portal interactions, leading to dark photons, dark Higgs bosons, and heavy neutral leptons (HNLs); light $B-L$ and $L_i - L_j$ gauge bosons; axion-like particles (ALPs) that are coupled dominantly to photons, fermions, and gluons through non-renormalizable operators; and dark pseudoscalars with Yukawa-like couplings.  A summary of the models discussed in this paper is given in \tableref{summary}.

\begin{table}
  \centering
  \begin{tabular}{|c|c|c|c|c|c|c|}
  \hline \hline
	{\bf Benchmark Model} & \, {\bf Label} \ & \ {\bf Section} \ & \, {\bf PBC} \ & \, {\bf Refs} \ & \, {\bf FASER} \ & \, {\bf FASER 2} \ \\ \hline \hline
    Dark Photons & V1 & \ref{sec:vanillaAprim} & BC1 & \cite{Feng:2017uoz} & $\surd$ & $\surd$ \\ 
    $B-L$ Gauge Bosons & V2 & \ref{sec:BminusLAprim} & ---  & \cite{Bauer:2018onh} & $\surd$ & $\surd$ \\ 
    $L_i - L_j$ Gauge Bosons & V3 & \ref{sec:LiminusLjAprim} & --- & \cite{Bauer:2018onh} & --- & --- \\ \hline
    Dark Higgs Bosons & S1 & \ref{sec:s1} & BC4 & \cite{Feng:2017vli,Batell:2017kty} & --- & $\surd$ \\ 
    Dark Higgs Bosons with $hSS$ & S2 & \ref{sec:s2} & BC5 & \cite{Feng:2017vli} & --- & $\surd$ \\ \hline
    HNLs with $e$ & F1 & \ref{sec:hnls} & BC6 & \cite{Kling:2018wct,Helo:2018qej} & --- & $\surd$ \\ 
    HNLs with $\mu$ & F2 & \ref{sec:hnls} & BC7 & \cite{Kling:2018wct,Helo:2018qej} & --- & $\surd$ \\ 
    HNLs with $\tau$ & F3 & \ref{sec:hnls} & BC8 & \cite{Kling:2018wct,Helo:2018qej} & $\surd$ & $\surd$ \\ \hline
    ALPs with Photon & A1 & \ref{sec:a1} & BC9 & \cite{Feng:2018noy} & $\surd$ & $\surd$ \\ 
    ALPs with Fermion & A2 & \ref{sec:a2} & BC10 & --- & --- & $\surd$ \\ 
    ALPs with Gluon & A3 & \ref{sec:a3} & BC11 & --- & $\surd$ & $\surd$ \\ 
    \hline 
    Dark Pseudoscalars & P1 & \ref{sec:pseudoscalarYukawa} & --- & \cite{Ariga:2018zuc} & --- & $\surd$\\
    \hline \hline
  \end{tabular}
  \caption{The benchmark models studied in this work, along with their labels, the sections in which they are discussed, their PBC labels, references in which they were previously studied, and the prospects for FASER and FASER 2 to probe new parameter space.  FASER and FASER 2 have discovery potential for all candidates with renormalizable couplings (dark photons, dark Higgs bosons, HNLs); ALPs with all types of couplings ($\gamma$, $f$, $g$); dark pseudoscalars with Yukawa-like couplings; and also other models that are not discussed here~\cite{Cheng:2018vaj, Hochberg:2018rjs, Berlin:2018jbm,Dercks:2018eua}.} 
\label{table:summary}
\end{table}

The paper is organized as follows. In \secref{FASERdetails} we present more details about the layout and sizes of FASER and FASER 2. This is followed by an overview of the various production processes of LLPs at the LHC in \secref{production}. The expected FASER reach is analyzed in \secref{vectors} for dark photons and other light gauge bosons, in \secref{scalars} for dark scalars, in \secref{hnls} for HNLs, in \secref{alps} for ALPs, and in \secref{pseudoscalarYukawa} for dark pseudoscalars.  \Secref{systematics} is devoted to a discussion of the impact of various systematic effects on FASER's reach in searches for new physics. We conclude in \secref{summary}. 

The models studied here have significant overlap with the benchmark models defined by the CERN Physics Beyond Colliders (PBC) study group~\cite{Beacham:2019nyx}.  One purpose of this paper is to provide a more detailed explanation of the underlying assumptions and analyses leading to the FASER results that are briefly summarized in the PBC study.  

\section{The FASER Detector}
\label{sec:FASERdetails}

In this section, we give a brief overview of FASER's location, signal and background, the detector components and layout, and the benchmark detector parameters we will assume in studying FASER's reach in the following sections.  These aspects of FASER have been presented at length in FASER's Letter of Intent~\cite{Ariga:2018zuc} and Technical Proposal~\cite{Ariga:2018pin}, and we refer readers to those documents for more details.

\subsection{Location}
\label{sec:location}

As noted in \secref{introduction}, FASER will be located in the cTempty and unused tunnel TI12, which connects the SPS and LEP/LHC tunnels.  This location is shown in \figref{maps}, and is roughly 480 m east of the ATLAS IP.  The beam collision axis passes along the floor of TI12, with its exact location depending on the beam crossing angle at ATLAS.  TI12 slopes upward when leaving the LHC tunnel to connect to the shallower SPS tunnel.  To place FASER along the beam collision axis, the ground of TI12 must be lowered roughly 45 cm at the front of FASER, where particles from the ATLAS IP enter.

\begin{figure}[tbp]
\centering
\includegraphics[width=0.59\textwidth]{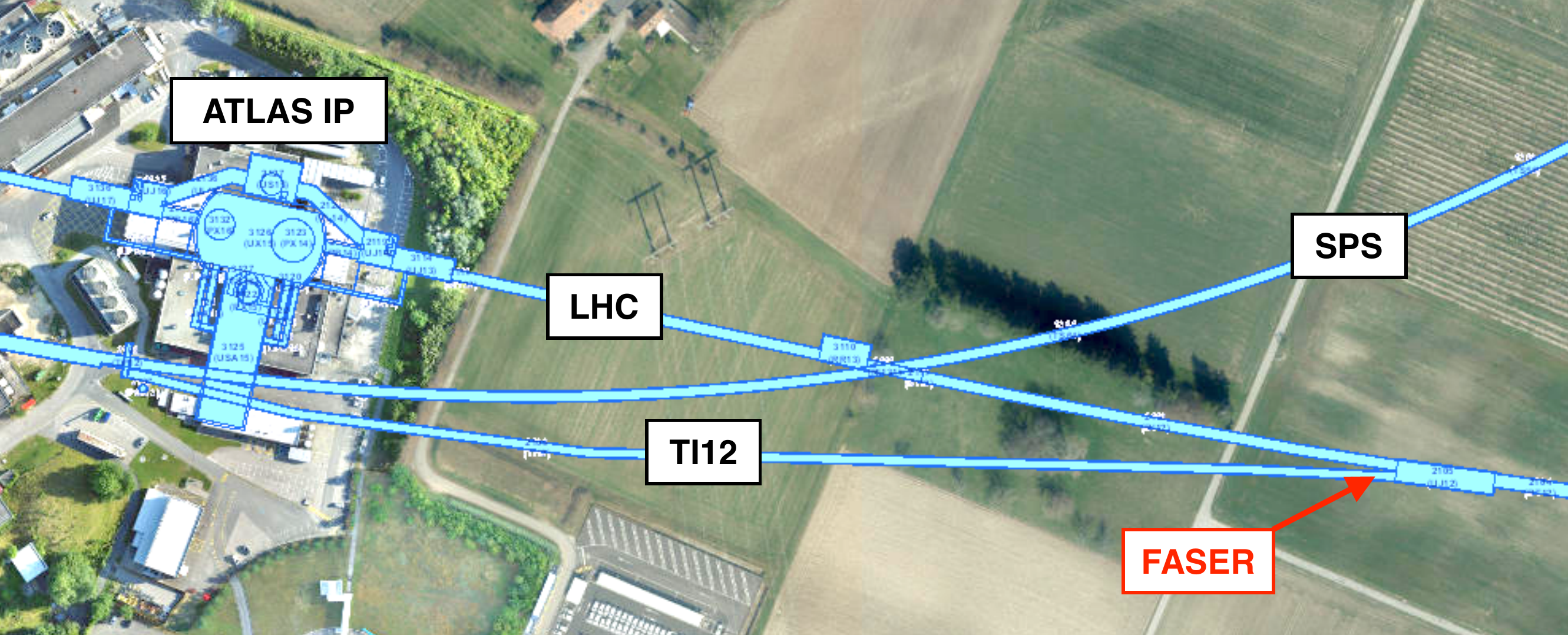} \hfill
\includegraphics[width=0.394\textwidth]{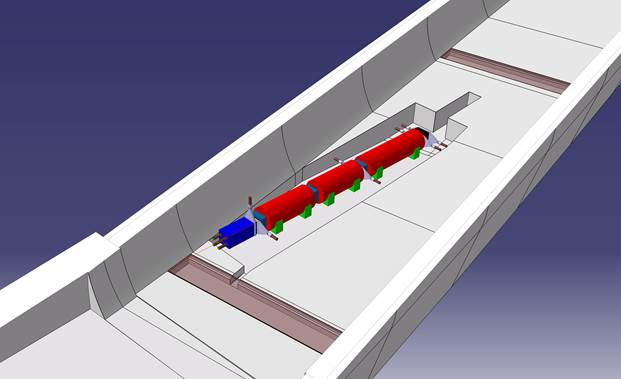}
\caption{
{\bf Left panel}: The arrow points to FASER's location in service tunnel TI12, roughly 480 m east of the ATLAS IP. Credit: CERN Geographical Information System.  {\bf Right panel}: View of FASER in tunnel TI12.  The trench lowers the floor by 45 cm at the front of FASER to allow FASER to be centered on the beam collision axis. Credit: CERN Site Management and Buildings Department. } 
\label{fig:maps}
\end{figure}

\begin{figure}[tbp]
\centering
\includegraphics[width=0.98\textwidth]{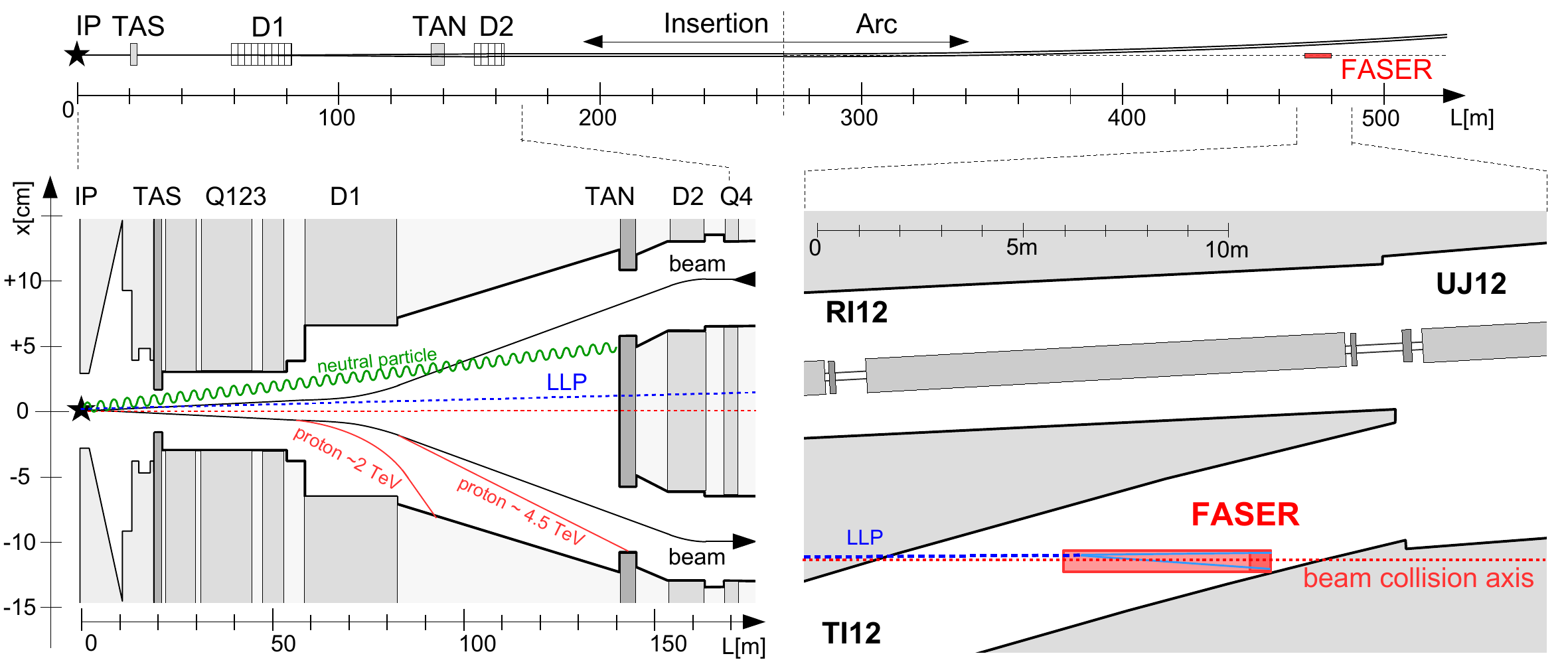} 
\caption{Schematic view of the far-forward region downstream of ATLAS and various particle trajectories. {\bf Upper panel}: FASER is located $480~\m$ downstream of ATLAS along the beam collision axis (dotted line) after the main LHC tunnel curves away.  {\bf Lower left panel}: High-energy particles produced at the IP in the far-forward direction.  Charged particles are deflected by LHC magnets, and neutral hadrons are absorbed by either the TAS or TAN, but LLPs pass through the LHC infrastructure without interacting. Note the extreme difference in horizontal and vertical scales.  {\bf Lower right panel}: LLPs may then travel $\sim 480~\m$ further downstream and decay within FASER in TI12.} 
\label{fig:infrastructure}
\end{figure}

A schematic view of the far-forward region downstream of ATLAS is given in \figref{infrastructure}.  From the ATLAS IP, the LHC beam passes through a 270 m-long straight ``insertion,'' and then enters an ``arc'' and bends.  Far-forward charged particles are bent by the beam optics, and neutral hadrons are typically stopped in the TAS or TAN absorbers, which are designed to protect the magnets.  To travel from the IP to FASER, particles must pass through roughly 10 m of concrete and 90 m of rock.  In the SM, only muons and neutrinos can reach FASER from the IP.   On the other hand, LLPs produced at or near the IP easily pass through all of the natural and man-made material without interacting and then can decay in FASER.  

\subsection{Signal}
\label{sec:signal}

At the LHC, light particles are typically produced with a characteristic transverse momentum comparable to their mass $p_T\sim m$. Consequently, LLPs that are produced within FASER’s angular acceptance, $\theta \lesssim 1~\mrad$, where $\theta$ is the angle with respect to the beam collision axis, tend to have very high energies $E\sim~\tev$, as can be inferred from
\begin{equation}
\theta\simeq \tan\theta = \frac{p_T}{p}\sim \frac{m}{E}\ll 1\ ,
\label{eq:theta}
\end{equation}
where for the lightest mesons the relevant mass scale $m$ can be replaced with $\Lambda_{\textrm{QCD}}$.

The characteristic signal events at FASER are, then,
\begin{equation}
  p p  \to \text{LLP} +X, \quad  \text{LLP travels} \ \sim 480~\text{m}, \quad \text{LLP} \to e^+ e^- , \mu^+ \mu^- , \pi^+ \pi^-, \gamma \gamma, \ldots ,
\end{equation}
where the LLP decay products have $\sim \tev$ energies. The target signals at FASER are therefore striking: two oppositely charged tracks or two photons with $\sim \tev$ energies that emanate from a common vertex inside the detector and have a combined momentum that points back through 100 m of concrete and rock to the IP. 

The decay products of such light and highly boosted particles are extremely collimated, with a typical opening angle $\theta \sim m/E$. For example, for an LLP with mass $m \sim 100~\mev$ and energy $E \sim 1~\tev$, the typical opening angle is $\theta \sim m/E \sim 100~\murad$, implying a separation of only $\sim 100~\micm$ after traveling through $1~\m$ in the detector.  To resolve the two charged tracks produced by a decaying LLP, FASER must include a magnetic field to split the oppositely-charged tracks.

\subsection{Detector Layout}
\label{sec:faser-layout}

To be sensitive to the many possible forms of light, weakly-interacting particles, and to differentiate signal from background, the FASER detector has several major components.  These components and the detector layout are shown in \figref{DetectorLayout}.
  
\begin{figure}[tbp]
\centering
\includegraphics[width=0.87\textwidth]{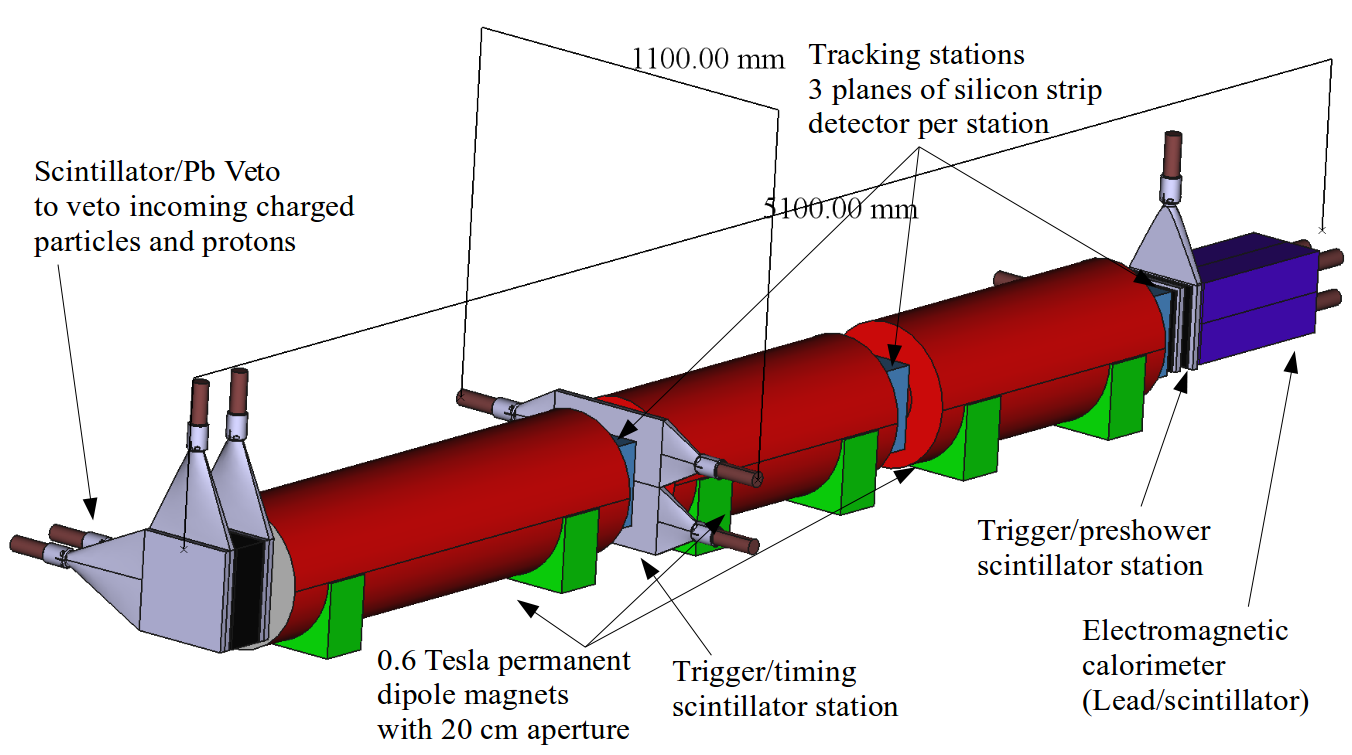} 
\caption{
Layout of the FASER detector. LLPs enter from the left and the entire length of the detector is roughly 5 m.  The detector components include scintillators (gray), dipole magnets (red), tracking stations (blue), a calorimeter (dark purple), and support structures (green). 
}
\label{fig:DetectorLayout}
\end{figure}

Particles produced at the ATLAS IP enter the detector from the left.  At the entrance to the detector is a double layer of scintillators (gray) to veto charged particles coming through the cavern wall from the IP, primarily high-energy muons.  Between the scintillation layers is a 20-radiation-lengths-thick layer of lead that converts photons produced in the wall into electromagnetic showers that can be efficiently vetoed by the scintillators.

The veto layer is followed by a $\Delta = 1.5~\text{m}$ long, 0.6 T permanent dipole magnet (red) with a $R= 10~\text{cm}$ aperture radius.  Such permanent magnets take up relatively little space and, unlike electromagnets, do not require high voltage power and cooling. The cylindrical volume enclosed by this magnet serves as the decay volume for the light, weakly-interacting particles, with the magnet providing a horizontal kick to separate oppositely-charged particles to a detectable distance. 

Next is a spectrometer consisting of two $1~\text{m}$-long, 0.6 T dipole magnets with three tracking stations (blue), each composed of layers of precision silicon strip detectors located at either end and in between the magnets. The primary purpose of the spectrometer is to observe the characteristic signal of two oppositely charged particles pointing back towards the IP, measure their momenta, and sweep out low-momentum charged particles before they reach the back of the spectrometer. Scintillator planes (gray) for triggering and precision time measurements are located at the entrance and exit of the spectrometer.  

The final component is an electromagnetic calorimeter (purple) to identify high energy electrons and photons and measure the total electromagnetic energy.  As the primary signals are two close-by electrons or photons, these cannot be resolved by the calorimeter.

\subsection{Background}
\label{sec:faser-background}

The natural (rock) and LHC infrastructure (concrete, magnets, and absorbers) shielding dramatically reduces the high-energy charged particle and photon flux in FASER.  To determine the background, the CERN STI group has performed FLUKA simulations~\cite{Ferrari:2005zk,Bohlen:2014buj} to estimate both the high-energy particle flux in FASER and the low-energy radiation levels that may impact detector electronics~\cite{FLUKAstudy}.  In addition, detectors that were installed in the TI18 and TI12 tunnels during LHC Technical Stops in 2018 now provide {\em in situ} measurements of the high-energy particle flux and radiation levels. Within the uncertainties in the FLUKA simulation and the detector efficiencies, these {\em in situ} measurements have validated the FLUKA results. The current simulations and most of the {\em in situ} measurements are for TI18, but the expectation is that the particle fluxes will be the same in TI12, and initial {\em in situ} measurements from TI12 demonstrate that this is the case.  Details of these studies have been presented in FASER's Letter of Intent~\cite{Ariga:2018zuc} and Technical Proposal~\cite{Ariga:2018pin} and are summarized here.

The FLUKA simulation tracks particle production, deflection, and energy loss with a detailed model of the geometry of the LHC tunnels, including the LHC material map and magnetic field layout. The simulation includes three potential sources of background at the FASER location:
\begin{itemize}
    \item Particles produced in the $pp$ collisions at the IP or by particles produced at the IP that interact further downstream, e.g., in the TAN neutral particle absorber.
    \item Particles from showers initiated by off-momentum (and therefore off-orbit) protons hitting the beam pipe in the dispersion suppressor region close to FASER.
    \item Particles produced in beam-gas interactions by the beam passing FASER in the ATLAS direction (for which there is no rock shielding).
\end{itemize}
The results show that the latter two sources do not contribute significantly to the high-energy particle flux in FASER and are therefore negligible backgrounds.

In the first category, as expected, only muons and neutrinos from the IP can pass through 100 m of concrete and rock to produce high-energy particles in FASER.  For neutrinos, preliminary estimates indicate that the flux of neutrino-induced background events in FASER would be too low to constitute an obstacle for LLP searches.  This is due to the small neutrino-material cross sections and the event kinematics, which is different from LLP decays~\cite{Feng:2017uoz}.

The dominant source of background, then, is radiative processes associated with muons coming from the IP, such as the production of photons or electromagnetic or hadronic showers. These can occur in the rock before FASER or inside the detector material.
Although the background from these processes depends on the details of the FASER design, kinematics assures that the opening angle between any high-energy ($E > 100~\gev$) secondary particle and its parent muon is at most a few $\mrad$~\cite{Groom:2001kq,VanGinneken:1986rf,Ariga:2018zuc}.  Consequently, such background may be rejected by vetoing events in which an LLP-like signature is accompanied by a high-energy muon that enters the detector from the direction of the IP.  The FLUKA results and {\em in situ} measurements imply that less than $10^5$ high-energy muon-induced background events are expected in FASER in Run 3~\cite{Ariga:2018zuc}.  By including two scintillator veto stations at the entrance to FASER (the side facing the IP), each able to detect at least $99.99\%$ of the incoming high-energy muons, the background can be reduced to negligible levels. 

We also note that cosmic ray backgrounds will be efficiently distinguished from LLP signals based on directionality and timing information. They are therefore not an obstacle to new physics searches at FASER.

In summary, given FLUKA simulation results for high-energy particle fluxes, validated by {\em in situ} measurements, and the ability to veto events with charged particles entering FASER from the outside, we expect that the characteristic LLP signatures will have extremely suppressed backgrounds.  In the remainder of this work, we present FASER sensitivity reaches assuming negligible background and requiring three signal events for discovery.

\subsection{Detector Benchmarks}
\label{sec:faser-benchmark}

In the following we will consider two detector benchmarks: FASER as described above and designed to collect data during LHC Run 3 from 2021-23; and FASER 2, which may collect data in the HL-LHC era from 2026-35. Following the FASER design, we assume these detectors have cylindrical shapes with depth $\Delta$ and radius $R$.  The parameters for these two detectors, and the assumed integrated luminosity for each of them, are 
\be
\textbf{FASER:} \ \  &\Delta = 1.5~\m, &R& = 10~\cm, &\mathcal{L}& =150~\ifb  \\
\textbf{FASER 2:} \ \  &\Delta = 5~\m,  &R& = 1~\m,     &\mathcal{L}& =3~\iab \ .
\label{eq:geomtery}
\ee
The collision energy is assumed to be 14 TeV in all cases. As with FASER, we assume FASER 2 will be located $L = 480$ m from the IP.  At present, the design of FASER 2 has not been carefully studied, and the FASER 2 parameters should only be taken as representative of a detector that is much larger than FASER.   We note that, with these parameters, FASER 2 will require significant excavation to extend either TI12 or TI18, or to widen the staging area UJ18 near TI18 or the cavern UJ12 near TI12.

In determining the physics reach for the various models below, we will further assume that FASER will be able to observe all decays of LLPs into visible final states within FASER's decay volume. We require a minimal visible energy of $100~\gev$, but note that this is typically already fulfilled for LLPs traveling close to the beam collision axis and sufficiently boosted to decay in FASER. Finally, we assume that FASER will be able to reduce possible high-energy backgrounds to a negligible level.

\section{Production of LLP\lowercase{s}}
\label{sec:production}

Depending on their couplings to the SM, new light particles can typically be produced at the LHC in several different processes. These include rare decays of SM hadrons, dark bremsstrahlung in coherent $pp$ collisions, and direct production in hard scatterings. In addition, particles produced at the IP may travel 140 m down the beam pipe and hit the TAN neutral particle absorber, effectively creating a beam dump experiment that may produce LLPs. In the following, we briefly discuss all of these production mechanisms.

\begin{figure}[tbp]
\centering
\includegraphics[width=0.99\textwidth]{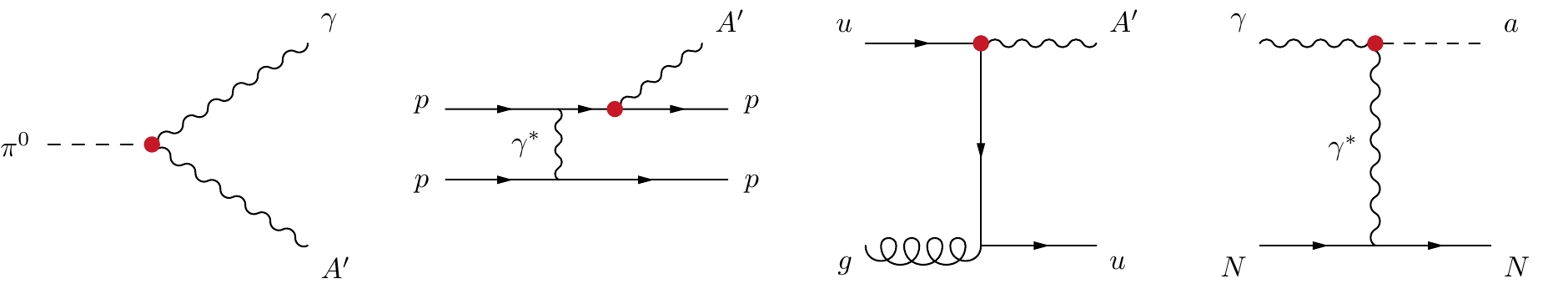} 
\caption{
Representative Feynman diagrams for the LLP production processes outlined in this section: dark photon production from pion decay (left), dark photon production via dark bremsstrahlung (center left), dark photon production in hard scattering (center right), and ALP production via the Primakoff process from photons scattering in the TAN (right).
}
\label{fig:feynman}
\end{figure}

\subsection{Rare Decays of SM Hadrons}

If LLPs couple to quarks, their most important production modes are often rare decays of SM hadrons. In particular, the leading production mechanism is typically the decays of the lightest mesons that are kinematically allowed to decay to the LLPs.  

Reliable estimates of the number of signal events in FASER require accurate modeling of the SM hadron spectra in the far forward region. This modeling has improved greatly in recent years, thanks to a number of experiments targeting the large pseudorapidity region of the LHC. (For a review, see Ref.~\cite{N.Cartiglia:2015gve}.) We exploit this progress and determine the hadron spectra for our estimates as follows: 

\begin{description}
\item [Light Hadrons] We use the Monte-Carlo event generator EPOS-LHC~\cite{Pierog:2013ria}, as implemented in the CRMC simulation package~\cite{CRMC}, to simulate the kinematic distributions of light mesons, such as pions and kaons. In particular, we obtain a production cross section in each hemisphere for neutral pions $\pi^0$ and $\eta$ mesons of $1.6 \times 10^{12}~\pb$ and $1.7 \times 10^{11}~\pb$, respectively. 
These particles are highly concentrated in the very forward , as noted previously in the discussion surrounding~\eqref{theta}. This is illustrated in \figref{spectrum} (left), where we show the production rate of neutral pions in the $(\theta, p)$ plane, where $\theta$ and $p$ are the meson's angle with respect to the beam axis and momentum, respectively. As noted in \secref{introduction}, approximately 0.6\% ($10\%$) of the pions are produced within $0.2~\mrad$ ($2~\mrad$) of the beam collision axis, the angular acceptance for FASER (FASER 2). If one focuses on high energy pions, the fraction that is in the very forward direction is even larger.

\begin{figure}[tbp]
\centering
\includegraphics[width=0.48\textwidth]{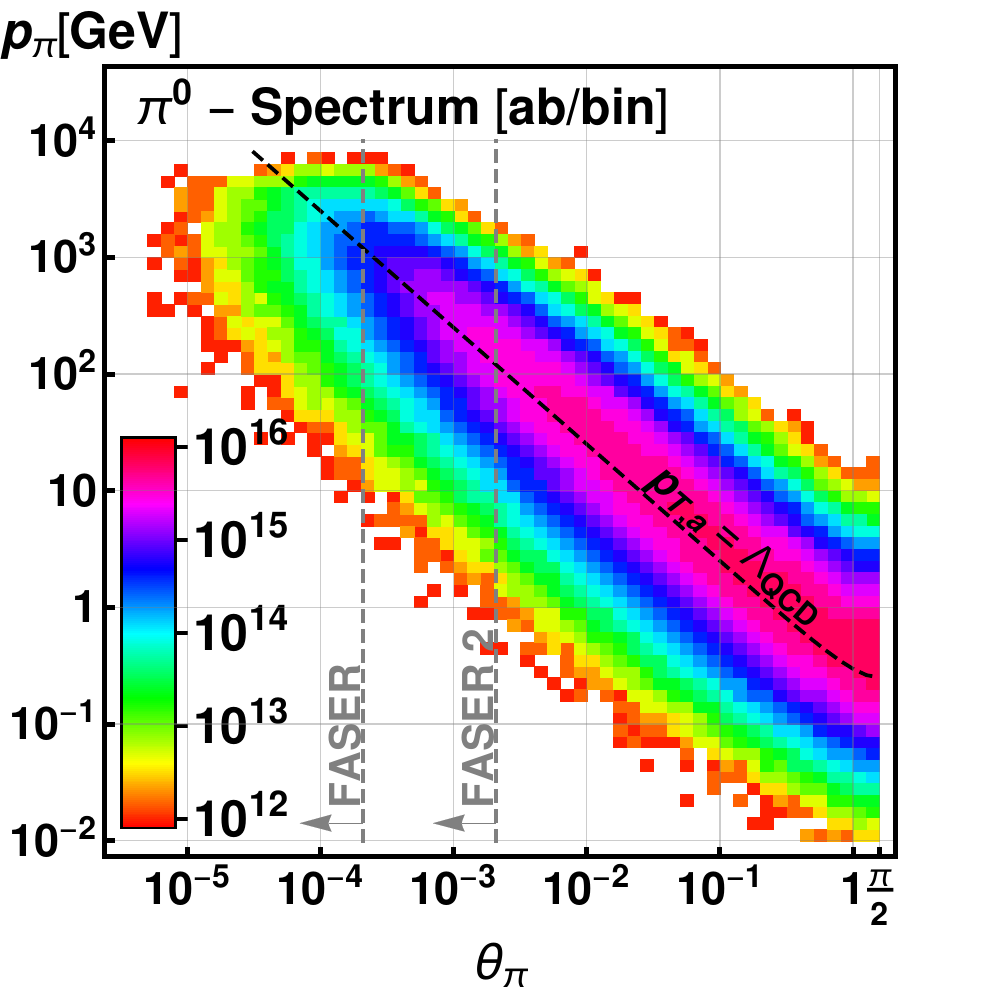} 
\includegraphics[width=0.48\textwidth]{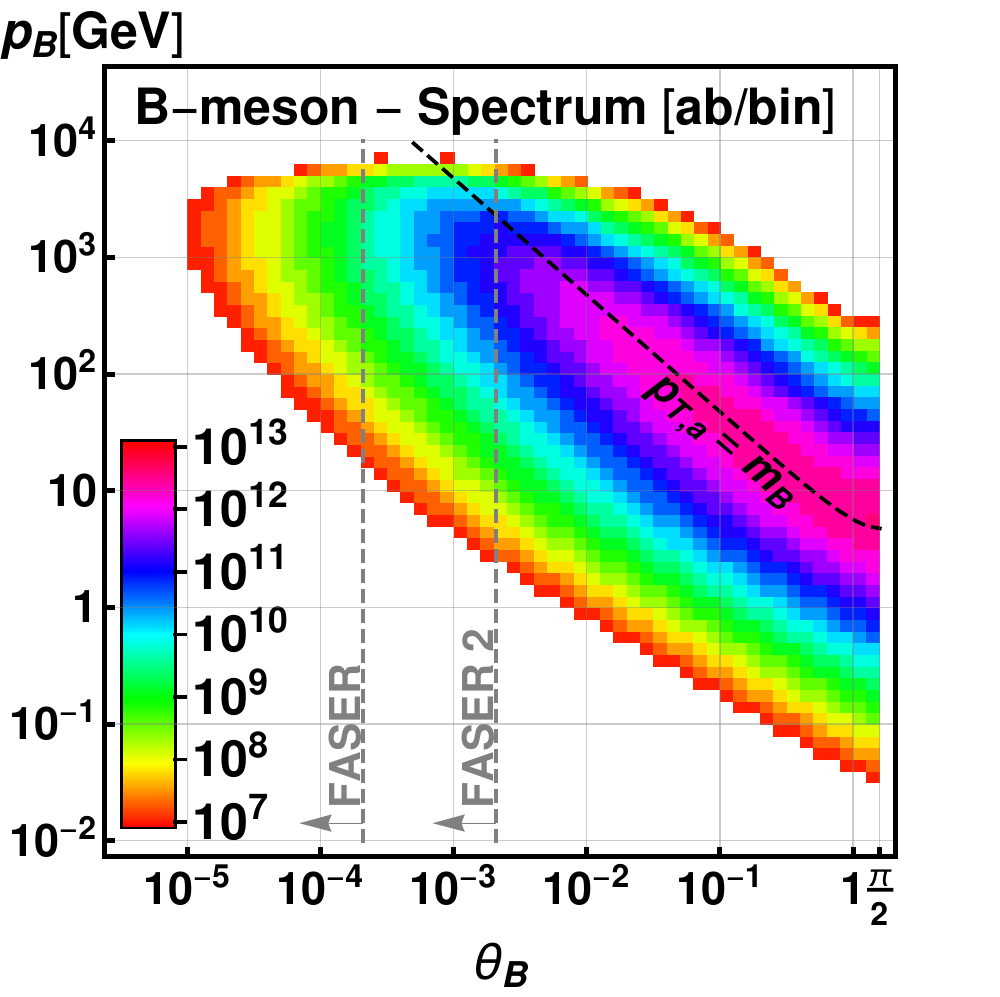} 
\caption{
Differential meson production rate in each hemisphere in the $(\theta,p)$ plane, where $\theta$ and $p$ are the meson's angle with respect to the beam axis and momentum, respectively. The bin thickness is 1/10 of a decade along each axis. We show the $\pi^0$ spectrum (left), obtained via EPOS-LHC~\cite{Pierog:2013ria}, and the $B$ meson spectrum (right), obtained using FONLL with CTEQ6.6~\cite{Nadolsky:2008zw}. The diagonal black dashed lines highlight the characteristic transverse momentum scale $p_T \sim \Lambda_{\text{QCD}} \sim 250~\mev$ for pions and $p_T \sim m_B$ for $B$ mesons. The angular acceptances for FASER and FASER 2 are indicated by the vertical gray dashed lines.
}
\label{fig:spectrum}
\end{figure}

\item [Heavy Hadrons]  We use the simulation tool FONLL~\cite{Cacciari:1998it,Cacciari:2001td} to calculate the differential cross section for charm and beauty hadrons. In particular, we take into account non-perturbative fragmentation functions to obtain the hadronic spectra: BCFY~\cite{Braaten:1994bz} for charmed hadrons and Kartvelishvili et al.~\cite{Kartvelishvili:1977pi,Cacciari:2005uk} with fragmentation parameter $\alpha=24.2$ for beauty hadrons. We use the CTEQ6.6~\cite{Nadolsky:2008zw} parton distribution functions (PDFs) with $m_b=4.75~\gev$ and $m_c=1.5~\gev$, and obtain production cross sections in each hemisphere of $D$-mesons and $B$-mesons of $7.4 \times 10^9~\pb$ and $4.7 \times 10^8~\pb$, respectively. The spectrum for $B$ mesons is illustrated in \figref{spectrum} (right).
\end{description}

In LHC Run 3 with an expected integrated luminosity of $150~\ifb$, we expect about $2.3 \times 10^{17}$ neutral pions, $2.5 \times 10^{16}$ $\eta$-mesons, $1.1 \times 10^{15}$ $D$-mesons, and $7.1 \times 10^{13}$ $B$-mesons to be produced in each hemisphere. More details about  LLP production in specific hadron decay channels can be found in Refs.~\cite{Feng:2017uoz,Feng:2017vli,Kling:2018wct}.

\subsection{Dark Bremsstrahlung}

Production of LLPs heavier than thresholds for the decays of the lightest mesons can be dominated by dark bremsstrahlung in coherent $pp$ scatterings, $pp\rightarrow pp+ \text{LLP}$ (see center left panel of \figref{feynman}). This is typically modeled using the Fermi-Weizsacker-Williams approximation~\cite{Blumlein:2013cua}; see, e.g., Ref.~\cite{deNiverville:2016rqh} for a recent discussion. In particular, for the case of dark vector bosons $V$, dark bremsstrahlung typically becomes the dominant production mode for masses $m_V > m_{\pi}$. On the other hand, for other LLP models considered below, bremsstrahlung plays a subdominant role with respect to, for example, the decays of heavy mesons.

\subsection{LLP Production in Hard Scatterings}

At the parton level, the production of LLPs can also go through a variety of hard scattering processes, as illustrated in the center right panel of \figref{feynman}. However, in the far forward region where FASER is, this production mode suffers from large uncertainties in the determination of PDFs at low momentum transfer $Q^2$ and low parton momentum fraction $x$. As a result, we will not take into account hard scattering processes when presenting the FASER reach for various LLP models. This difficulty can be overcome for $m_{\textrm{LLP}}\gtrsim 2~\gev$, where, for example, the Drell-Yan process can become the dominant production mechanism, as discussed in \cite{Berlin:2018jbm}.

\subsection{``Beam Dump'' Production from SM Particles Hitting the TAN}

Interestingly, particles produced at the IP that then hit the TAN can effectively produce fixed-target beam dump experiments that can produce LLPs.  In particular, this has been illustrated in Ref.~\cite{Feng:2018noy} for the case of ALPs coupling to two photons. Such ALPs can be dominantly produced in the Primakoff process, $\gamma N\rightarrow aN$, through the exchange of a virtual photon (see right panel of \figref{feynman}), when high-energy photons produced at the IP travel $\sim 140~\m$ and hit the TAN. Given the $\sim 10^{16}$ forward-going photons that will hit the TAN during LHC Run 3, a large number of boosted forward-going ALPs could be produced. LLPs produced at the TAN travel only 340 m to FASER, which can also boost event rates.  Similarly, dark gauge bosons $V$ can be produced in photon collisions with the TAN through dark Compton scattering $\gamma e^-\rightarrow V e^-$ (see, e.g., Ref.~\cite{Liu:2017htz}), but this process is subdominant with respect to other production mechanisms.

\subsection{Number of Signal Events}

For an LLP with mass $m$ produced at the IP with momentum $p$ and angle $\theta$ with respect to the beam axis, the  probability that it will decay within the detector volume of FASER is
\be
\mathcal{P}(p,\theta) = \left( e^{-(L-\Delta)/d }- e^{-L/d} \right) \Theta(R-\tan\theta L) \approx \frac{\Delta}{d} e^{-L/d}  \Theta(R- \theta L) \ ,
\label{eq:decayinvolume}
\ee
where $\Theta$ is the Heaviside step function, $L$, $R$, and $\Delta$ define the geometry of the detector, as discussed in \secref{faser-benchmark}, and $d=c\tau\beta\gamma = c \tau p/m$ is the LLP's decay length in the lab frame, where $\tau$ is the LLP's lifetime. 
The first term in the brackets corresponds to the probability that LLP will decay within the $(L-\Delta,L)$ interval, and the second term enforces the angular acceptance of the detector. Given this probability, the total number of LLP decays inside FASER is
\be
N = \mathcal{L}  \int dp \ d\theta \ \frac{d\sigma_{p p \to \llp+X}}{dp \ d\theta}  \times  \mathcal{P}(p,\theta)\ .
\ee
In the following, we assume that possible decays of LLPs into invisible dark sector particles are either absent, e.g., due to kinematics, or suppressed, so that they do not affect the visible event rate in the detector. We will also assume a $100\%$ detection efficiency for all the visible decay modes for a better comparison with other experiments. An extended discussion of this issue is provided in \secref{signalefficiency} and will be a subject of future studies.

\section{FASER Reach for Dark Vectors}
\label{sec:vectors}
\setlength{\abovecaptionskip}{-15pt}

Among the best motivated LLPs with renormalizable couplings are those predicted in models with an additional U(1) symmetry and a corresponding vector field $X_\mu$ that couples through kinetic mixing to the hypercharge gauge boson or, at low energies, effectively to the SM photon~\cite{Holdom:1985ag}. The resulting new gauge boson is called the dark photon. Such a scenario can be motivated by simple extensions of the SM that involve dark matter~\cite{Battaglieri:2017aum}. 

Alternatively, new gauge bosons are predicted if one of the anomaly-free global symmetries of the SM is gauged.  (See Ref.~\cite{Bauer:2018onh} for a recent review.) These can be the U(1)$_{B-L}$ or U(1)$_{L_i-L_j}$ gauge bosons, where $B$, $L$, and $L_i$ are baryon, lepton, and lepton family number, respectively, with $i = e,\mu,\tau$. In the $B-L$ case, right-handed neutrinos are required to cancel the anomaly. In all of these cases, a new gauge boson $X_{\mu}$ couples with coupling $g_X$ to the SM current $j^X_\mu$, where $j^X_\mu$ involves SM fermions charged under the appropriate U(1) symmetry. 
 
In general, new gauge bosons can couple to SM currents and also kinetically mix with the hypercharge gauge boson.  A general Lagrangian for interactions between vectors $X_{\mu}$ and the SM is, then,
\be
\mathcal{L} = \mathcal{L}_{\text{SM}}
+\mathcal{L}_{\text{DS}} 
+ \frac12 \textcolor{Red}{m_{X}^{\textcolor{black}{2}}} X^\mu X_\mu\, - \red{g_X} j_\mu^X X^\mu - \frac{\red{\epsilon}}{2 \cos\theta_W} B_{\mu\nu} X^{\mu\nu}  \ ,
\label{eq:LgeneralAprim}
\ee
where  $\mathcal{L}_{\text{SM}}$ is the SM Lagrangian, $\mathcal{L}_{\text{DS}}$ is the dark sector Lagrangian involving only non-SM states, $m_X$ is the mass parameter of the new gauge boson, $g_X$ parametrizes the coupling to SM currents, and $\epsilon$ parametrizes the kinetic mixing term. Note that, even if the kinetic mixing term is absent at tree level, it can be loop-induced by fields charged under both gauge groups. Importantly, even if the kinetic mixing term is forbidden by, for example, embedding $U(1)_X$ in a larger, non-Abelian gauge group, non-zero values of $\epsilon$ can be induced at loop level when the larger gauge group is broken. 

In the following, we present FASER's reach for new light gauge bosons in three simple cases. We begin in \secref{vanillaAprim} with dark photons, where the only coupling between the new gauge boson and the SM is through kinetic mixing. We then discuss scenarios with $U(1)_{B-L}$ and $U(1)_{L_i-L_j}$ gauge bosons, where there is no kinetic mixing at tree-level in \secsref{BminusLAprim}{LiminusLjAprim}, respectively.

\subsection{Benchmark V1: Dark Photons}
\label{sec:vanillaAprim}

The dark photon Lagrangian extends the SM Lagrangian with the following terms:
\be
\mathcal{L} \supset -\frac{\epsilon'}{2} F_{\mu\nu}F'^{\mu\nu} + \frac{1}{2} {m'}^2 X^2\ ,
\ee
where $F_{\mu\nu}$ and $F'_{\mu\nu}$ are the field strength tensors for the SM photon and a new gauge boson $X$, respectively. After rotating to the mass basis, the dark photon--SM fermion coupling parameter is given by $\epsilon=\epsilon'\cos\theta_W$, cf. \eqref{LgeneralAprim}. (See, e.g., Appendix A of Ref.~\cite{Bauer:2018onh} for a detailed discussion.) The kinetic mixing parameter is naturally small if it is induced by loops of new heavy charged particles. After a field re-definition to remove the kinetic mixing term, the dark photon $A'$ emerges as a physical mass eigenstate that couples to the charged SM fermions proportional to their charges through
\be
\mathcal{L} \supset  \frac12 \textcolor{Red}{m_{A'}^{\textcolor{black}{2}}} A'^2
- \textcolor{Red}{\epsilon} \, e \sum_f q_f \bar{f} \slas{A'}   f \ .
\ee
The parameter space of the model is spanned by the dark photon mass $m_{A'}$ and the kinetic mixing parameter $\epsilon$. 

\begin{figure}[tbp]
\centering
\vspace*{-0.6cm}
\includegraphics[width=0.98\textwidth]{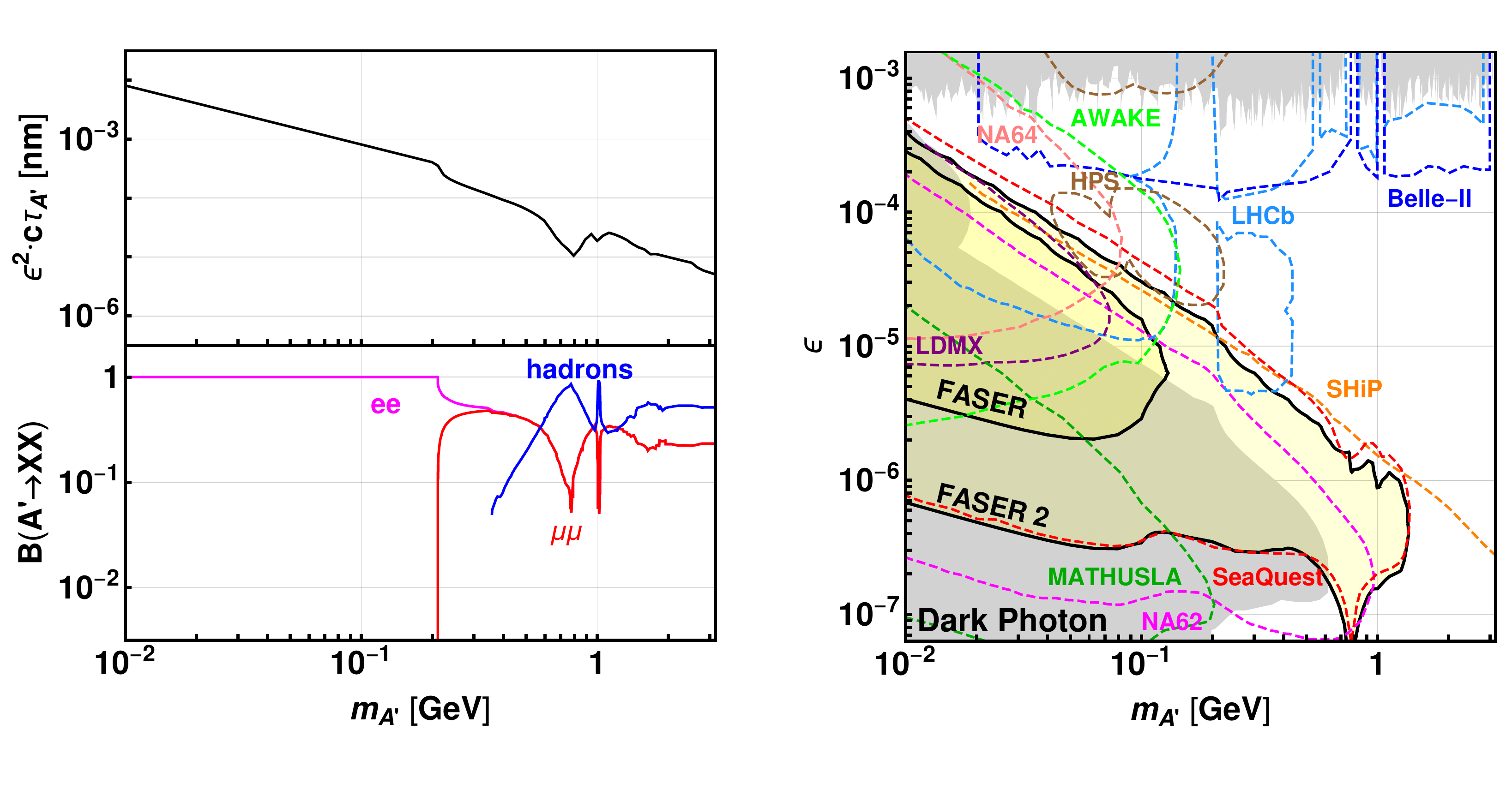} 
\caption{
{\bf Benchmark Model V1.} The dark photon decay length (top left panel), its branching fractions  into hadronic and leptonic final states (bottom left panel) and FASER's reach (right panel). In the right panel, the gray-shaded regions are excluded by current bounds, and the projected future sensitivities of other experiments are shown as colored contours.  See the text for details.}
\label{fig:v1}
\end{figure}

\begin{description}

\item [Production] Light dark photons are mainly produced through decays of light mesons, $\pi,\eta \to \gamma A'$ and through dark bremsstrahlung. To a good approximation, these processes are suppressed by $\epsilon^2$ relative to their SM counterparts.

\item [Decay and Lifetime] Dark photons can decay into all kinematically accessible light charged states, but, especially for $m_{A'}$ below a few hundred \mev, they mainly decay into $e^+e^-$ and $\mu^+\mu^-$ pairs. Heavier $A'$s have various hadronic decay modes, but they are typically dominated by decays into $\pi^+\pi^-$. 
The decay width is proportional to $\epsilon^2$. Thanks to this, dark photons naturally have decay lengths that are large enough for them to be observed in FASER, especially when they are highly boosted by the large energies they inherit from $pp$ collisions at the LHC. The dark photon decay length and branching fractions into leptonic and hadronic final states are shown in the left panel of \figref{v1}, following Refs.~\cite{Bauer:2018onh, Buschmann:2015awa}. 

\item [Results] The projected dark photon sensitivity reaches for FASER at LHC Run 3 with $150~\ifb$ and FASER 2 at HL-LHC with $3~\iab$ are shown in the right panel of \figref{v1}. The gray-shaded regions are excluded by current bounds; see Refs.~\cite{Bauer:2018onh,Beacham:2019nyx} and references therein. For comparison we also show the projected sensitivities of other experiments: NA62 assumes $10^{18}$ protons on target (POT) while running in a beam dump mode that is being considered for LHC Run 3~\cite{Dobrich:2018ezn}; SeaQuest assumes $1.44 \times 10^{18}$ POT, which could be obtained in two years of parasitic data taking and requires additionally the installation of a calorimeter~\cite{Berlin:2018pwi}; the proposed beam dump experiment SHiP assumes $\sim 2 \times 10^{20}$ POT collected in 5 years of operation~\cite{Alekhin:2015byh}; the proposed electron fixed-target experiment LDMX during Phase II with a beam energy of $8~\gev$ and $10^{16}$ electrons on target (EOT)~\cite{Berlin:2018bsc}; Belle-II and LHCb assume the full expected integrated luminosity of $50~\iab$~\cite{Dolan:2017osp} and $15~\ifb$~\cite{Ilten:2015hya, Ilten:2016tkc}, respectively; HPS assumes 4 weeks of data at JLab at each of several different beam energies~\cite{Moreno:2018xxx, Battaglieri:2017aum}; NA64~\cite{Gninenko:2320630} corresponds to $5\times 10^{12}$ EOT with $100~\gev$ energy; and AWAKE~\cite{Caldwell:2018atq} is assumed to be working as a fixed-target experiment with a 10-m-long decay volume and $10^{16}$ EOT accelerated in a $50-100~\m$ long plasma cell to the energy $\mathcal{O}(50~\gev)$.

As can be seen, already during LHC Run 3, FASER will be able to probe interesting regions of the dark photon parameter space. In the HL-LHC era, FASER 2 will extend the reach to masses above a GeV and explore a large swath of parameter space with $\epsilon\sim 10^{-7}-10^{-4}$. 

Combining the dependence on $\epsilon$ in both the production rate and the decay width, one can see that in the regime of large lifetime, the low $\epsilon$ boundary, the total number of signal events in the detector scales as $\epsilon^4$. On the other hand, for lower lifetime, which corresponds to the high $\epsilon$ boundary of the region covered by FASER, the number of signal events becomes exponentially suppressed once the $A'$ decay length drops below the distance to the detector. As a result, in this region of the parameter space, the reach of FASER is similar to other, even much larger, proposed detectors.

\end{description}

\subsection{Benchmark V2: $B-L$ Gauge Bosons}
\label{sec:BminusLAprim}

In the absence of kinetic mixing, the $B-L$ gauge boson Lagrangian is 
\be
\mathcal{L} \supset  \frac12 \textcolor{Red}{m_{A'}^{\textcolor{black}{2}}} A'^2-  \red{g_{B-L}} \sum_f Q_{B-L,f} \bar{f}  \slas{A'}   f\ ,
\ee
where $Q_{B-L,f}$ is the $B-L$ charge of fermion $f$. The parameter space is spanned by the gauge boson mass $m_{A'}$ and the coupling $g_{B-L}$. 

\begin{description}
\item [Production] As in the case of the dark photon, a light $B-L$ gauge boson is mainly produced through light meson decays and dark bremsstrahlung. The corresponding production rates are proportional to $g_{B-L}^2 Q_{B-L}^2$.

\item [Decay and Lifetime] $B-L$ gauge bosons decay into all kinematically accessible states with $B-L$ charge.  Light $B-L$ gauge bosons decay mainly into neutrinos, $e^+e^-$, $\mu^+\mu^-$ and $\pi^+\pi^-$, with the decay widths proportional to $g_{B-L}^2 Q_{B-L}^2$. When deriving the results presented below, we use the decay width obtained in Refs.~\cite{Bauer:2018onh,Ilten:2018crw} and include only the visible final states (not the  neutrino final states) in the signal event rates presented below. We show the decay width and branching fractions in the left panel of \figref{v2}.

\item [Results] The projected $B-L$ gauge boson sensitivity reaches for FASER at LHC Run 3 with $150~\ifb$ and FASER 2 at HL-LHC with $3~\iab$ are shown in the right panel of \figref{v2}. Here we only consider the decays into visible final states, while decays into neutrinos do not contribute to the sensitivity. Both the existing constraints (gray shaded areas, see Ref.~\cite{Bauer:2018onh} and references therein) and the projected sensitivities of other proposed searches have been adapted from Refs.~\cite{Bauer:2018onh,Ilten:2018crw}. Besides recasting the dark photon search sensitivity at Belle-II~\cite{Dolan:2017osp}, LHCb~\cite{Ilten:2015hya, Ilten:2016tkc}, SeaQuest~\cite{Gardner:2015wea} and SHiP~\cite{Alekhin:2015byh}, they include additionally search strategies utilizing the $A' \to \nu\nu$ decay channel at Belle-II and NA64~\cite{Gninenko:2018tlp}. In particular, NA64-$\mu$ is a modified version of NA64 that assumes an upgraded muon beam at the CERN SPS delivering up to $10^{12}$ muons.  Additionally, a search utilizing $A' \to \nu\nu$ has been suggested for the proposed electron fixed target experiment LDMX during Phase II with a beam energy of $8~\gev$ and $10^{16}$ EOT~\cite{Berlin:2018bsc}. Furthermore, $B-L$ gauge bosons may be probed by the coherent neutrino scattering experiment MINER, assuming a germanium target with an exposure of $10^4~\kg \cdot \text{days}$, an energy threshold of $100~\ev$, and an assumed background of approximately 100 events per day per kg per keV~\cite{Dent:2016wcr}.

As can be seen, as in the dark photon case, both FASER and FASER 2 can probe currently unconstrained regions of the parameter space with FASER 2 extending the reach above $m_{A'}\sim 1~\gev$.

\end{description}

\begin{figure}[tbp]
\centering
\vspace*{-0.6cm}
\includegraphics[width=0.98\textwidth]{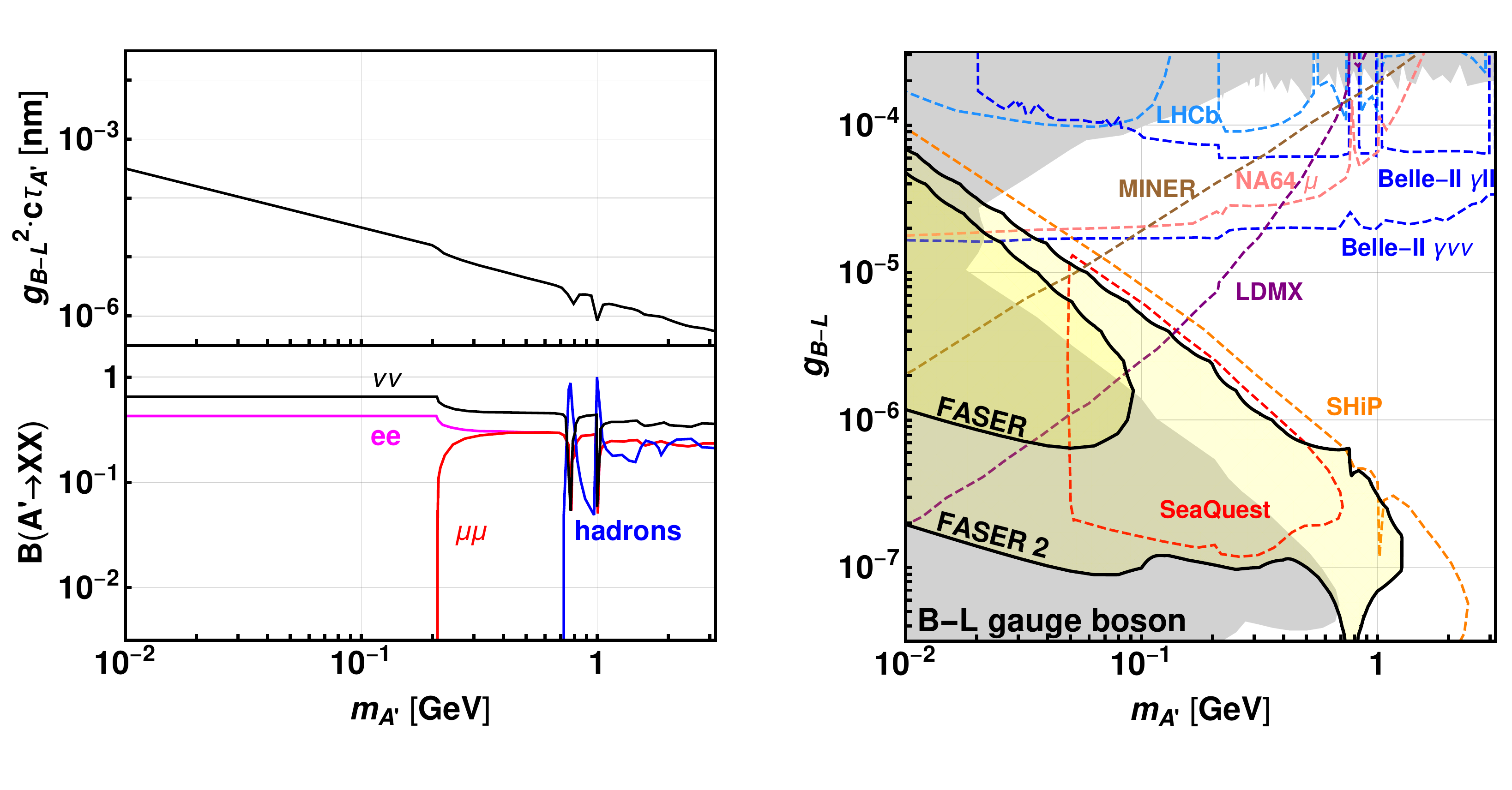} 
\caption{
{\bf Benchmark Model V2.} As in \figref{v1}, but for the $B-L$ gauge boson. In the right panel, projected future sensitivities of other experiments are shown following Ref.~\cite{Bauer:2018onh}. }
\label{fig:v2}
\end{figure}

\subsection{Benchmark V3: $L_i-L_j$ Gauge Bosons}
\label{sec:LiminusLjAprim}

In the absence of tree-level kinetic mixing, the $L_i - L_j$ gauge boson Lagrangian is 
\be
\mathcal{L} \supset  \frac12 \textcolor{Red}{m_{A'}^{\textcolor{black}{2}}} A'^2-  \red{g_{ij}} \sum_{\ell=i,j}  \bar{\ell}  \slas{A'} \ell \ .
\ee
At tree level, there is, of course, no coupling to hadrons.  However, since hadron decays are among the leading production mechanisms at the LHC, it is important to include the coupling to hadrons induced at loop level, unlike the $B-L$ case.  Because the new gauge boson couples to charged SM leptons, it also mixes with the photon at one-loop level. The resulting effective kinetic mixing parameter is~\cite{Bauer:2018onh} 
\be
\epsilon_{ij}(g_{ij},m_{A'})= \frac{e g_{ij}}{4 \pi^2} \int_0^1 dx \ 3  x (1-x) \log\left(  \frac{m_i^2 + m_{A'}^2 x(1-x)}{m_j^2 + m_{A'}^2 x(1-x)}\right) .
\ee
This non-zero kinetic mixing then generates couplings of the new gauge boson to all the SM fermions. Note that the loop-induced kinetic mixing parameter is suppressed with respect to the $g_{ij}$ coupling, since $\epsilon_{ij}^2 \sim (e/4\pi^2)^2\, g_{ij}^2 \sim (\alpha/4\pi^3)\, g_{ij}^2$.  An effective Lagrangian is therefore given by
\be
\mathcal{L} \supset  \frac12 \textcolor{Red}{m_{A'}^{\textcolor{black}{2}}} A'^2-  \red{g_{ij}} \sum_{\ell=i,j}  \bar{\ell}  \slas{A'} \ell 
- \epsilon_{ij}(\red{g_{ij}},\red{m_{A'}}) \, e \sum_{f}  q_f \bar{f} \slas{A'} f \ .
\ee
The parameter space is spanned by the gauge boson mass $m_{A'}$ and the couplings $g_{ij}$. 

\begin{description}
\item [Production] The production of $L_i-L_j$ gauge bosons at the LHC proceeds similarly to the dark photon; that is, it is mainly produced through decays of the light neutral mesons, $\pi^0,\eta \to \gamma A'$, and dark bremsstrahlung. However, as discussed above, the couplings to quarks are suppressed. The resulting production rate is proportional to $\epsilon_{ij}^2 e^2 \sim (\alpha/\pi)^2 g_{ij}^2$ and therefore significantly reduced with respect to the dark photon scenario discussed in \secref{vanillaAprim}. Additionally, the $L_i-L_j$ gauge bosons can also be produced in charged mesons decay $\pi^\pm,K^\pm \to \ell \nu A'$, in which case the gauge boson is radiated off the lepton or neutrino and the decay width is proportional to $g_{ij}$~\cite{Ibe:2016dir}. The largest contribution is provided by the decay $K^\pm \to \ell \nu A'$, which is sizable but still subleading compared to the light meson decays because of the small probability of the kaon to decay before being deflected by the first quadrupole magnet.

\item [Decay and Lifetime] A light $L_i-L_j$ gauge boson decays mainly into the charged leptons $i,j$ and the corresponding neutrinos. The decay widths are proportional to $g_{ij}^2$. In the following, we only take into account leptonic decays that are dominant with respect to the SM hadronic ones. The relevant decay lengths and branching fractions as functions of $m_{A'}$ for $L_\mu-L_e$ and $L_e-L_\tau$ gauge bosons are shown in the left panels of \figref{v3m-e} and \figref{v3e-t}, respectively.

\item [Results] The expected reaches for $L_\mu-L_e$ and $L_e-L_\tau$ gauge bosons are shown in the right panels of \figref{v3m-e} and \figref{v3e-t}, respectively. Here we only consider the decays into electrons, while decays into neutrinos remain invisible. Both the existing constraints (gray shaded area) and the projected sensitivities of SHiP, Belle-II and NA64-$\mu$ have been adapted from Ref.~\cite{Bauer:2018onh} and references therein. (See also the discussion in~\secref{BminusLAprim}.) 
Given the highly suppressed production rate for these gauge bosons with only lepton couplings at tree level, FASER 2 does not probe new parameter space.

\end{description}

\begin{figure}[tbp]
\centering
\vspace*{-0.6cm}
\includegraphics[width=0.98\textwidth]{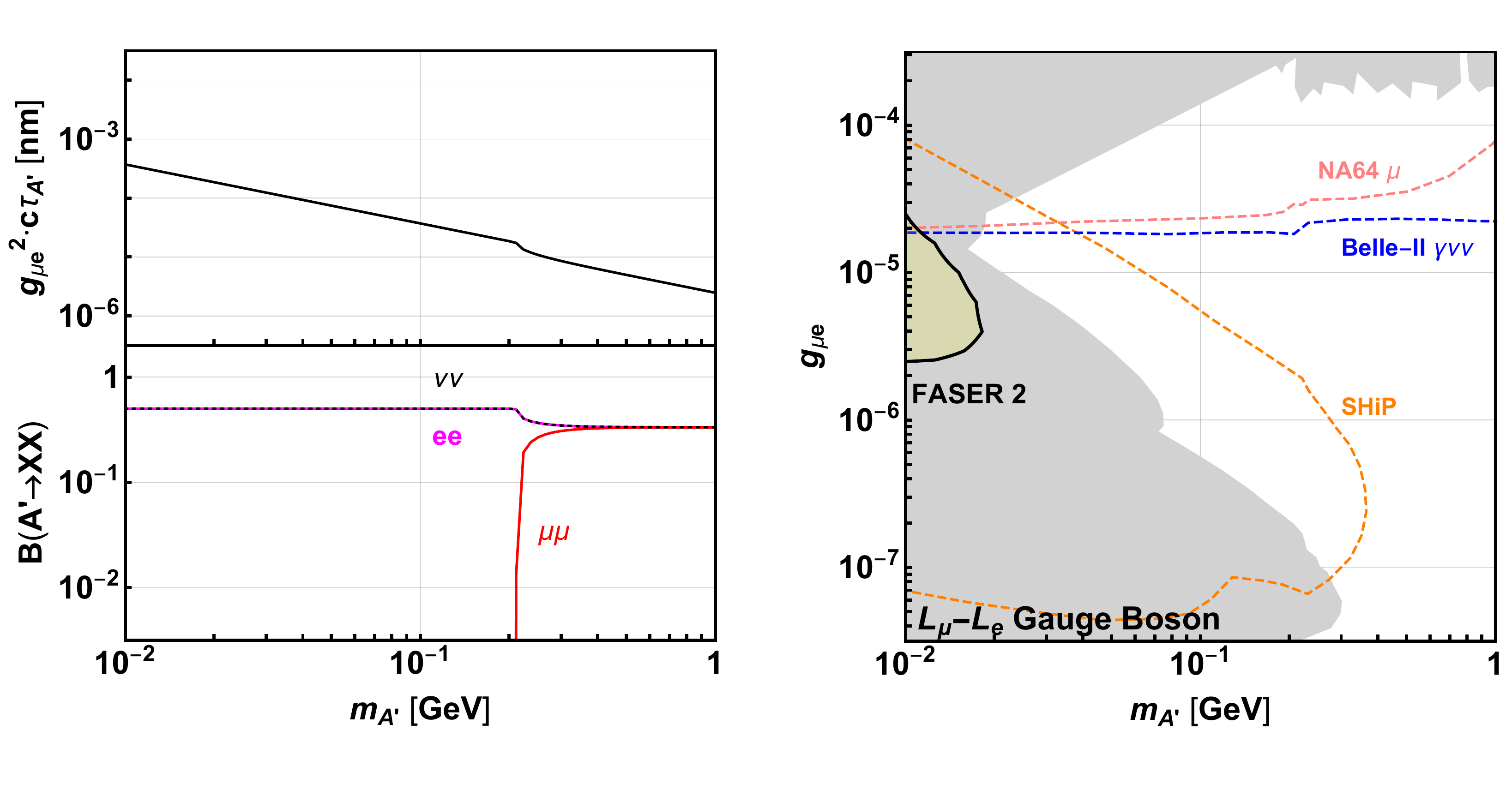} 
\caption{
{\bf Benchmark Model V3.} As in \figref{v1}, but for the $L_\mu-L_e$ gauge boson. In the right panel, the gray-shaded regions excluded by current bounds and projected future sensitivities of other experiments are adapted from Ref.~\cite{Bauer:2018onh}. 
}
\label{fig:v3m-e}
\end{figure}
\begin{figure}[tbp]
\centering
\vspace*{-0.6cm}
\includegraphics[width=0.98\textwidth]{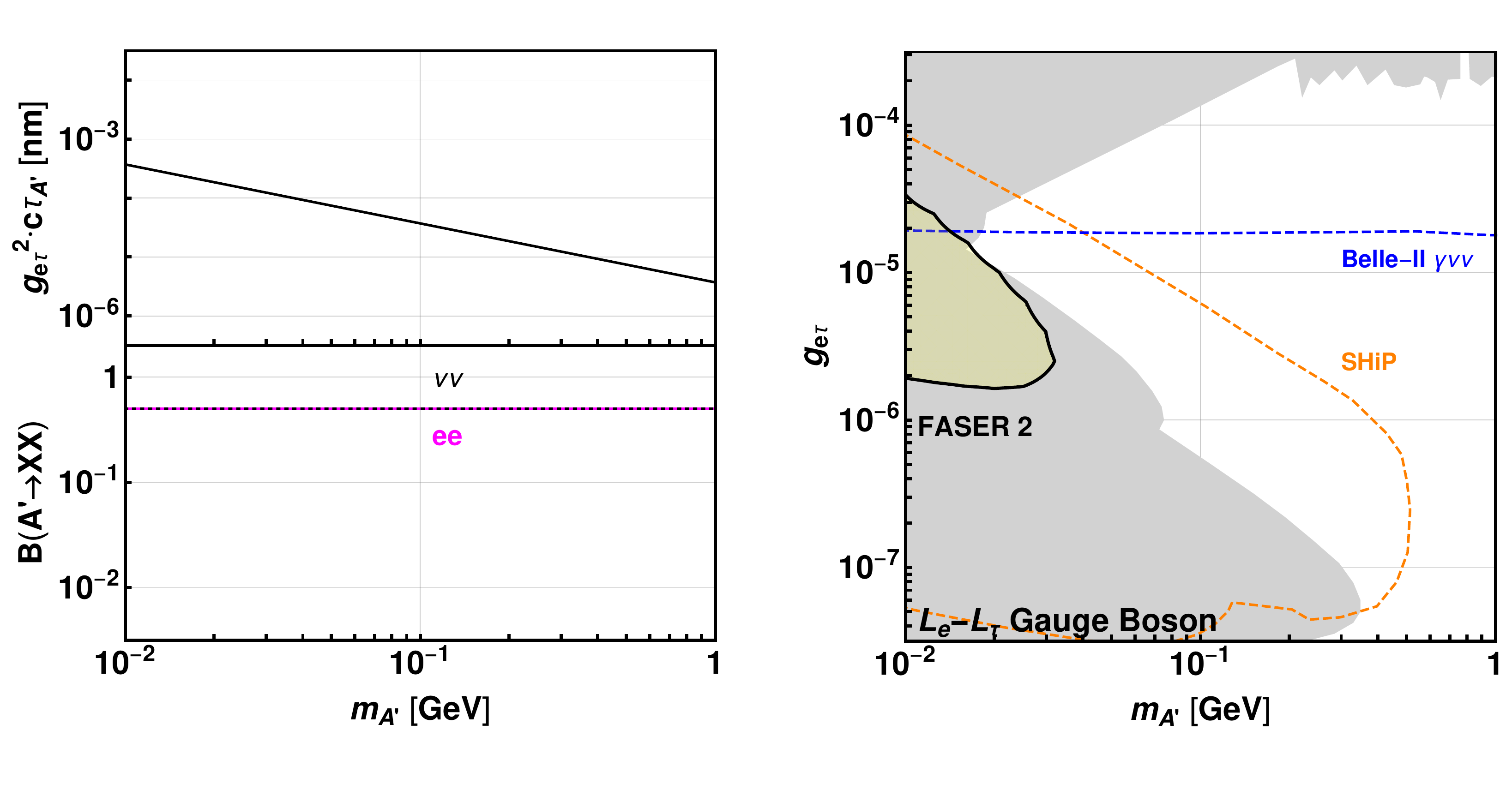} 
\caption{
{\bf Benchmark Model V3.} As in \figref{v1}, but for the $L_e-L_\tau$ gauge boson.  In the right panel, the gray-shaded regions excluded by current bounds and projected future sensitivities of other experiments are adapted from Ref.~\cite{Bauer:2018onh}.  
}
\label{fig:v3e-t}
\end{figure}

\section{FASER Reach for Dark Scalars}
\label{sec:scalars}

Another widely discussed renormalizable portal between the dark sector and the SM is a scenario with a new scalar particle $S$ with quartic couplings to the SM Higgs, $H$. A simple corresponding Lagrangian is
\be
\mathcal{L} = \mathcal{L}_{\text{SM}}+\mathcal{L}_{\text{DS}} +\mu_S^2 S^2 - \frac{1}{4} \lambda_S S^4 - \epsilon S^2 |H|^2 \ ,
\label{eq:LdarkHiggs}
\ee
where terms with an odd number of dark scalars $S$ are assumed suppressed, for example, by a discrete symmetry. The quartic term in \eqref{LdarkHiggs} induces mixing between the dark scalar and the SM Higgs boson once both get non-zero vacuum expectation values (vevs) and $S=(v_S + s)/\sqrt{2}$ and $H=(v_h + h_{\textrm{SM}})/\sqrt{2}$, where $v_S$ and $v_h$ correspond to vevs of the $S$ and $H$ fields, respectively. After diagonalization, the physical fields are the 125 GeV SM Higgs boson, $h$, and a scalar $\phi$, often called the dark Higgs boson.  In terms of the gauge eigenstates, the physical fields are
\be
h_{SM} = \phi \sin\theta + h \cos\theta  \quad \text{and} \quad s = \phi \cos\theta - h \sin\theta \ , 
\label{eq:darkHiggsmixing}
\ee
where the mixing angle $\theta \sim v_h/v_S\ll 1$ must be small to satisfy current experimental constraints. This can be achieved by assuming large $v_S$, while the dark Higgs boson can be made light with $\mphi\ll m_h$ by suppressing the coupling $\mu_S$ and tuning the quartic couplings to be $\epsilon, \lambda_S\ll 1$~\cite{Bezrukov:2009yw}.  Alternatively, if $S$ does not get a non-zero vev and the trilinear term $\delta_1 S |H|^2$ is explicitly introduced in the Lagrangian along with the quartic term $\epsilon S^2\,|H|^2$, a small mixing angle $\theta\simeq \delta_1\,v_h/(m_h^2-\mphi^2)$ can be achieved by suppressing the coupling $\delta_1$~\cite{OConnell:2006rsp}, where $m_h$ and $\mphi$ are the SM Higgs and dark Higgs masses, respectively. 

The Higgs-dark Higgs mixing generates Yukawa-like couplings between the SM fermions and the dark Higgs boson. In addition, there can appear a non-negligible trilinear interaction term between $\phi$ and $h$ with the corresponding coupling denoted by $\lambda$. The effective Lagrangian can, then, be written as
\be
\mathcal{L} = -  \textcolor{Red}{\mphi^{\textcolor{black}{2}}} \phi^2 - \sin  \textcolor{Red}{\theta} \, \frac{m_f}{v} \, \phi \bar{f} f  
-  \textcolor{Red}{\lambda} \, v \,h \phi \phi\ +\ldots \ ,
\label{eq:darkHiggseffL}
\ee
where cubic and quartic terms involving $\phi$ and $h$ have been omitted. Note that the dark scalar coupling to SM fermions can also be generated in other ways, e.g., by coupling the dark scalar to additional vector-like fermions that mix with the SM ones.

In the following, we analyze FASER's sensitivity to dark Higgs bosons. We consider cases with vanishing and sizable values of $\lambda$ in \secsref{s1} {s2}, respectively.

\subsection{Benchmark S1: Dark Higgs Bosons}
\label{sec:s1}

We first focus on the dark Higgs boson with trilinear coupling $\lambda=0$. The parameter space of the model is then spanned by the dark Higgs mass $m_{\phi}$ and mixing angle $\theta$. 

\begin{description}
\item [Production] For FASER, a light dark Higgs is mainly produced through rare $B$-meson decays 
with the corresponding branching fraction given by~\cite{Grinstein:1988yu, Chivukula:1988gp, Feng:2017vli}
\be
B (B \to X_s \phi) = 5.7 \left(1-\frac{m_\phi^2}{m_b^2}\right)^2 \theta^2 \ . 
\ee
In the following, we neglect additional contributions from kaon decays that are sizable only in the region of the parameter space that is already strongly constrained by other experiments. Decays of D-mesons into scalars are further suppressed due to the absence of top loops mediating such process. 

\item [Decay and Lifetime] The dark Higgs boson mainly decays into the heaviest kinematically-available SM states $f$ with decay widths proportional to $\theta^2 m_f^2/v^2$. This induces sharp threshold effects in both the decay width and branching fractions, which are shown in the left panel of \figref{s1}.  There are large uncertainties in the modeling of the corresponding hadronic decay widths in the few GeV mass range. In the following, we adopt the numerical results of Ref.~\cite{Bezrukov:2013fca}. For the low-mass range, $2m_\pi < m_\phi  < 1~\gev$, these employ the results of chiral perturbation theory~\cite{Donoghue:1990xh}, for the large-mass range, $m_\phi >  2.5~\gev$, they use the spectator model~\cite{Gunion:1989we, McKeen:2008gd}, and in the intermediate-mass range, $1~\gev < \mphi<  2.5~\gev$, the hadronic branching fraction is obtained by interpolating between these two. A recent evaluation of the decay width and branching fractions of a light scalar~\cite{Winkler:2018qyg} shows good agreement with the description used in this work.

\item [Results] The expected reach of FASER for dark Higgs bosons is shown in the right panel of \figref{s1} along with the current bounds (see Ref.~\cite{Beacham:2019nyx} and references therein) and the projected sensitivities of other ongoing and future experiments. As discussed in~\secref{vanillaAprim}, the sensitivity line for NA62 assumes $10^{18}$ protons on target (POT) for the experiment running in a beam dump mode that is being considered for LHC Run 3~\cite{Dobrich:2018ezn}; SeaQuest assumes $1.44 \times 10^{18}$ POT, which could be obtained in two years of parasitic data taking and requires additionally the installation of a calorimeter~\cite{Berlin:2018pwi}; and the proposed beam dump experiment SHiP assumes $\sim 2 \times 10^{20}$ POT collected in 5 years of operation~\cite{Alekhin:2015byh}. The projected sensitivity line for LHCb follows Ref.~\cite{Gligorov:2017nwh} and assumes a zero-background search with the full expected integrated luminosity of  $300~\ifb$. This is also the case for the proposed CODEX-b detector~\cite{Gligorov:2017nwh}, while the corresponding line further assumes that the whole $10~\m\times 10~\m\times 10\m$ fiducial volume is hidden behind 25 radiation lengths of lead shielding to suppress background. The reach for the proposed MATHUSLA experiment~\cite{Evans:2017lvd,Curtin:2018mvb} assumes a $200~\m\times 200~\m\times 20\m$ size detector collecting $3~\iab$ of integrated luminosity at the HL-LHC. A portion of the currently unconstrained parameter space below the $m_K-m_\pi$ threshold can also be covered by the proposed KLEVER experiment~\cite{Ambrosino:2019qvz}, for which the corresponding reach~\cite{Beacham:2019nyx} assumes $5\times10^{19}$ POT from the $400~\gev$ SPS beam.

Since dark Higgs bosons are produced mainly in rare decays of $B$ mesons, they have a larger angular spread than dark vectors. As a result, the sensitivity reach for dark Higgs bosons is greatly improved by increasing the detector radius from 10 cm at FASER to 1 m at FASER 2.  At $\mphi \sim 1 \gev$, FASER 2 is sensitive to $\theta \sim 10^{-5} - 10^{-4}$ and is highly complementary to other proposed experiments, such as MATHUSLA, Codex-b, and SHiP.\

\begin{figure}[tbp]
\centering
\vspace*{-0.6cm}
\includegraphics[width=0.98\textwidth]{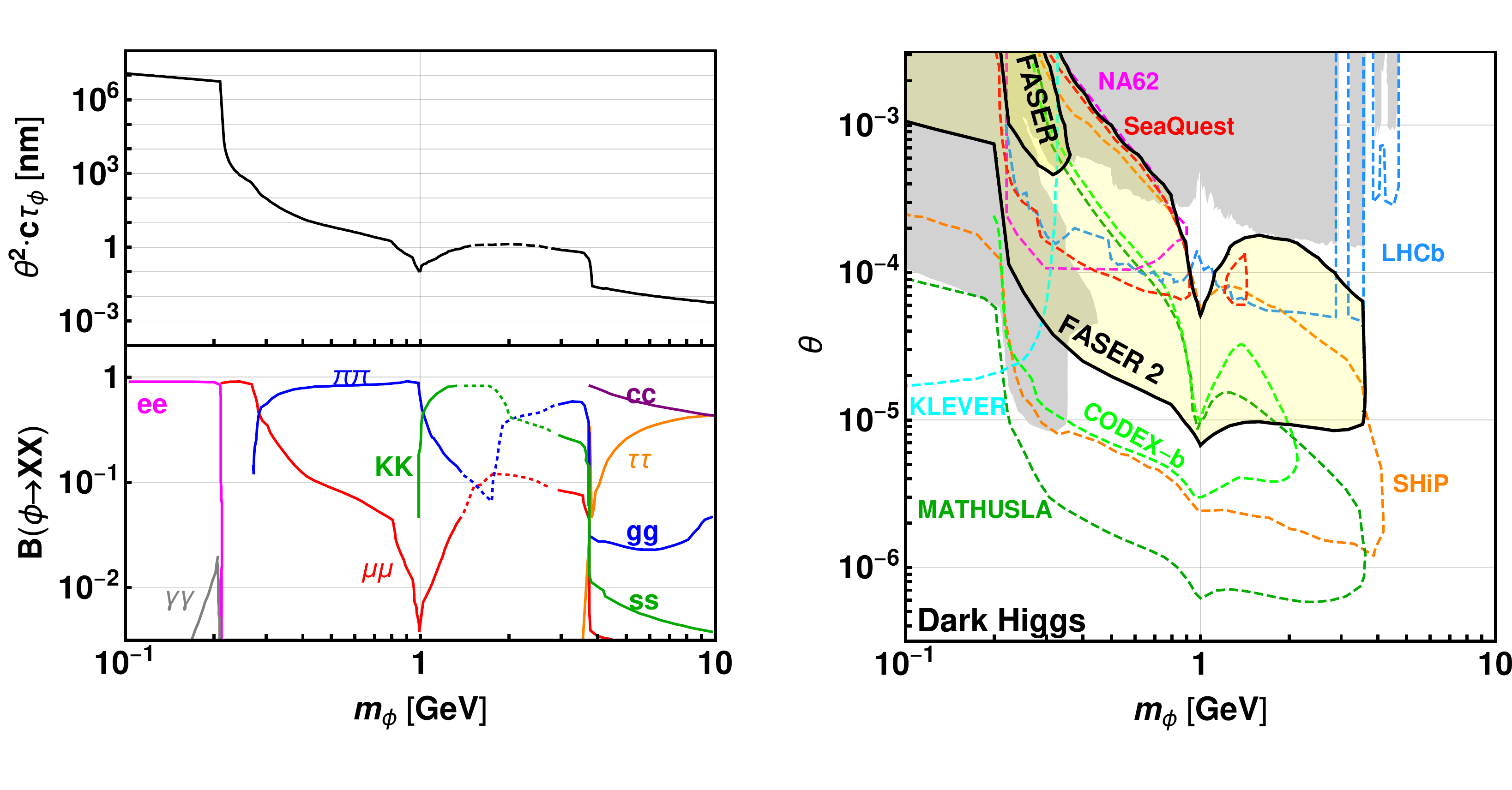} 
\caption{
{\bf Benchmark Model S1.} The decay length (top left panel), decay branching fractions (bottom left panel), and FASER's reach (right panel) for the dark Higgs boson with negligible trilinear coupling to the SM Higgs. The gray shaded regions are excluded, and the colored contours are the projected sensitivities of other proposed experiments; see text for details.}
\label{fig:s1}
\end{figure}
\end{description}

\subsection{Benchmark S2: Dark Higgs Bosons with Large Trilinear Couplings}
\label{sec:s2}

If the trilinear coupling $\lambda$ in \eqref{darkHiggseffL} is large, the dark Higgs boson can also be produced in pairs from intermediate real or virtual SM Higgs bosons, $h^{(\ast)} \to \phi \phi$. The parameter space of the model is then spanned by the dark Higgs mass $m_{\phi}$, the mixing angle $\theta$, and the trilinear coupling $\lambda$. 

\begin{description}
\item [Production] The dark Higgs boson in this model can be still be produced in rare meson decays, as in \secref{s1}, but now it can also be pair-produced by on- and off-shell SM Higgs bosons. For the latter mechanism, SM Higgs bosons can decay through $h \to \phi \phi$, yielding a signal of invisible Higgs decays that can be discovered at ATLAS or CMS or Higgs bosons decaying to LLPs, which can be discovered by MATHUSLA, for example.  However, the trilinear coupling also yields a new production mechanism for FASER, namely, rare $B$ decays to strange hadrons and an off-shell Higgs boson, leading to $B \to X_s h^* \to X_s \phi \phi$. The corresponding decay branching fraction is given by~\cite{Bird:2004ts,Altmannshofer:2009ma} 
\be
B(B \to X_s \phi \phi) = \frac{C^2 \lambda^2}{\Gamma_B} \frac{m_b^5}{256 \pi^3} \ f\left( \frac{m_\phi}{m_b} \right) , 
\ee
where $C=4.9 \times 10^{-8}~\gev^{-2}$, and $f$ is given by~\cite{Feng:2017vli}
\begin{equation}
f(x) = \frac{1}{3}  \sqrt{1-4x^2} (1 + 5x^2 - 6 x^4) -4 x^2 (1 - 2 x^2 + 2 x^4 ) \log\left[  \frac{1}{2x} \left(1 +\sqrt{1-4 x^2}\right) \right] .
\end{equation}

\item [Decay and Lifetime] If $\theta>0$, the dark Higgs can decay into SM fermions, and its decay width and branching fractions are as discussed in \secref{s1}. 
 
\item [Results] The expected reach of FASER 2 for dark Higgs bosons with sizable trilinear couplings is shown in the right panel of \figref{s2}. The shaded contours show results, the reach obtained from the dark Higgs pair production process only, for $\lambda = 0.0046, 0.0015$ corresponding to $B(h \to \phi\phi)\approx 4700 \lambda^2 = 10\%, 1\% $. The larger value is currently allowed.  The smaller value will be very challenging to probe through invisible Higgs decays even at the HL-LHC, but could be probed by other future colliders, such as the ILC~\cite{Aihara:2019gcq} and FCC~\cite{Mangano:2018mur}. 

As can be seen, the additional production mechanism through off-shell SM Higgs boson $B \to X_s \phi \phi$ allows FASER to probe parameter space reaching to lower values of the mixing angle $\theta$. One can probe values even as low as $\theta \sim 10^{-6}$ for $\mphi\simeq 1~\gev$ and $B(h \to \phi\phi)=0.1$. Of course, FASER 2 can also still see dark Higgs bosons produced through $B \to X_s \phi$; this region is also shown in \figref{s2} as the area enclosed by the dashed line.

The projected sensitivities for dark Higgs bosons without trilinear couplings, shown in \figref{s1}, also apply to this scenario. In \figref{s2} we therefore only show the projected sensitivities of proposed searches utilizing the trilinear coupling~\cite{Beacham:2019nyx,Evans:2017lvd}. In particular, both MATHUSLA and CODEX-b are expected to probe this scenario through the decay of a SM Higgs boson to two dark Higgs bosons, $h \to \phi\phi$. For sufficiently large mixing angles, this process allows these experiments to probe dark Higgs boson masses as large as $m_\phi = m_h/2$.

\end{description}

\begin{figure}[tbp]
\centering
\vspace*{-0.6cm}
\includegraphics[width=0.98\textwidth]{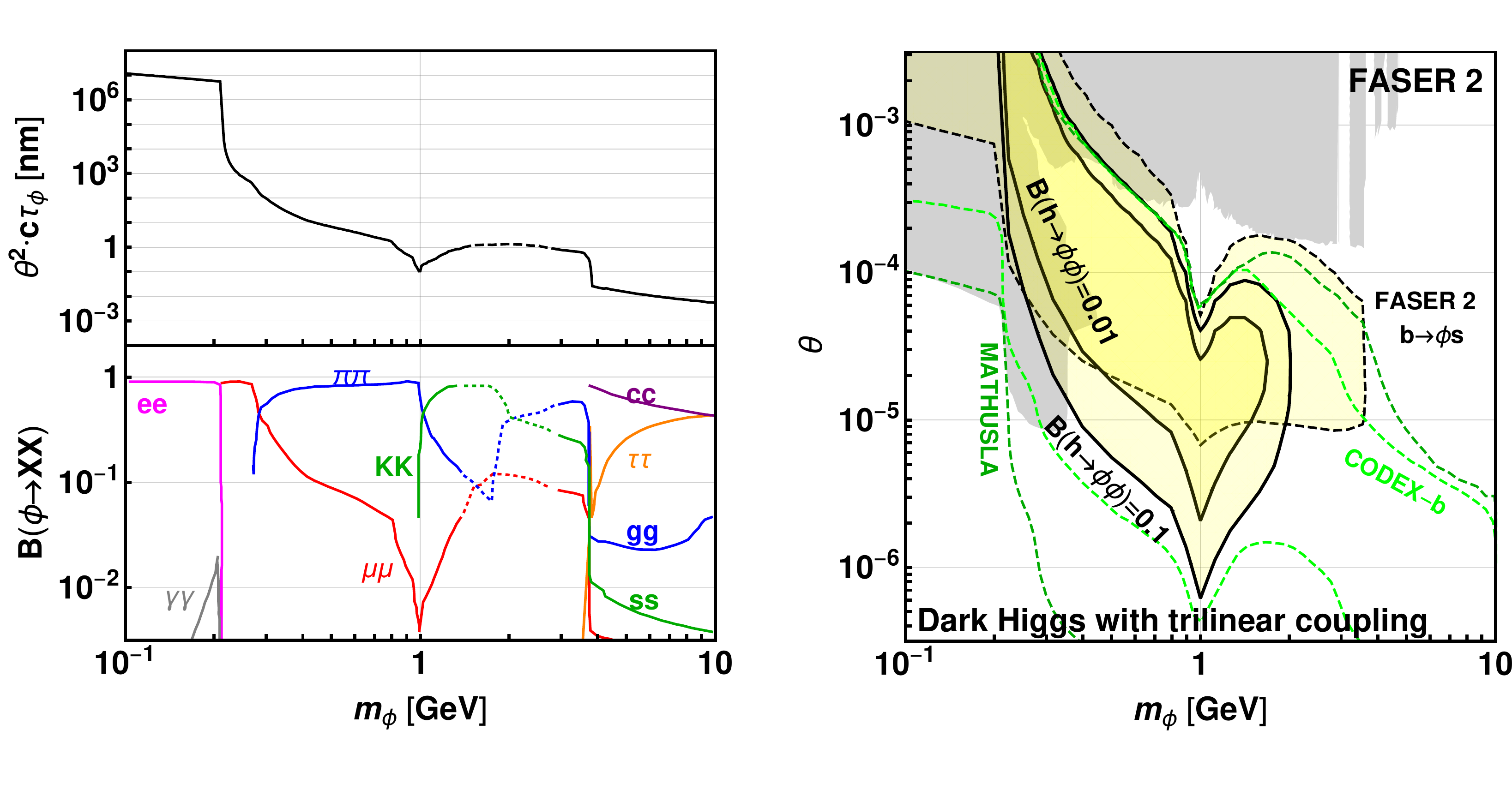} 
\caption{
{\bf Benchmark Model S2.} As in \figref{s1}, but the reach shown in the right panel is for dark Higgs bosons pair-produced through $B \to X_s \phi \phi$ with trilinear couplings $\lambda = 0.0046, 0.0015$ corresponding to $B(h \to \phi\phi) \approx 4700 \lambda^2 = 10\%, 1\%$, as indicated.  The region probed by $B \to X_s \phi$ is also shown by the dashed black line. The projected sensitivities of MATHUSLA and Codex-b to the trilinear couplings through the SM Higgs decay $h \to \phi \phi$ are also shown for $\lambda = 0.0046$. Note that the projected sensitivities of other experiments for vanishing trilinear coupling, $\lambda=0$, also apply; they are not shown in this figure, but can be found in \figref{s1}.} 
\label{fig:s2}
\end{figure}

\section{FASER Reach for Heavy Neutral Leptons}
\label{sec:hnls}

One of the best motivated candidates for new particles are new SM-singlet heavy neutral leptons (HNLs), or sterile neutrinos. See, for example, Ref.~\cite{Drewes:2015iva} for a recent review, and Refs.~\cite{Shrock:1978ft,Gallas:1994xp} for examples of early work on searches for HNLs at beam dumps.  In the minimal such case, the interaction Lagrangian can be written as 
\be
\mathcal{L} = \mathcal{L}_{\text{SM}}+\mathcal{L}_{\text{DS}} -  \sum y_{\alpha I} (\bar{L}_\alpha H) N_I\ ,
\ee
where the $y_{\alpha I}$ are Yukawa couplings, and the sum is over the three SM lepton doublets $L_\alpha$ and HNL fields $N_I$.  The dark sector might additionally contain both Dirac and Majorana mass terms for the HNL fields. 

After electroweak symmetry breaking and diagonalization of the mass terms, one finds a mixing of the SM neutrinos and HNLs.  This leads to a coupling of the HNLs to the $W$ and $Z$ bosons, with an effective Lagrangian
\be
\mathcal{L} \supset \bar{N}_I (i \slass{\partial}-\textcolor{Red}{m_{N,I}}) N_I - (g/\sqrt{2}) W_\mu \bar{\ell}_{L,\alpha} \gamma^\mu   \textcolor{Red}{U_{\alpha I}} N_I\ - (g/\sqrt{2} c_W) Z_\mu \bar{\nu}_{L,\alpha} \gamma^\mu   \textcolor{Red}{U_{\alpha I}} N_I\ .
\label{eq:LF}
\ee

\subsection{Benchmarks F1, F2, F3: HNLs Coupled to $e$, $\mu$, $\tau$}
\label{sec:F123}

We will now focus on a single HNL that couples to only one of the SM lepton doublets, either $L_e$, $L_\mu$ or $L_\tau$, resulting in three benchmarks: F1, F2, and F3. These models are described by only two parameters: the HNL mass $m_N$ and its non-zero mixing angle with the respective SM lepton doublet, $U_{N\alpha}$, where $\alpha=e\,\mu,\tau$. The reach for more general scenarios with more than one HNL or more complicated mixing patterns can be derived from these results.

\begin{description}
\item [Production] HNL production at FASER mainly occurs through heavy meson and $\tau$ decay~\cite{Gorbunov:2007ak}. In particular, the most relevant HNL production mechanisms are semi-leptonic $D$ decays $D \to K \ell N$ for masses $m_N<m_D-m_K$, leptonic $D$ decays $D^\pm \to \ell^\pm N$ for $m_N<m_D$, semi-leptonic $B$ decays $B \to D \ell N$ for $m_N<m_B-m_D$, and leptonic $B$ decays $B^\pm \to \ell^\pm N$ for $m_N<m_B$.  Among these, since there are far more $D$ mesons produced at the LHC than $B$ mesons, typically HNLs with masses $m_N<m_D$ are primarily produced in $D$ decay, while heavier HNLs with $m_D < m_N < m_B$ are only produced in $B$ decay. In addition, for HNLs mixing with $\nu_\tau$ and masses $m_N < m_\tau$, the dominant production mode is due to decays of $\tau$ leptons. A full list of the production modes we include are described in Ref.~\cite{Kling:2018wct}. 

\item [Decay and Lifetime] Heavy HNLs have a multitude of possible decay channels. These include the invisible decay mode into three neutrinos; various decay modes with two charged particles in the final state that most closely resemble the LLP signals described above for other models (e.g., $N\to \pi^{\pm} \ell^{\mp}, \ell\ell\nu, \pi^+ \pi^- \nu$); and, for larger $m_N$, other decay modes with more particles (especially pions) in the final state. A detailed discussion is given in Ref.~\cite{Kling:2018wct} and references therein. In the following we will assume $100\%$ efficiency for detection of all the channels beside the invisible one, while detailed discussion of the FASER efficiency for the various visible decay modes is left for future studies. The corresponding decay lengths and branching fractions into different final states are shown in the left panels of Figs.~\ref{fig:n1}--\ref{fig:n3}.

\begin{figure}[tbp]
\centering
\vspace*{-0.6cm}
\includegraphics[width=0.98\textwidth]{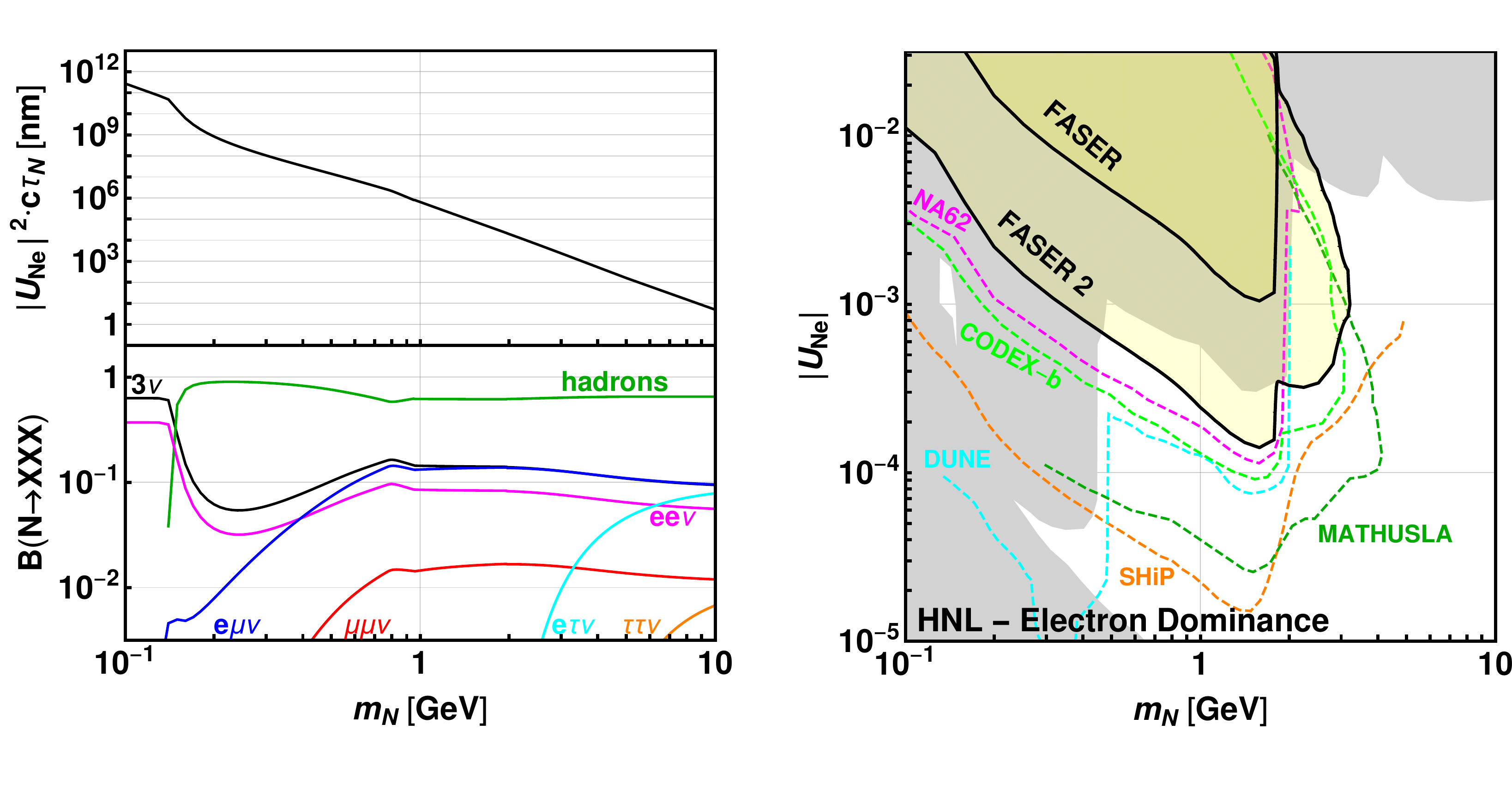} 
\caption{
{\bf Benchmark Model F1.} The decay length (top left panel), decay branching fractions (bottom left panel), and FASER's reach (right panel) for the HNL that mixes only with the electron neutrino $\nu_e$. The gray shaded regions are excluded by current limits, and the colored contours are the projected sensitivities for other proposed experiments.  See the text for details. }
\label{fig:n1}
\end{figure}

\begin{figure}[tbhp]
\centering
\vspace*{-0.6cm}
\includegraphics[width=0.98\textwidth]{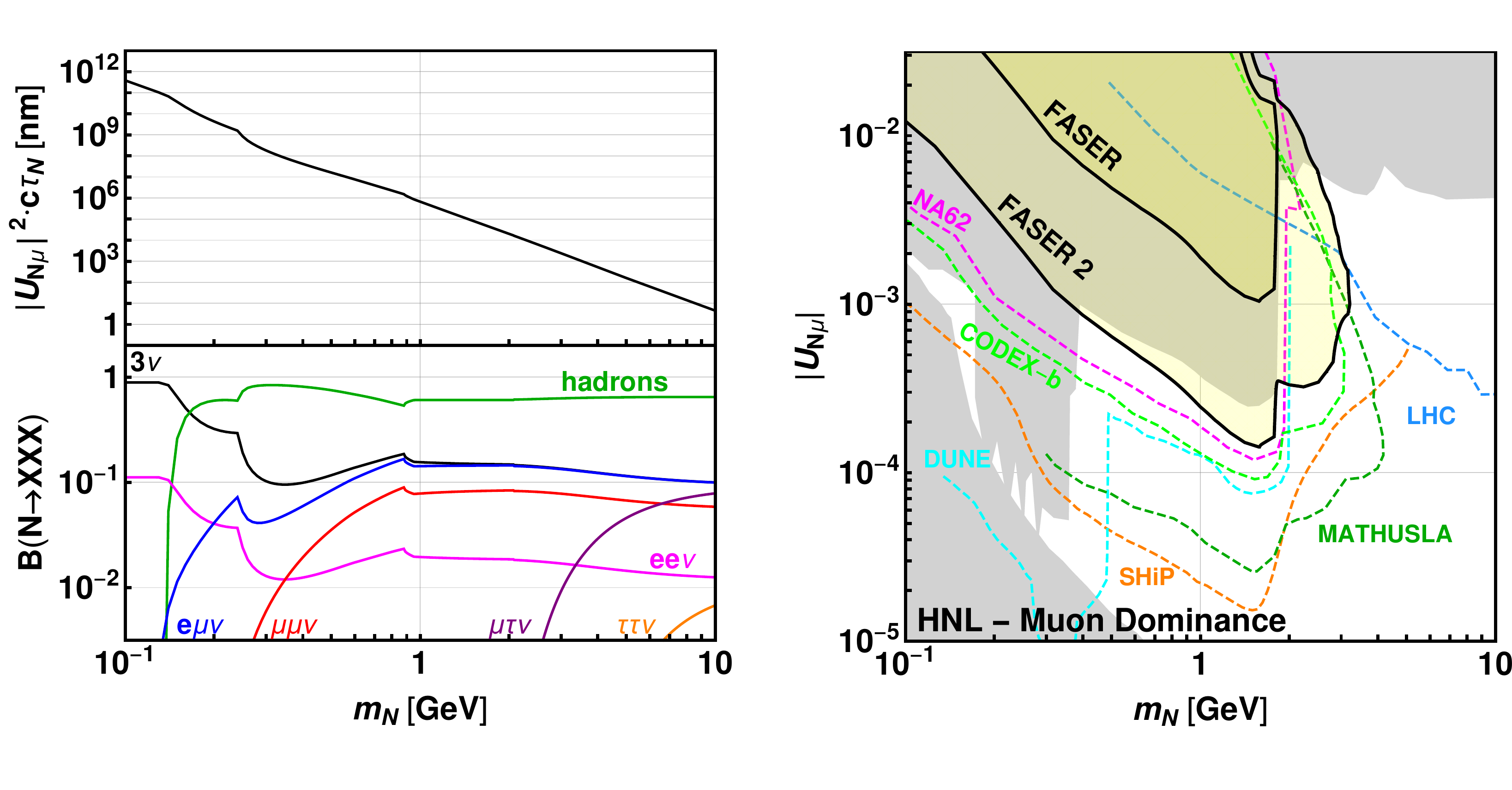} 
\caption{
{\bf Benchmark Model F2.} As in \figref{n1}, but for an HNL that only mixes with $\nu_{\mu}$.}
\label{fig:n2}
\end{figure}

\begin{figure}[tbp]
\centering
\vspace*{-0.6cm}
\includegraphics[width=0.98\textwidth]{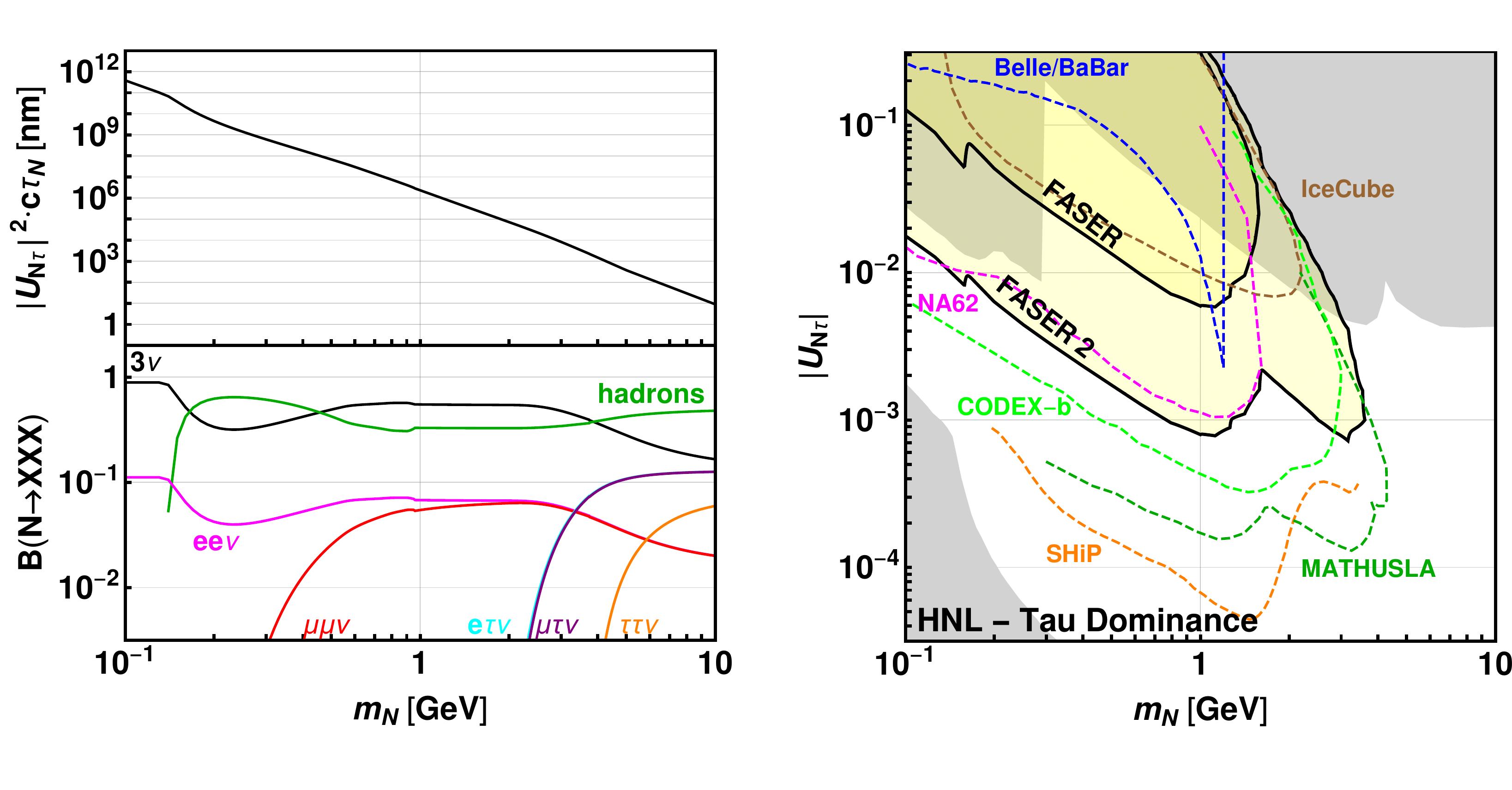} 
\caption{
{\bf Benchmark Model F3.} As in \figref{n1}, but for an HNL that only mixes with $\nu_{\tau}$.}
\label{fig:n3}
\end{figure}

\item [Results] the projected HNL sensitivity reaches for FASER at LHC Run 3 with $150~\ifb$ and FASER 2 at HL-LHC with $3~\iab$ are shown for the cases of mixing only with $\nu_e$, $\nu_\mu$, and $\nu_\tau$ in Figs.~\ref{fig:n1}, \ref{fig:n2}, and \ref{fig:n3}, respectively.  The gray-shaded regions are excluded by current bounds~\cite{Kling:2018wct,Drewes:2018irr} (see Ref.~\cite{Beacham:2019nyx} and references therein). For comparison, we also show the sensitivities of other proposed experiments: NA62 assumes $10^{18}$ POT while running in a beam dump mode that is being considered for LHC Run 3~\cite{Drewes:2018irr}; the DUNE limit assumes a normal hierarchy of neutrinos and corresponds to the five years of data-taking by the $30~\m$ long LBNE near detector with $5\times 10^{21}$ protons on target~\cite{Adams:2013qkq}; SHiP assuming $\sim 2 \times 10^{20}$ POT collected in 5 years of operation~\cite{Alekhin:2015byh}; the LHC searches for a prompt lepton plus a single displaced lepton jet for $\sqrt{s}=13~\tev$ and $300~\ifb$ of integrated luminosity~\cite{Izaguirre:2015pga} (see also Ref.~\cite{Antusch:2017hhu} for sensitivity in displaced vertex searches at LHCb); the proposed MATHUSLA experiment assumes a large-scale $200~\m\times 200~\m\times 20\m$ detector located on the surface above ATLAS or CMS and operating during the HL-LHC era to collect full $3~\iab$ of integrated luminosity~\cite{Curtin:2018mvb}; and the proposed CODEX-b detector assumes a $10~\m\times 10~\m\times 10\m$ fiducial volume close to LHCb and $300~\ifb$ to be collected by the HL-LHC~\cite{Helo:2018qej,Beacham:2019nyx}. For the $\nu_{\tau}$ mixing scenario, one of the future projected limits comes from searches for $\tau$ production in $B$ factories like Belle-II, with their subsequent decay into HNLs~\cite{Kobach:2014hea} under the assumption that $\sim 10$M tau decays will be analyzed. In addition, we show the sensitivity line for the proposed search for double-bang events at IceCube for 6 years of data taking~\cite{Coloma:2017ppo}. Interestingly, HNLs can also be succesfully searched for in heavy-ion collisions at the LHC when lighter-than-Pb nuclei are employed~\cite{Drewes:2018xma}.

As can be seen in the right panels of \figsref{n1}{n2}, in the $\nu_e$ and $\nu_\mu$ cases, FASER 2 will probe unconstrained regions of parameter space both below and above the threshold for HNL production in $D$-meson decays. Notably, due to the typically large lifetimes of HNLs, their decay rate in FASER simply scales as $U^2$, similarly to the production rate, so that the total number of events scales as $U^4$. In this long-lifetime regime, the reach can be significantly improved by increasing the size and luminosity of the experiment, as can be seen by comparing the FASER and SHiP detectors in \figssref{n1}{n2}{n3}. Importantly, however, in the region above the $D$-meson threshold, the prospect of detecting HNLs in these detectors can be comparable, while many other experiments lose their sensitivity due to the large energy required for efficient $B$ meson production. In particular, although the number of $D$ mesons produced at SHiP is 10 times the number produced at the HL-LHC, the number of $B$ mesons is 100 times more at the HL-LHC than at SHiP, because the SHiP rate is suppressed by the large $B$ mass. Last, but not least, for the case of mixing with the tau neutrino, and where current bounds are relatively weak, there is a large unconstrained region of parameter space that will be covered by both FASER and FASER 2. 
\end{description}

\section{FASER Reach for Axion-Like Particles}
\label{sec:alps}

Unlike the previous models, axion-like particles (ALPs) couple to the SM through dimension-5 operators. They are pseudoscalar SM-singlets that can appear as pseudo-Nambu-Goldstone bosons in theories with broken global symmetries in analogy to the QCD axion~\cite{Peccei:1977hh,Peccei:1977ur,Wilczek:1977pj,Weinberg:1977ma}. In the most general case, ALPs can have arbitrary couplings to photons, gluons, and fermions, with a mass $m_a$ that is an independent parameter~\cite{Jaeckel:2010ni}.  (See also Ref.~\cite{Bauer:2017ris} for a recent review.) A general Lagrangian for an ALP $a$ defined at a scale $\Lambda$ is
\be
\mathcal{L} = \mathcal{L}_{\text{SM}}+\mathcal{L}_{\text{DS}} 
-  \frac{1}{2} \red{m_a^{\textcolor{black}{2}}} a^2
-  \frac{a}{4 \red{f_\gamma}} F_{\mu\nu}  \widetilde F^{\mu\nu}
-  \frac{a}{4 \red{f_G}} \text{Tr} \, G_{\mu\nu}  \widetilde G^{\mu\nu}
+ \frac{ \partial^\mu a}{\red{f_f}} \sum_{f} \bar{f} \gamma_\mu \gamma_5 f \ .
\label{eq:ALPLagrangian}
\ee
The ALP-fermion interaction may be re-written by integrating by parts  and employing the equations of motion: 
\be
\frac{ \partial^\mu a}{ 2f_f} \bar{f} \gamma_\mu \gamma_5 f 
= - i \frac{ m_f}{ f_f}   a \bar{f}  \gamma_5 f  
+  \frac{ N_C^f Q^2_f e^2}{ 16 \pi^2 } \frac{a}{f_f}    F_{\mu\nu}  \widetilde F^{\mu\nu} + \dots \ .
\label{eq:ALPL}
\ee
Here the first part corresponds to the coupling of a pseudoscalar to fermions, and the second part is an additional contribution to the coupling of the ALP to photons. 

To describe the phenomenology of ALPs at the LHC, we need to consider the running of the coupling constants $f_i$ between the scale $\Lambda$ and the relevant low-energy scale~\cite{Bauer:2017ris}. The resulting effective Lagrangian at the one-loop level is
\be
\mathcal{L} = \mathcal{L}_{\text{SM}}+\mathcal{L}_{\text{DS}} 
-  \frac{1}{2} \red{m_a^{\textcolor{black}{2}}} a^2
-  \frac{1}{4} \red{g_{a\gamma\gamma}} a F_{\mu\nu}  \widetilde F^{\mu\nu}
-  \frac{g_s^2}{8} \red{g_{agg}} a  G_{\mu\nu}^A  \widetilde G^{A\mu\nu}
- i   \sum_f \red{g_{aff}} \frac{m_f}{v} a \bar{f} \gamma_5 f \ ,
\label{eq:ALPeffL}
\ee
where new symbols for the coupling constants, $g_{aii}$, have been introduced for clarity. Note that, in principle, each of these coefficients depends on all the coefficients defined at the scale $\Lambda$, that is, $g_{aii}=g_{aii}(f_\gamma,f_G,f_f,\Lambda)$. 

In the following sections we will consider simple cases in which, at the high-energy scale $\Lambda$, only one of the couplings is non-vanishing: that is, either $f_\gamma^{-1} \neq 0$ (\secref{a1}), $f_f^{-1} \neq 0$ (\secref{a2}), or $f_G^{-1} \neq 0$ (\secref{a3}).

\subsection{Benchmark A1: Photon Dominance}
\label{sec:a1}

Let us first consider the case in which the ALPs only couple to photons at the high-energy scale $\Lambda$. At the low energy scale, the coupling to photons is simply given by $g_{a\gamma\gamma} = 1/f_\gamma$, up to $\mathcal{O}(\alpha)$ corrections. Additionally, the ALP obtains loop-induced couplings to all charged SM fermions $g_{aff} \sim Q_f^2 \alpha^2 / f_{\gamma}$. Since these couplings are suppressed by $\alpha^2$, they typically have negligible effect on the phenomenology of ALPs at FASER when compared to the dominant di-photon coupling, and hence they can be ignored in the following discussion. One can therefore write an effective low-energy Lagrangian
\be
\mathcal{L} \supset -\frac12 \,\red{m_a^{\textcolor{black}{2}}}\, a^2 -\frac14\, \red{g_{a\gamma\gamma}} \,a F^{\mu\nu} \widetilde F_{\mu\nu} \ ,
\label{eq:La1}
\ee
for which the parameter space is spanned by the ALP mass, $m_a$, and its di-photon coupling $g_{a\gamma\gamma}$. 

\begin{description}
\item [Production] ALPs with dominantly di-photon couplings can be produced by photon fusion (see, e.g., Ref.~\cite{Dobrich:2015jyk}), rare decays of light mesons, and the Primakoff process. For highly boosted ALPs in the far forward region of the LHC, the dominant production mechanism is the Primakoff process, in which high-energy, forward-going photons produced at the IP convert into ALPs when interacting with matter. In particular, efficient conversion can take place when the photons hit the neutral particle absorber (TAN) about $140~\m$ away from the IP~\cite{Feng:2018noy}.  The rate is proportional to $g_{a\gamma\gamma}^2$.

\item [Decay and Lifetime] ALPs with dominantly di-photon couplings mainly decay into a pair of photons; decays into pairs of SM fermions are highly suppressed. A subleading decay channel, in which one of the photons is produced off-shell and converts into an electron-positron pair, has a branching fraction of $B(a \to \gamma e^+ e^-) \approx B(\pi^0 \to \gamma e^+ e^-) \sim 1\%$. 
The total decay width of the ALP is given by 
\be   
\Gamma(a\to\gamma\gamma) = \frac{g_{a\gamma\gamma}^2 m_a^3}{64\pi} \ .
\label{eq:A1_gamma_photon}
\ee
In the left panel of \figref{a1} we show the ALPs decay length and its branching fractions to $\gamma\gamma$ and $\gamma\,e^+e^-$ as a function of $m_a$.

\item [Results] The projected ALP sensitivity reaches for FASER at LHC Run 3 with $150~\ifb$ and FASER 2 at HL-LHC with $3~\iab$ are shown in the right panel of \figref{v1}. The gray-shaded regions are excluded by current bounds~\cite{Feng:2018noy}. (See also Refs.~\cite{Beacham:2019nyx,Dolan:2017osp} and references therein.) For comparison, the colored contours show projections for other experiments: NA62 assumes $10^{18}$ protons on target (POT) while running in a beam dump mode that is being considered for LHC Run 3~\cite{Dobrich:2015jyk}; SeaQuest assumes $1.44 \times 10^{18}$ POT, which could be obtained in two years of parasitic data taking and requires additionally the installation of a calorimeter~\cite{Berlin:2018pwi}; the proposed beam dump experiment SHiP assumes $\sim 2 \times 10^{20}$ POT collected in 5 years of operation~\cite{Dobrich:2015jyk}; the proposed electron fixed-target collisions experiment LDMX during Phase II with a beam energy of $8~\gev$ and $10^{16}$ electrons on target (EOT)~\cite{Berlin:2018bsc}; Belle-II assumes the full expected integrated luminosity of $50~\iab$~\cite{Dolan:2017osp}; and NA64~\cite{Gninenko:2320630} corresponds to $5\times 10^{12}$ EOT with $100~\gev$ energy.

As can be seen, both FASER and FASER 2 can probe currently unconstrained regions of parameter space with the potential for discovery in the mass range $m_a\sim 30-400~\mev$.

\end{description}

\begin{figure}[tbp]
\centering
\vspace*{-0.6cm}
\includegraphics[width=0.98\textwidth]{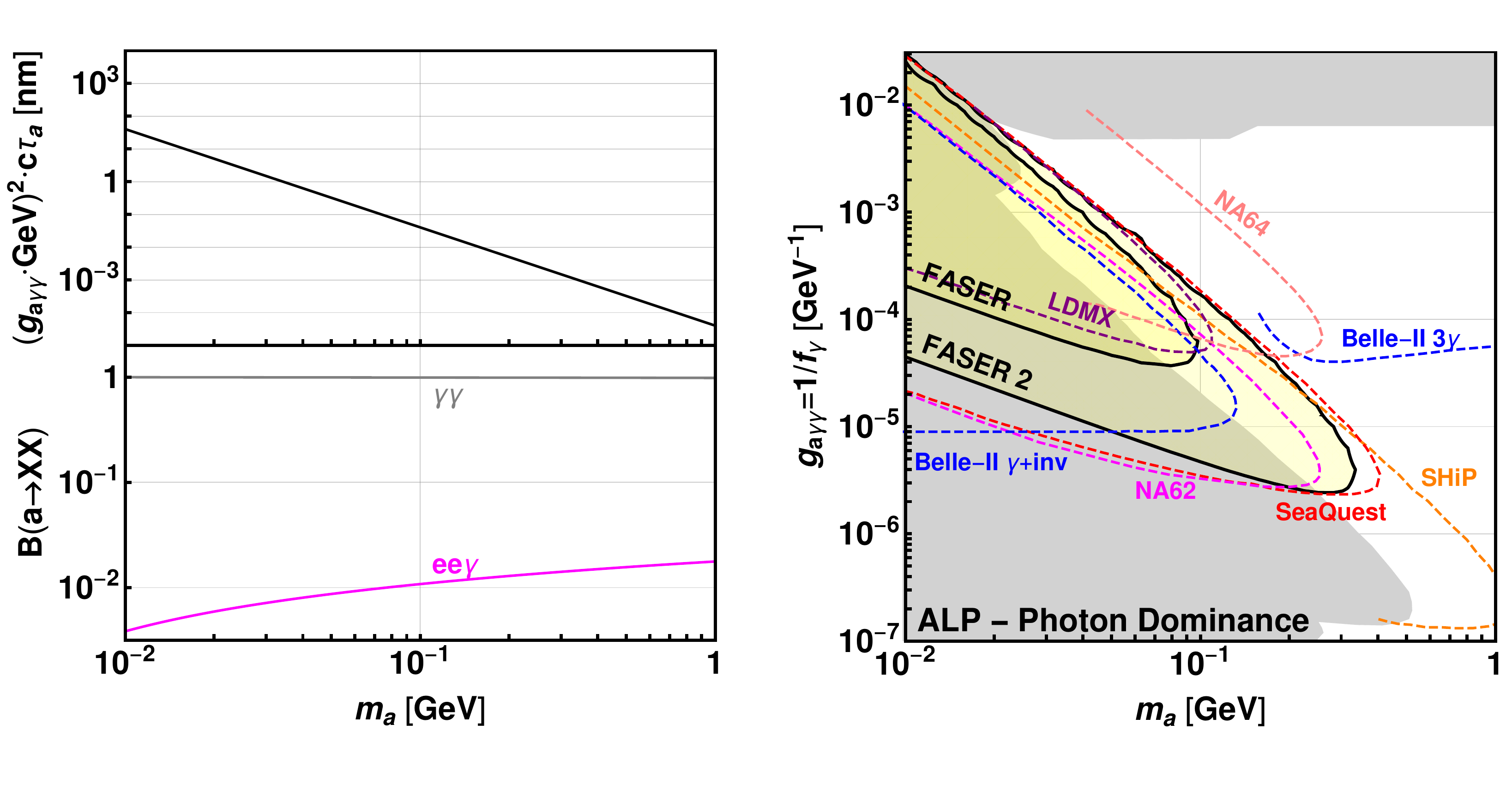} 
\caption{
{\bf Benchmark Model A1.} The decay length (top left panel), decay branching fractions (bottom left panel) and FASER's reach (right panel) for ALPs with dominantly di-photon couplinga. The gray-shaded regions are excluded by current limits, and the colored contours give the projected sensitivities of several other proposed experiments. See the text for details.}
\label{fig:a1}
\end{figure}

\subsection{Benchmark A2: Fermion Dominance}
\label{sec:a2}

Let us now consider the case in which the ALP only couples to fermions at a scale $\Lambda$. At the low energy scale, the coupling to fermions is (up to $\mathcal{O}(\alpha,\alpha_s)$ corrections) given by $g_{aff} = 2v/f_f$.  We will assume that all fermion coupling constants $g_{aff}$ are identical at the low-energy scale (or equivalently that all the SM fermions carry the same PQ charge). This then implies Yukawa-like couplings of the ALP to the SM fermions.

Additional ALP couplings are induced at loop level. In particular, a flavor-changing $a-s-b$ coupling arises through a $W$-boson and top-quark loop, inducing an effective coupling~\cite{Beacham:2019nyx, Batell:2009jf} 
\footnote{The axion considered here shares many properties with a pseudoscalar mediator with Yukawa-like couplings.  However, because of the different way in which electroweak symmetry is broken, the loop-induced couplings are not identical. Most importantly, the flavor-changing $a-s-b$ coupling differs by a factor $1/4$, as discussed in Ref.~\cite{Dolan:2014ska}. The pseudoscalar model has been investigated, e.g., in Refs.~\cite{Dolan:2014ska, Dobrich:2018jyi}, and FASER's reach will be presented below in \secref{pseudoscalarYukawa}.}
\be
g_{asb}= g_{aff} \ \frac{ m_t^2 m_b V_{ts}^* V_{tb} }{16 \pi^2 v^3 }  \log\left(\frac{\Lambda^2}{m_t^2} \right) \ .
\label{eq:gasb-alpfermion}
\ee
Furthermore, the ALP will obtain small couplings to photons and gluons, $g_{a\gamma \gamma}$ and $g_{agg}$, respectively.  These couplings scale as
\be
g_{a\gamma\gamma}\sim\frac{n_f\alpha}{\pi v}g_{aff}\sim \frac {n_f~g_{aff}}{ 10^5~\gev}
\quad \text{and} \quad
g_{agg} \sim  \frac{n_f } {2 \pi^2 v} g_{aff} \sim \frac {n_f~g_{aff}}{ 10^4~\gev } \ ,
\ee
where $n_f$ is the number of light fermions with $m_f \alt m_a$ contributing to the loop-induced coupling. These couplings do not have any significant effect on the phenomenology at FASER and are therefore ignored below. 

The effective low-energy Lagrangian can be written as 
\be
\mathcal{L}
 \supset -\frac12 \red{m_a^{\textcolor{black}{2}}} a^2 
- i  \red{g_{aff}} a  \sum_f  \frac{m_f}{v} \bar{f} \gamma_5 f \ 
+ \left[ g_{asb}(\red{g_{aff}}) a  \bar{s}_L b_R + \text{h.c.} \right] \ ,
\label{eq:a2}
\ee
where $g_{asb}(g_{aff})$ is proportional to $g_{aff}$ and given in \eqref{gasb-alpfermion}. The parameter space is spanned by the ALP mass, $m_a$, and a universal coupling to fermions $g_{aff}$.

\begin{description}
\item [Production] Since ALPs with dominantly fermion couplings have Yukawa-like couplings, they are mainly produced through the flavor-changing heavy meson decay $B \to X_s a$.  The corresponding branching fraction is~\cite{Batell:2009jf}
\be
B(B \to X_S\, a) \approx  \left[3.1 \left(1-\frac{m_a^2}{m_B^2}\right) + 3.7 \left(1-\frac{m_a^2}{m_B^2}\right)^3 \right] \times  g_{aff}^2 \ ,
\ee
where we have used $m_K^{(*)} \ll m_B$. Note that this branching fraction depends on the high-energy cutoff scale $\Lambda$ in \eqref{gasb-alpfermion}, which we assume to be $\Lambda=1~\tev$~\cite{Beacham:2019nyx,Dolan:2014ska}. In the following, we neglect additional contributions from kaon decays that are sizable only in the region of the parameter space that is already strongly constrained by other experiments. Decays of D-mesons into scalars are further suppressed due to the absence of top loops mediating such process. 

The ALPs can also be produced through their mixing with pions~\cite{Alekhin:2015byh}, which could enhance the reach of FASER at low masses. However, this mixing vanishes if the up-quark and down-quark couplings are equal to each other $g_{auu}=g_{add}$~\cite{Bauer:2017ris}. Given our assumptions, this contribution therefore plays a negligible role in setting the FASER sensitivity reach. 

\item [Decay and Lifetime] Given the Yukawa-like fermion couplings, the dominant decay modes are typically pairs of the heaviest kinematically available SM fermions; decays into lighter fermions and two photons are typically sub-dominant. The decay width of the ALP into leptons and quarks is given by
\be
\Gamma(a\to ff)= N_c^f g_{aff}^2 \  \frac{m_a m_f^2}{8 \pi v^2}\  \sqrt{ 1-\frac{4 m_f^2}{m_a^2} } \ ,
\label{eq:A2_gamma_fermions}
\ee
where $N_c^f$ denotes the fermion's color multiplicity. 

Of course, for $m_a \alt 500~\mev$, one must consider decays not into quarks, but into hadrons.  Possible decays into light hadrons are notoriously hard to calculate, but they are also suppressed~\cite{Dolan:2014ska,Domingo:2016yih}.  For example, decays into two pseudoscalars, such as $a \to \pi \pi$, or into a single pion and a photon, $a \to \pi \gamma$, are not allowed by $CP$ invariance and conservation of angular momentum.  Decays into 3-body final states are phase-space suppressed, and, in fact, the decay to the lightest allowed hadronic final state, $a \to \pi \pi \pi$, vanishes in the case of $g_{auu}=g_{add}$~\cite{Bauer:2017ris}. For light ALPs, we therefore neglect hadronic decay modes in the following, and consider only $f = e, \mu, \tau, c, b$ in \eqref{A2_gamma_fermions}.  We show the ALP decay length and its branching fractions in the left panel of \figref{a2}.

\item [Results] The expected FASER and FASER 2 reaches are shown in the right panel of \figref{a2}. In particular, FASER 2 will be able to explore regions of parameter space that are currently unconstrained (see Ref.~\cite{Beacham:2019nyx} and references therein) and extend sensitivities by up to one order of magnitude in the coupling constant.

For comparison, following Ref.~\cite{Beacham:2019nyx}, we also show the expected sensitivity reach for other proposed experiments: the reach for Codex-b~\cite{Gligorov:2017nwh} corresponds to $300~\ifb$ of data collected by a $10\times 10\times 10~\m^3$ detector situated $25~\m$ away from the LHCb IP; the reach for KLEVER~\cite{Ambrosino:2019qvz} assumes $5\times10^{19}$ POT from the $400~\gev$ SPS beam; the expected sensitivity of MATHUSLA~\cite{Evans:2017lvd,Curtin:2018mvb} assumes $3~\iab$ of data collected by a $200\times 200\times 20~\m^3$ detector placed at the ground level $\sim 100~\m$ away from the ATLAS or CMS IP; the sensitivity of REDTOP~\cite{Gatto:2016rae} has been obtained for $10^{17}$ POT with low energy $\sim 1.7-1.9~\gev$ and assuming that the LLP will be produced in the rare decays of $\sim 10^{13}$ $\eta$ mesons; and the reach of SHiP~\cite{Alekhin:2015byh} corresponds to $2\times 10^{20}$ POT from the 400 GeV SPS beam collected during 5 years of operation.
\end{description}

\begin{figure}[tbp]
\centering
\vspace*{-0.6cm}
\includegraphics[width=0.98\textwidth]{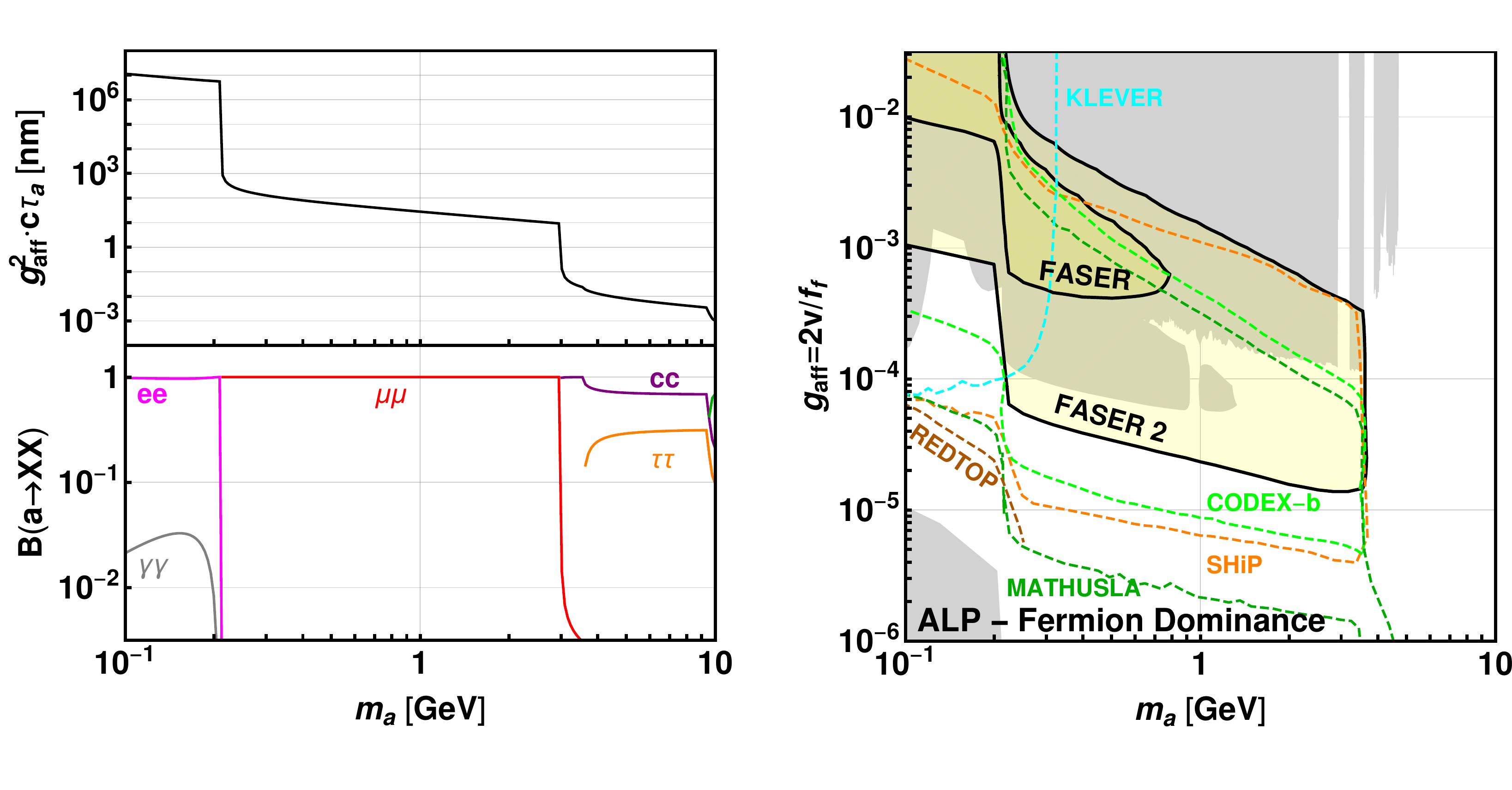} 
\caption{
{\bf Benchmark Model A2.} As in \figref{a1}, but for ALPs with dominantly fermion couplings.}
\label{fig:a2}
\end{figure}

\subsection{Benchmark A3: Gluon Dominance}
\label{sec:a3}

Let us now consider the case in which the ALP only couples to gluons at the scale $\Lambda=1~\tev$.  At the low-energy scale, the coupling to gluons is then given by $g_{agg} = 1/f_G$, where we have explicitly taken into account the running of the strong coupling and replaced $g_s(\Lambda)G_{\mu\nu} \sim G_{\mu\nu} \rightarrow g_s G_{\mu\nu}$ in \eqref{ALPeffL}. But the ALP's gluon coupling also induces loop-level couplings to quarks, which are given by 
\be
g_{aqq}= - 2\,\alpha_s^2\, v \ g_{agg}  \left[ \log\left( \frac{\Lambda^2}{m_q^2}\right) -\frac{11}{3}+ g\left(\frac{4 m_q^2}{m_a^2}\right) \right]\, ,
\label{eq:gaqq}
\ee
where the function $g(\tau)$ is defined in Ref.~\cite{Bauer:2017ris} and approaches $g(\tau) \to 7/3$ in the limit of large fermion masses. Couplings to SM leptons are also induced, but at the three-loop level, and so can be neglected. Furthermore, the ALP will obtain  
a flavor-changing $a-s-b$ coupling at the two-loop level, inducing an effective coupling~\cite{Beacham:2019nyx,Aloni:2018vki}
\be
g_{asb}= g_{agg} \ \alpha_s^2(m_t) \ \frac{ m_t^2 m_b V_{ts}^* V_{tb} }{8 \pi^2 v^2 } \times \mathcal{UV}\ .
\label{eq:gasb-alpgluon}
\ee
Here the loop factor $\mathcal{UV}$ encodes the dependence on the ultraviolet physics.

If the ALP is sufficiently light, $m_a < 2 \pi \Lambda_{\text{QCD}}$, its interactions can be described using chiral perturbation theory. In this case, the ALP mixes with the neutral pion, the $\eta$-meson, and the $\eta'$-meson~\cite{Beacham:2019nyx}:
\be
\pi^0 &= \pi^0_{\text{phys}} + \theta_{a\pi}\, a_{\text{phys}} &\text{with}& \
\theta_{a\pi} =  2 \pi^2\, f_\pi\, g_{agg} \ \frac{1-m_u/m_d}{1+m_u/m_d} \ \frac{m_a^2}{m_a^2 - m_\pi^2 + i m_\pi \Gamma_\pi(m_a)}\, , \\
\eta &= \eta_{\text{phys}} + \theta_{a\eta}\, a_{\text{phys}} &\text{with}& \
\theta_{a\eta} =  2 \pi^2\, f_\pi\, g_{agg} \ \cos\theta_p \ \frac{m_a^2}{m_a^2 - m_\eta^2 + i m_\eta \Gamma_\eta(m_a)} \, , \\
\eta' &= \eta'_{\text{phys}} + \theta_{a\eta'}\, a_{\text{phys}} &\text{with}& \
\theta_{a\eta'} =  2 \pi^2\, f_\pi\, g_{agg} \ \sin\theta_p \  \frac{m_a^2}{m_a^2 - m_\eta'^2 + i m_{\eta'} \Gamma_{\eta'}(m_a)} \, .
\label{eq:A3_mixing}
\ee
Here $f_\pi=0.13~\gev$ is the pion decay constant, $\sin\theta_p\approx 0.8$ and $\cos\theta_p\approx0.6$ characterize the $\eta$-$\eta'$-mixing, and $m_u/m_d=0.483$ is the up-to-down-quark mass ratio. This mixing also introduces an effective coupling of the ALP to the photon, given by
\be
g_{a\gamma\gamma} = 4 \pi\, \alpha\, g_{agg} \left(\frac{4+m_u/m_d}{3+3m_u/m_d} - \frac{1}{2}\frac{m_a^2}{m_a^2-m_\pi^2} \frac{1-m_u/m_d}{1+m_u/m_d} \right)\, ,
\label{eq:A3_photon_coupling}
\ee
where we have omitted additional contributions coming from the ALP mixing with the $\eta$ and $\eta'$ mesons.

On the other hand, if the ALP is sufficiently above the hadronic scale, one can describe its decays using perturbation theory. In this case, ALPs obtain a coupling to photons at the two-loop level, but these are unimportant for the values of $m_a$ that can be probed at FASER. 

The effective low energy Lagrangian takes the form 
\be
\mathcal{L}  \supset
&-  \frac{1}{2} \red{m_a^{\textcolor{black}{2}}} a^2
-  \frac{1}{4} g_{a\gamma\gamma}(\red{g_{agg}})\, a\, F_{\mu\nu}  \widetilde F^{\mu\nu}
-  \frac{g_s^2}{8} \red{g_{agg}}\, a\, \text{Tr} G_{\mu\nu}  \widetilde G^{\mu\nu} \\
&- i   \sum_q g_{aqq}(\red{g_{agg}}) \frac{m_q}{v}\, a\, \bar{f} \gamma_5 f
+ \left[ g_{asb}(\red{g_{agg}}) a  \bar{s}_L b_R + h.c. \right] \, ,
\ee
where $g_{aqq}(g_{agg})$, $g_{asb}(g_{agg})$, and $g_{a\gamma\gamma}(g_{agg})$ are proportional to $g_{agg}$ and are given in Eqs.~(\ref{eq:gaqq}), (\ref{eq:gasb-alpgluon}), and (\ref{eq:A3_photon_coupling}), respectively. The parameter space of the model is spanned by $m_a$ and $g_{agg}$. 
\begin{description}
\item [Production] Because the ALP mixes with the neutral pseudoscalar mesons, it is produced in any process that produces such mesons and we can estimate its production cross section as
\be
\sigma(a) = |\theta_{a\pi}|^2 \sigma(\pi) +  |\theta_{a\eta}|^2 \sigma(\eta) +  |\theta_{a\eta'}|^2 \sigma(\eta') \ . 
\ee
We use the $\pi^0$, $\eta$, and $\eta'$ spectra obtained from EPOS-LHC, re-weighted by the corresponding mixing angles. Note that this approach is just an approximation, for example it does not take into account interference effects between the different pseudoscalars or a possible ALP-mass dependence in hadronization. Additional ALPs can be produced in flavor-changing decays of heavy quarks, $B \to a X_s$. The corresponding decay branching fraction is given by~\cite{Aloni:2018vki,Beacham:2019nyx}
\be
B(B \to X_S\ a) \approx  \left[ 33 \left(1-\frac{m_a^2}{m_B^2}\right) + 40 \left(1-\frac{m_a^2}{m_B^2}\right)^3 \right] \times \mathcal{UV} \times (g_{agg}\cdot \gev)^2   \ ,
\ee
where we have used $m_K^{(*)} \ll m_B$. Following the suggestions of the authors of Ref.~\cite{Aloni:2018vki}, we assume that the UV-physics dependent factor $\mathcal{UV} \sim \log{\Lambda^2/m_t^2}+ \mathcal{O}(1)$, originating from loop integrals, can be taken to be unity: $\mathcal{UV} \rightarrow 1$. Note that this choice for the UV-factor induces an $\mathcal{O}(1)$ arbitrariness in the constraints.

\item [Decay and Lifetime] the dominant decay modes are into pairs of photons at low ALP mass and into hadronic final states for heavier ALPs, while leptonic decays only arise at three loop at do not play a significant role. 

At low mass $m_a < 3 m_{\pi}$ the ALP mainly decays into photon pairs. The corresponding decay width is given in \eqref{A1_gamma_photon}, where the photon coupling is $g_{a\gamma\gamma}$ is induced through the mixing with the pions and given in \eqref{A3_photon_coupling}. The lightest allowed hadronic decay mode is $a \to 3 \pi$, and the corresponding decay width has been estimated using chiral perturbation theory to be~\cite{Bauer:2017ris}
\be
\Gamma_{a \to 3 \pi} = \frac{\pi}{6} \frac{m_a m_\pi^4 g_{agg}^2}{64 f_\pi^2 } \left( \frac{m_a^2}{m_a^2-m_\pi^2} \frac{1-m_u/m_d}{1+m_u/m_d} \right)^2 I\left(\frac{m_\pi^2}{m_a^2} \right) ,
\ee
where
\be
I(r) =  \int_{4r}^{(1-\sqrt{r})^2} \!\!\! dz \sqrt{1-\frac{4r}{z}} \ \sqrt{(2-r-z)^2-4 rz} \ \left[12 (r-z)^2 +2 \right] . 
\ee
Although the diphoton and hadronic decay widths are of similar size below $m_a = 2m_\pi + m_\eta$, many new decay channels open up at larger masses, and hence hadronic decays will dominate. This includes 3-body decays, such as $a \to \eta\pi\pi$, as well as 2-body decays, such as $a \to \rho \pi, f_0 \pi,  a_0 \pi, K K^*$, which will quickly increase the hadronic decay width. At large masses, $m_a > 2 \pi \Lambda_{\text{QCD}} \approx 1.5~\gev$, the hadronic decay width is expected to approach the partonic decay width for $a \to gg$, which can be calculated using perturbation theory to be
\be
\Gamma(a \to gg) = \frac{1}{2} \pi \alpha_s^2 \ m_a^3 \ g_{agg}^2  \ .
\ee
The decay width in the intermediate regime for ALP masses in the range $2m_\pi+m_{\eta} < m_a < 2 \pi \Lambda_{\text{QCD}}$ is notoriously hard to calculate. We therefore interpolate the decay width, following the strategy proposed in Ref.~\cite{Beacham:2019nyx}, using a cubic function $\Gamma = \Gamma^* (m_a- m^*)^3$. Here the constants $m^*$ and $\Gamma^*$ are chosen to match the ALP decay width into pions and photons at a low mass matching point $m_a=2m_\pi+m_\eta$ and the decay width into gluons at a high-mass matching point $m_a=2 \pi \Lambda_{\text{QCD}}$. Additionally, we include resonant contributions from ALP meson mixing for ALP masses close to $m_\eta$ and $m_\eta'$. Following Ref.~\cite{Beacham:2019nyx}, the corresponding decay widths are given by 
\be
\Gamma(a \to \eta^* \to XX) = |\theta_{a\eta}|^2 \Gamma_\eta (m_a) 
\quad \text{and} \quad 
\Gamma(a \to \eta'^* \to XX) = |\theta_{a\eta'}|^2 \Gamma_{\eta'} (m_a)
\ee
where the mixing angles $\theta_{a\eta}$ and $\theta_{a\eta'}$ have been defined in \eqref{A3_mixing}. Finally, at masses above $m_a>2m_c$ and $m_a > 2 m_b$, decay channels into heavy mesons open up whose decay width can be estimated using \eqref{A2_gamma_fermions}. 

The branching fractions and lifetime for this scenario are shown in the left panel of \figref{a3}. The three resonant features are due to the mixing of the ALP with the $\pi^0$, $\eta$, and $\eta'$ mesons. 

\item [Results] The expected FASER reach is shown in the right panel of \figref{a3}. The existing constraints are shown as the gray shaded region (see Ref.~\cite{Beacham:2019nyx} and references therein). At large couplings, they are mainly due to flavor constraints which we have adapted from Ref.~\cite{Aloni:2018vki}. Additionally, we have recast the search for LLPs decaying into photons at CHARM~\cite{Bergsma:1985qz}, assuming that ALPs are produced through ALP-meson mixing. We also show the expected sensitivity reach for the Codex-b~\cite{Gligorov:2017nwh} and MATHUSLA~\cite{Evans:2017lvd,Curtin:2018mvb} experiments, following~\cite{Beacham:2019nyx}. The former assumes $300~\ifb$ data collected by a $10\times 10\times 10~\m^3$ detector placed $25~\m$ away from the LHCb IP, while the latter corresponds to $3~\iab$ of data and a $200\times 200\times 20~\m^3$ detector on the surface about $100~\m$ away from the ATLAS or CMS IP. In addition, the expected sensitivity reach~\cite{Beacham:2019nyx} for the proposed REDTOP experiment~\cite{Gatto:2016rae} is shown; this corresponds to $10^{17}$ POT with energies of about $1.7-1.9~\gev$, which is enough to produce about $10^{13}$ $\eta$ mesons.

Both FASER and FASER 2 can probe currently unconstrained regions of parameter space, with FASER 2's reach extending from $m_a\sim 20~\mev - 1~\gev$ and $g_{agg}\sim 10^{-8} - 10^{-2}$. 

\end{description}

\begin{figure}[tb]
\centering
\vspace*{-0.6cm}
\includegraphics[width=0.98\textwidth]{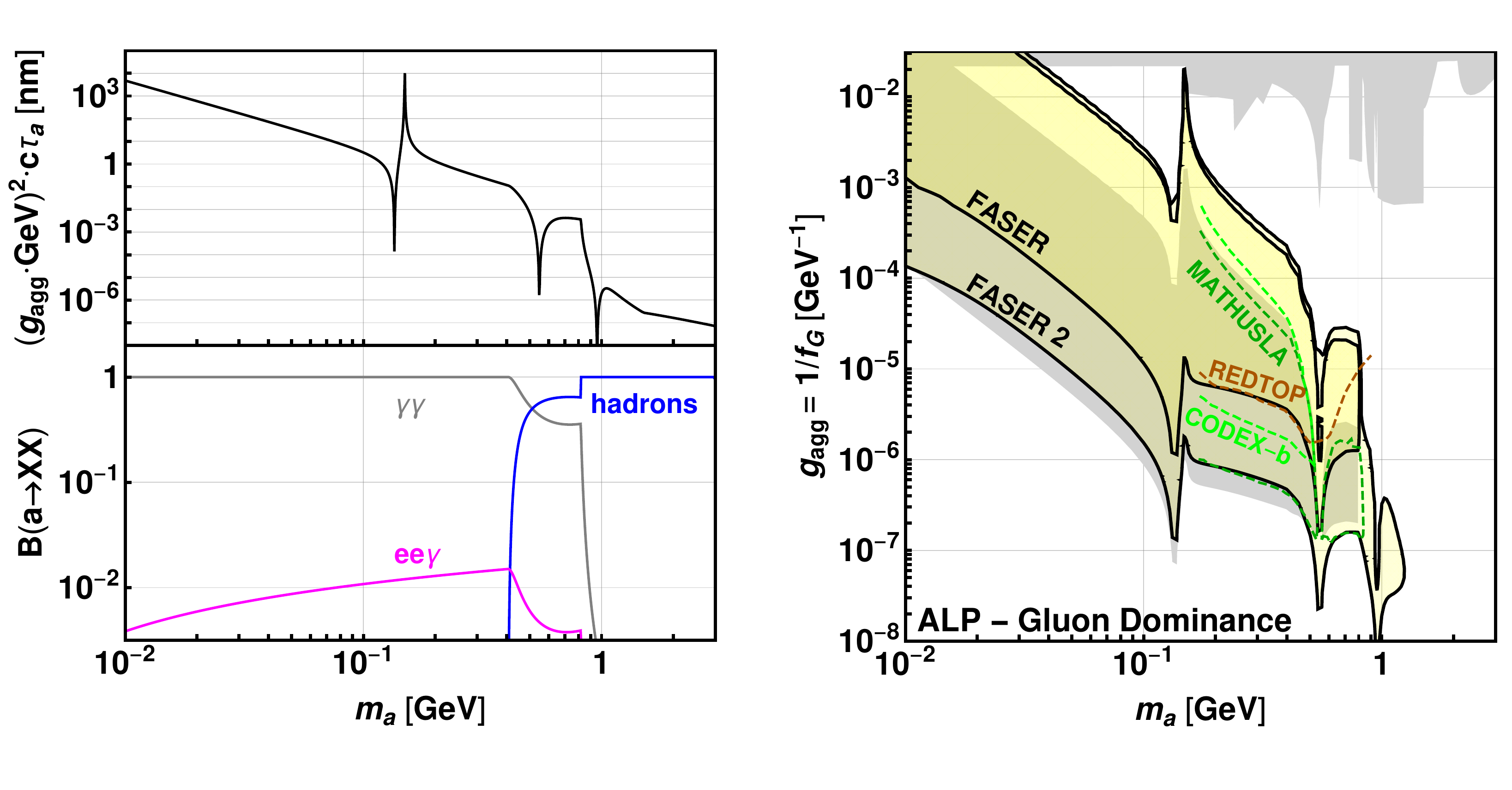}  
\caption{
{\bf Benchmark Model A3.} As in \figref{a1}, but for ALP with dominantly gluon couplings. 
}
\label{fig:a3}
\end{figure}

\section{FASER Reach for Dark Pseudoscalars}
\label{sec:pseudoscalarYukawa}

In the previous section we have focused on FASER's reach in several benchmark scenarios with pseudoscalar ALPs derivatively-coupled to the SM, including one with dominant couplings to the SM fermions. Similar, but not identical, phenomenology can be obtained for a light pseudoscalar $a$ with Yukawa-like couplings to the SM fermions, which we consider here.

\subsection{Benchmark P1: Pseudoscalar with Yukawa-like Couplings}

The Lagrangian for a pseudoscalar with Yukawa-like couplings is
\be
\mathcal{L} \supset -\red{m_a}^2 a^2 + i\,\red{g_Y}\,a\,\sum_f  \frac{m_f}{v} \bar{f} \gamma_5 f\ .
\ee
Note that this Lagrangian does not respect the unbroken SM gauge symmetries and therefore should be seen as a low-energy effective theory of a more complete theory, for example a two-Higgs doublet model. A recent discussion of this model and its properties can be found, e.g., in Refs.~\cite{Dobrich:2018jyi, Dolan:2014ska}, while FASER's reach has also been presented in Ref.~\cite{Ariga:2018zuc}. 

At one-loop level, additional flavor-changing couplings of the pseudoscalar are induced. Of particular interest for the pseudoscalar's phenomenology at FASER is the effective $a-s-b$ coupling
\be
g_{asb}= g_{Y} \ \frac{ m_t^2 m_b V_{ts}^* V_{tb} }{4 \pi^2 v^3 }  \log\left(\frac{\Lambda^2}{m_t^2} \right) \ ,
\ee
which is induced through a top-quark loop~\cite{Dolan:2014ska, Dobrich:2015jyk}.\footnote{As noted before, this benchmark scenario shares many properties with a pseudoscalar mediator with derivative couplings as discussed in~\secref{a2}. However, because of the different way in which electroweak symmetry is broken, the loop-induced couplings are not the same. Most importantly, the flavor-changing $a-s-b$ coupling differs by a factor $4$, as discussed in Ref.~\cite{Dolan:2014ska}. }  The effective theory breaks down at a high-energy scale $\Lambda$, which has been introduced to regularize the generally divergent loop integral. We further set $\Lambda=1~\tev$, resulting in $g_{asb}=3.28 \times 10^{-5} \ g_Y$.

\begin{description}
\item [Production] In analogy to ALPs with derivative couplings to the SM fermions described in \secref{a2}, pseudoscalar $a$ is dominantly  produced in rare decays of $B$ mesons, $B\rightarrow X_s\,a$. When modeling this, we employ the relevant branching ratio calculated at the quark level~\cite{Dolan:2014ska}
\be
B(b \to s a) 
=  \frac{m_b}{\Gamma_B}\frac{|g_{asb}|^2}{32\pi} \left( 1-\frac{m_\phi^2}{m_b^2}\right)^2 
= 122 \times \left( 1-\frac{m_\phi^2}{m_b^2}\right)^2 g_{Y}^2\ .
\label{eq:BtosaYuk}
\ee 
We have also checked that this method is in a good agreement with a data-driven approach discussed in Ref.~\cite{Aloni:2018vki}, which assumes $B(b \to s a)\simeq 5\,\left[B(B\to Ka)+B(B\to K^* a)\right]$.
In the following, we neglect additional contributions from kaon decays $K \to \pi a$ that are sizable only in regions of parameter space with $m_a<m_K-m_\pi$, which are already strongly constrained by other experiments. 

\item[Decay and Lifetime] As dictated by the Yukawa-like nature of the couplings to the SM, the pseudoscalar $a$ decays dominantly into the heaviest accessible SM fermions. In addition, further suppression of the hadronic decay widths discussed in \secref{a2} leads to the dominant decays into the SM leptons. The corresponding branching ratios into fermions $f=e,\mu,\tau,b,c$ are given by~\cite{Dolan:2014ska}
\be
\Gamma(a\to ff)= N_c^f g_Y^2 \, \frac{m_a m_f^2}{8\pi v^2}\,\sqrt{1-\frac{4 m_f^2}{m_a^2}} \ .
\ee
For the hadronic and photonic branching ratios, we adopt the recent results of Ref.~\cite{Domingo:2016yih},\footnote{Note that Ref.~\cite{Domingo:2016yih} estimates the decay width of a pseudoscalar in the NMSSM. However, given that the hadronic branching fraction mainly originates from the pseudoscalar coupling to strange quarks, the branching fractions are roughly independent of $\tan\beta$ and can be applied to this work.} in which the hadronic decay widths were estimated employing the chiral Lagrangian for $m_a\lesssim 1~\gev$ and using the spectator model for larger masses. For even larger masses $m_a>3~\gev$, we estimate the hadronic decay width through the  partonic width into strange and charm quarks. We show the resulting branching ratios and lifetime as a function of the pseudoscalar mass in the left panel of \figref{p1}.

\item[Results] The expected FASER reach in this model is shown in the right panel of \figref{p1}. As can be seen, FASER 2 can cover some currently unconstrained regions in the parameter space reaching up to about $m_a \simeq 2 m_\tau$ and values of $g_{Y}$ below $10^{-5}$. Current bounds on this model exclude the gray-shaded region, following Ref.~\cite{Dobrich:2018jyi}. For comparison, we also show the expected reach~\cite{Dobrich:2018jyi} of the proposed SHiP detector obtained for $10^{20}$ POT, as well as for the NA62 experiment assuming $10^{18}$ POT and pseudoscalars produced in $B$ meson decays in the up stream copper beam collimator.
\end{description}

\begin{figure}[tb]
\centering
\vspace*{-0.6cm}
\includegraphics[width=0.98\textwidth]{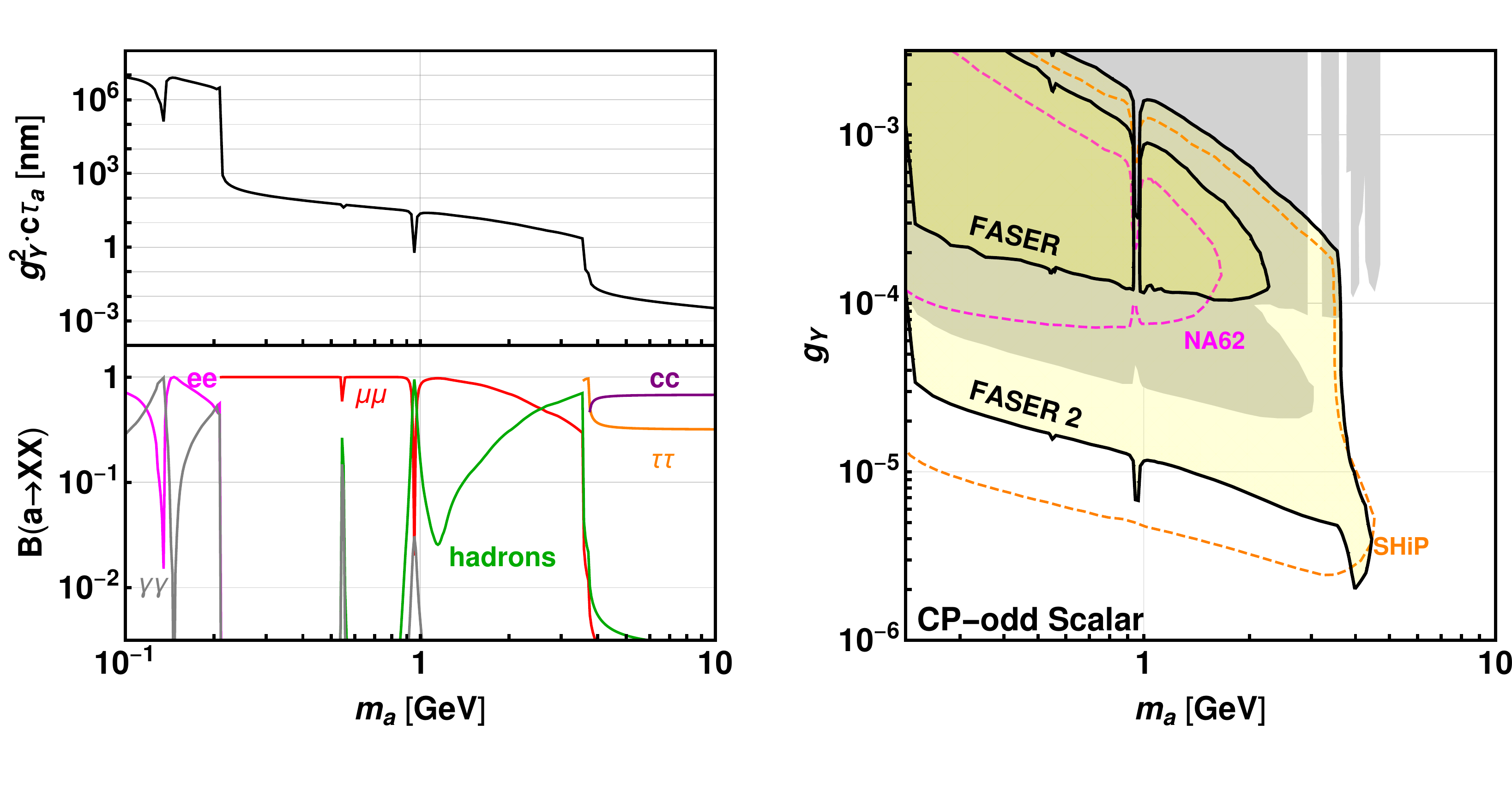}  
\caption{
{\bf Benchmark Model P1.} The decay length (top left panel), decay branching fractions (bottom left panel) and FASER’s reach (right panel) for a CP-odd scalar with Yukawa-like couplings.  The gray-shaded regions are excluded by current limits, and the colored contours give the projected sensitivities of several other proposed experiments. See the text for details.
}
\label{fig:p1}
\end{figure}

\section{Dependence on Beam Offset, Monte-Carlo Generators, Energy Threshold, and Signal Efficiency}
\label{sec:systematics}

\setlength{\abovecaptionskip}{0pt}

In the previous sections, we have presented the expected reach of FASER and FASER 2 in searches for several popular candidates for light and long-lived new particles. The obtained results correspond to the detector setups and modeling of particle production that have been outlined in \secsref{FASERdetails}{production}, respectively. 

In this section, we explore how robust these results are to variations in some of our underlying assumptions.  In \secref{beamoffset}, we determine how sensitive our results are to the assumption that the detector is perfectly centered on the beam collision axis. In \secref{mcpdf}, we investigate the dependence of our reach plots on the choice of Monte-Carlo generator and PDFs used in modeling particle production. In \secref{energythreshold}, we investigate the change in sensitivity when imposing a threshold on the energy of the long-lived particle. Finally, in \secref{signalefficiency}, we briefly comment on the dependence on signal efficiency.  As we will see, for reasonable variations in all of these assumptions, the sensitivity reaches vary little, and in some cases, almost imperceptibly.  

To illustrate these dependences, we will consider two representative models of new physics: the dark photon model V1 discussed in \secref{vanillaAprim}, and the ALP with fermion couplings model A2 discussed in \secref{a2}. These are representative in the sense that dark photons are mainly produced through light meson decays and dark bremsstrahlung and so are highly collimated, whereas ALPs with dominantly fermion couplings are typically produced in heavy meson decays and have larger $p_T$.  These two models therefore bracket the possible dependences on the exact position of the detector relative to the beam collision axis, and they also sample all the different production models used to determine signal rates throughout this study.

\subsection{Dependence on Beam Collision Axis Offset}
\label{sec:beamoffset}

In the previous sections we have assumed that the beam collision axis passes through the center of FASER's cylindrical decay volume. The beam collision axis has been mapped out by the CERN survey team in both the TI18 and TI12 tunnels to $\mm$ precision, assuming no crossing angle between the beams at IP1. However, to avoid long range beam-beam effects and parasitic collisions inside the common beam pipe, the LHC currently runs with a crossing half-angle that can be as large as $160~\mu$rad at IP1. At the FASER location, this crossing angle corresponds to a shift of the collision axis of roughly 7.2 cm compared to the nominal line of sight assuming no crossing angle. The crossing angle varies in time, and both the orientation and size of the beam crossing angle have not been fixed yet for the upcoming runs of the LHC. Indeed, at IP1, there are plans to flip the crossing angle from up to down in the vertical plane periodically (e.g., once per year) to distribute the collision debris or possibly to switch to horizontal cross angles.  In addition, the half-crossing angles may be reduced to a minimum of $\sim 120~\mu$rad during fills to increase the deliverable luminosity. The crossing angle may also be larger for the HL-LHC.  All these effects will lead to an offset $d$ between the center of the detector and the beam collision axis. 

The impact on the sensitivity reach of such an offset is analyzed in \figref{eff1} for offset parameters similar to the detector radius: $d=5, 10, 20~\cm$ for FASER and $d=0.5, 1, 2 ~\m$ for FASER 2. In particular, for a dark photon with mass $m_{A'}=100~\mev$ and $\epsilon=10^{-5}$, the expected number of events at FASER decreases from 8.4 for no offset to 7.6 (4.9, 1.2) for an offset of $5~\cm$ ($10~\cm$, $20~\cm$). We see that the impact of a beam offset is tiny as long as $d<R$, i.e., the offset is small enough that the beam axis still goes through the detector. This implies that a possible shift in the actual position of the beam collision axis of $d \approx 7.2~\cm$ due to variations of the beam crossing angle will not change the physics potential of the FASER detector, even for particles like the dark photon, that are very collimated around the beam axis. In the case of the larger FASER 2 detector, one can see that even much larger displacements are possible without affecting the physics reach.

\begin{figure}[tbp]
\centering
\includegraphics[width=0.48\textwidth]{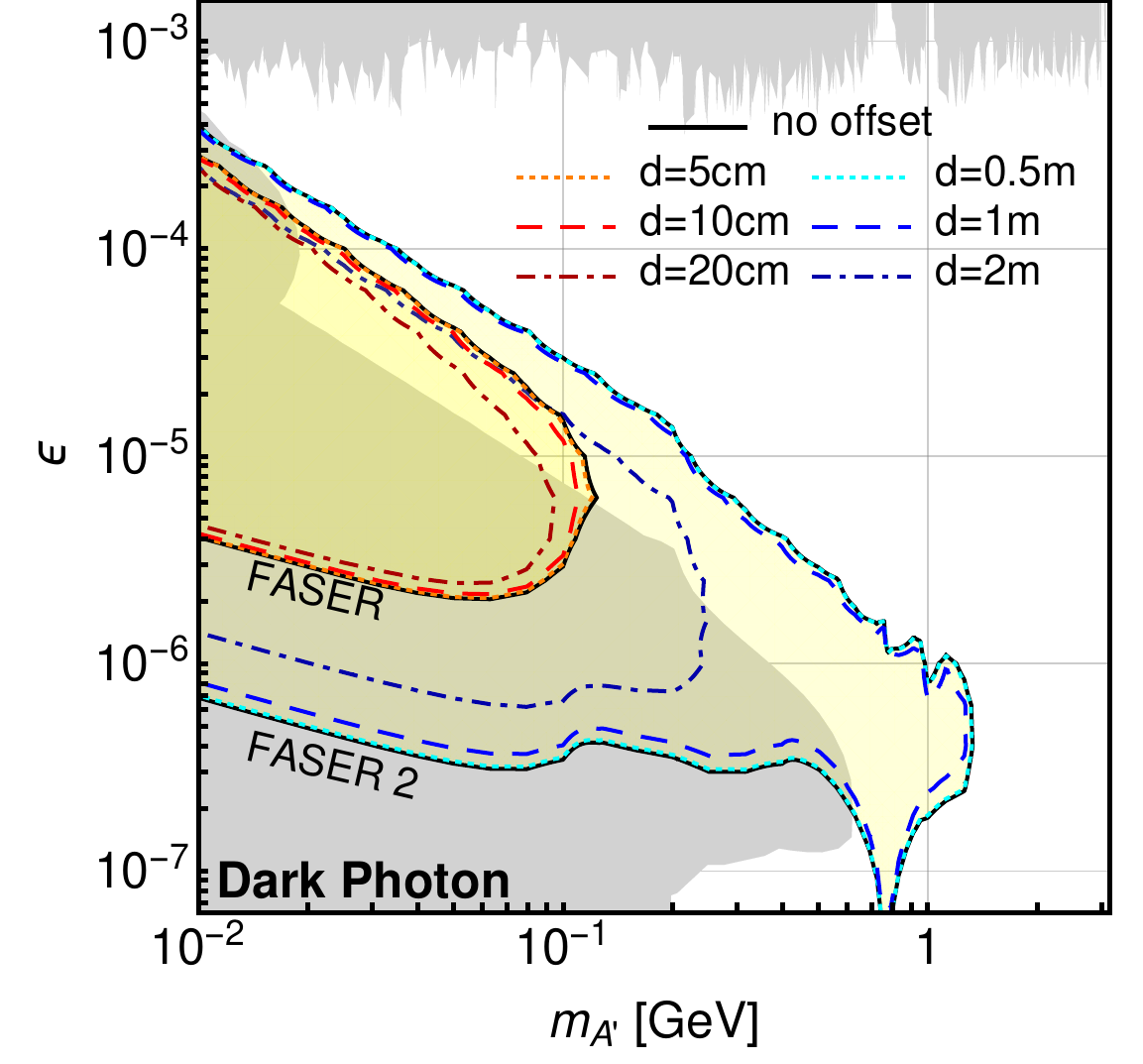}  \hfill
\includegraphics[width=0.48\textwidth]{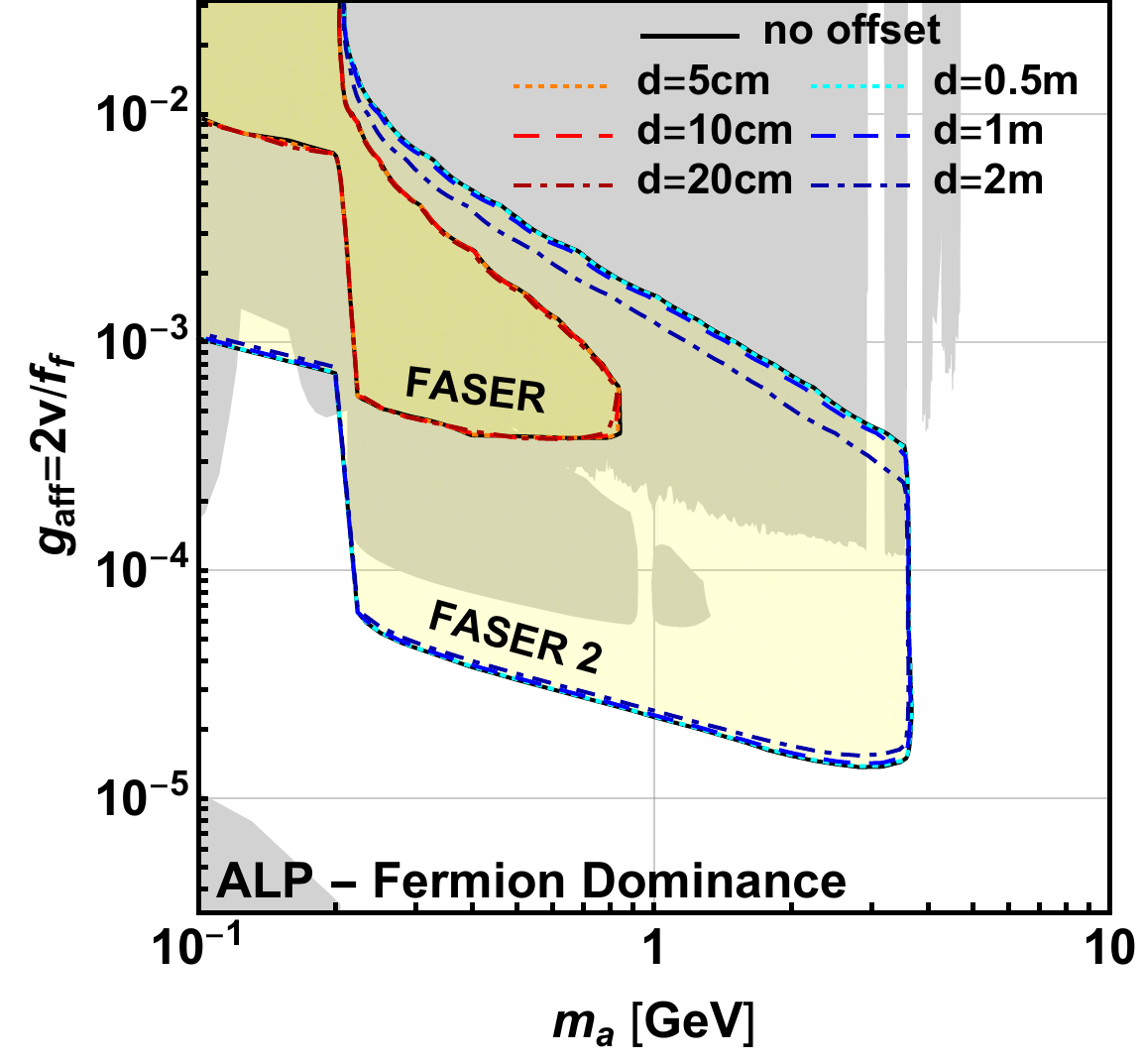}  
\caption{FASER reach for dark photons (left) and ALPs with dominantly fermion couplings (right) for different offsets $d$ between the beam collision axis and the center of FASER.
}
\label{fig:eff1}
\end{figure}

Particles produced in $B$ decay, such as ALPs, dark Higgses and HNLs, typically have a broader $p_T$ spectrum. Hence even large offsets $d$ up to a few meters only have a mild effect on the sensitivity of FASER 2. This implies that FASER 2 need not be built perfectly centered on the beam collision axis, as may be necessary, for example, to accommodate the geometry of the tunnels TI18 and TI12.

\subsection{Dependence on Monte-Carlo Generators and PDFs}
\label{sec:mcpdf}

Although rates for electroweak physics at the LHC have often been calculated with percent level precision, predictions for particle fluxes in the forward direction suffer from larger uncertainties. We therefore study the effect of modeling uncertainties for the production of light and heavy mesons in the far forward region on FASER's sensitivity for LLP searches, as illustrated in \figref{eff2}. 

In the left panel of \figref{eff2}, we show the sensitivity reaches for dark photons at FASER and FASER 2. The red lines correspond to dark photons produced in the decays of light mesons, $\pi^0, \eta \to A' \gamma$. Different lines correspond to several publicly available Monte-Carlo generators used to estimate the spectrum of $\pi^0$ and $\eta$ mesons produced in the far forward region: EPOS-LHC~\cite{Pierog:2013ria}, QGSJET II-04~\cite{Ostapchenko:2010vb}, and SIBYLL 2.3~\cite{Ahn:2009wx, Riehn:2015oba}. As can be seen, using various generators leads to almost imperceptible differences in the final sensitivities. 

\begin{figure}[tbp]
\centering
\includegraphics[width=0.48\textwidth]{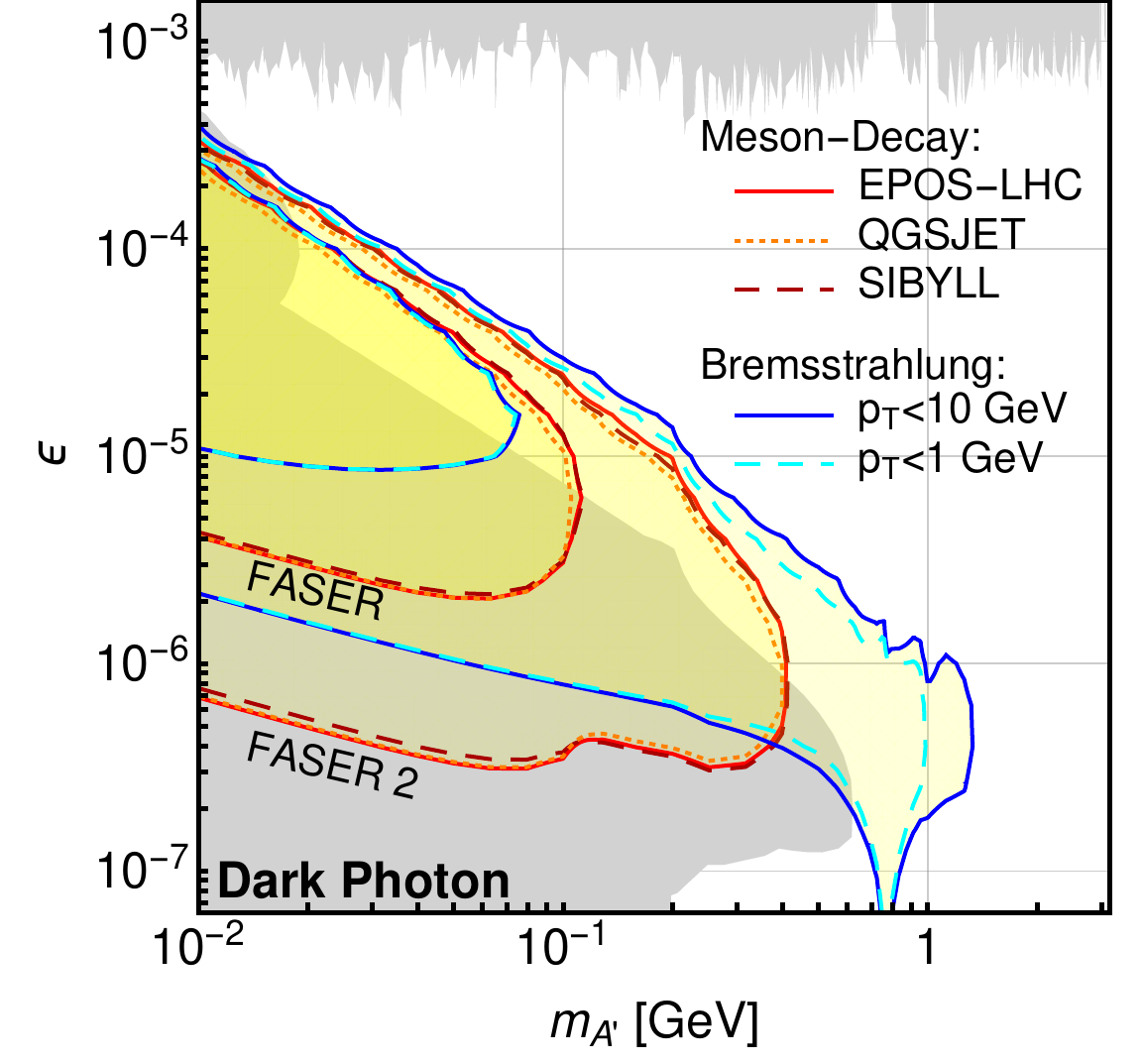} \hfill
\includegraphics[width=0.48\textwidth]{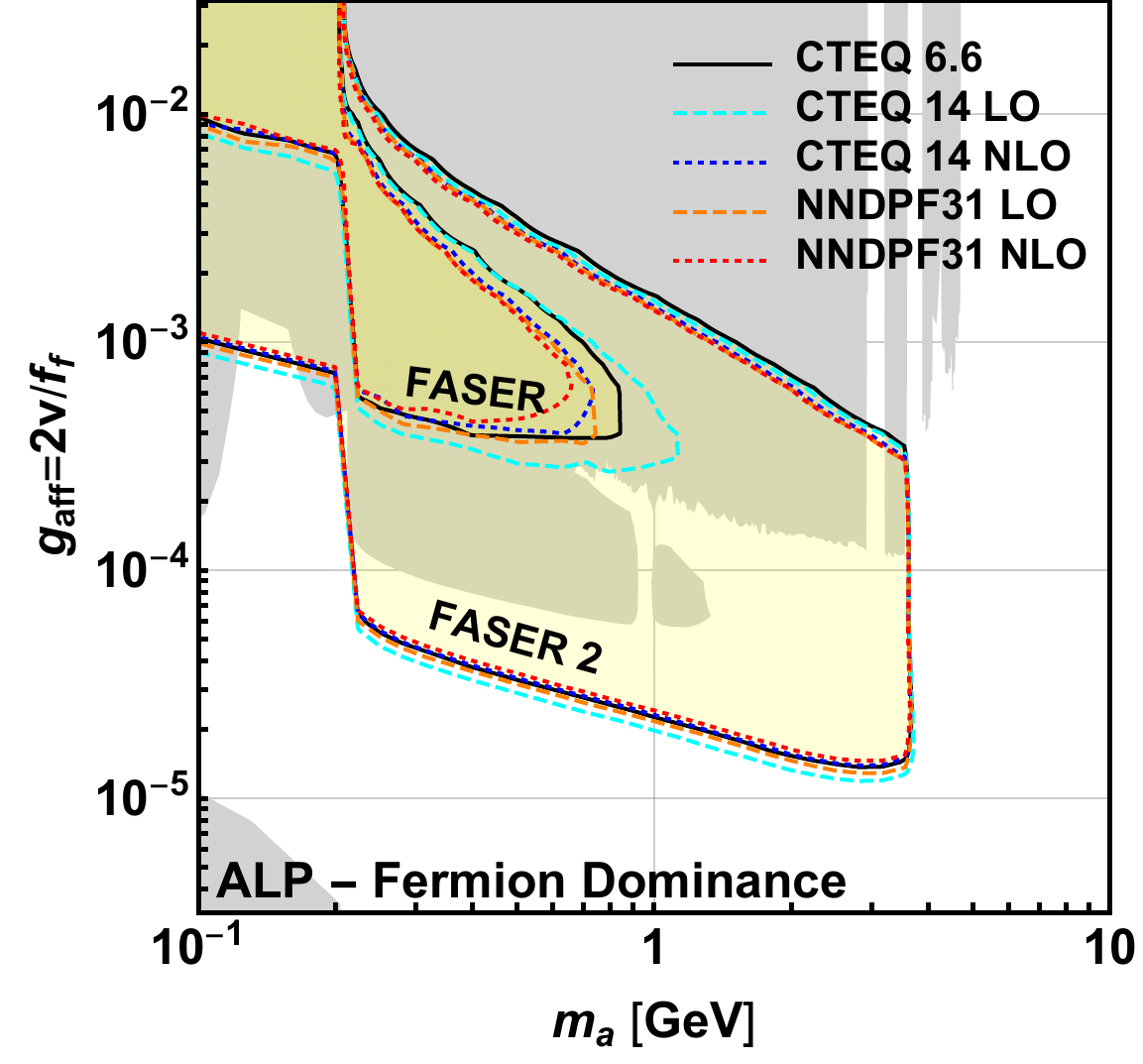}  
\caption{FASER reach for dark photons (left) and ALPs with dominant couplings to fermions (right). For the dark photon, we vary the forward Monte-Carlo generators used to produce the light meson spectrum as well as the validity on the transverse momentum of the dark photon used in the bremsstahlung approximation. For the ALPs, we change the PDF used to estimate the forward $B$-meson spectra in FONLL.  
}
\label{fig:eff2}
\end{figure}

The blue curves in the left panel of \figref{eff2} correspond to varying the cut-off scale for the transverse momentum of the dark photon up to which the Fermi-Weizsacker-Williams approximation for dark bremsstrahlung production of dark photons can be safely used. Although we use $p_{T,A'} < 10~\gev$ as our default choice, a more conservative threshold of $p_{T,A'} < 1~\gev \approx m_p$ does not change FASER's reach significantly. Only a small region of the parameter space corresponding to larger values of $m_{A’}$ and, therefore, typically larger spread in the transverse momentum, is affected by reduction of the maximum allowed value of $p_T$ from $10~\gev$ to $1~\gev$. 

In the right panel of \figref{eff2}, we compare the FASER and FASER 2 reaches in searches for ALPs with dominant couplings to fermions employing different PDFs. Although throughout this paper we use CTEQ 6.6~\cite{Nadolsky:2008zw} as our default choice, here we also consider more recent PDFs sets: CT14~\cite{Dulat:2015mca} and NNPDF3.1~\cite{Ball:2017nwa} in both their LO and NLO implementations. We can see that all of these PDF sets give similar physics reaches. While LO implementations typically lead to slightly enhanced rates and sensitivities, we have checked that the NNLO implementations of both CT14 and NNPDF3.1 given almost indistinguishable results compared to the NLO implementations. We have also analyzed the effect of changing the scale choice by a factor of two and found that the resulting rate variations are smaller than the variations due to the PDF choice. Finally we checked that the modeling of fragmentation has a negligible effect on the reach.

\subsection{Dependence on the Energy Threshold}
\label{sec:energythreshold}

To obtain FASER's sensitivity in the previous sections, we have applied an energy threshold of $E_{A'}>100~\gev$ to reduce the trigger rate and to remove possible low-energy backgrounds. This choice is mainly determined by the LLP's kinematics and FASER's geometry, as shown in \figref{spectrum}. On the one hand, the typical transverse momentum scale of LLPs produced in meson decay is given by the meson mass $p_T \sim m_{\text{meson}} \sim \gev$. On the other hand, FASER only covers the very forward direction with $\theta \lesssim \mrad$, where $\theta$ denotes the angle with respect to the beam axis.  Therefore, the energy of an LLP traveling in the direction of FASER is typically large, with $E \sim p_T / \theta \sim  \tev$ (cf. \eqref{theta}), well above the chosen threshold. 

The above argument shows that a higher minimal energy could be chosen without reducing FASER's physics sensitivity. The impact on the sensitivity reach of requiring different minimum energies for the LLP is presented in \figref{eff3} for energy thresholds $E_{\text{LLP}}> 100$, $200$, $500$ and $1000~\gev$. Requiring a larger LLP energy reduces the reach in the low coupling regime, in which the LLP production rates are small and the LLP lifetime is long, with $c\tau\gamma \gg 480~\m$. However, even imposing a very large energy threshold $E_{\text{LLP} }> 1~\tev$ only has a mild impact on FASER's reach. In particular, note that for dark photons, a larger energy threshold only effects the reach in a region of parameter space that is already excluded by previous experiments.

\begin{figure}[tbp]
\centering
\includegraphics[width=0.48\textwidth]{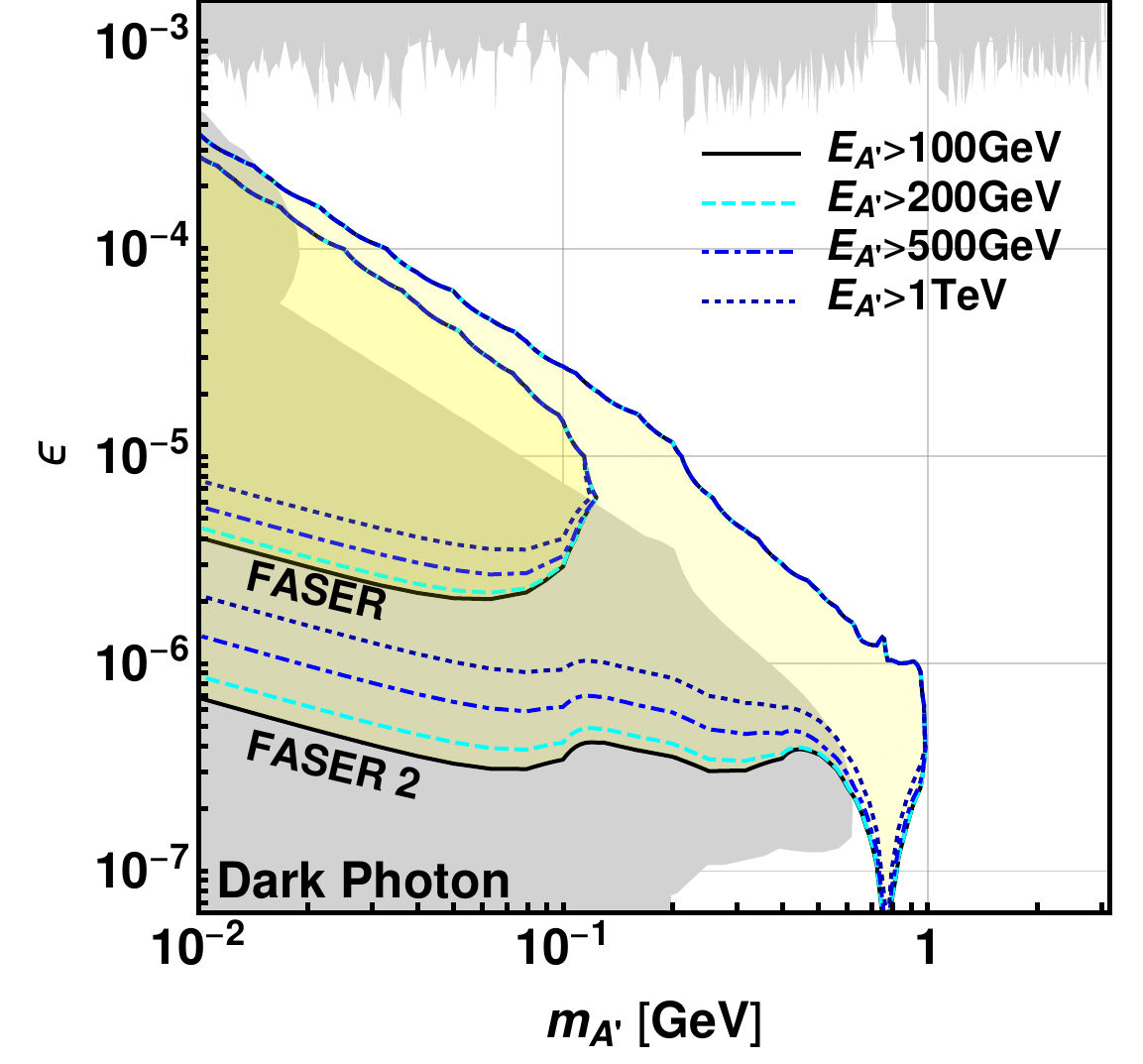} \hfill
\includegraphics[width=0.48\textwidth]{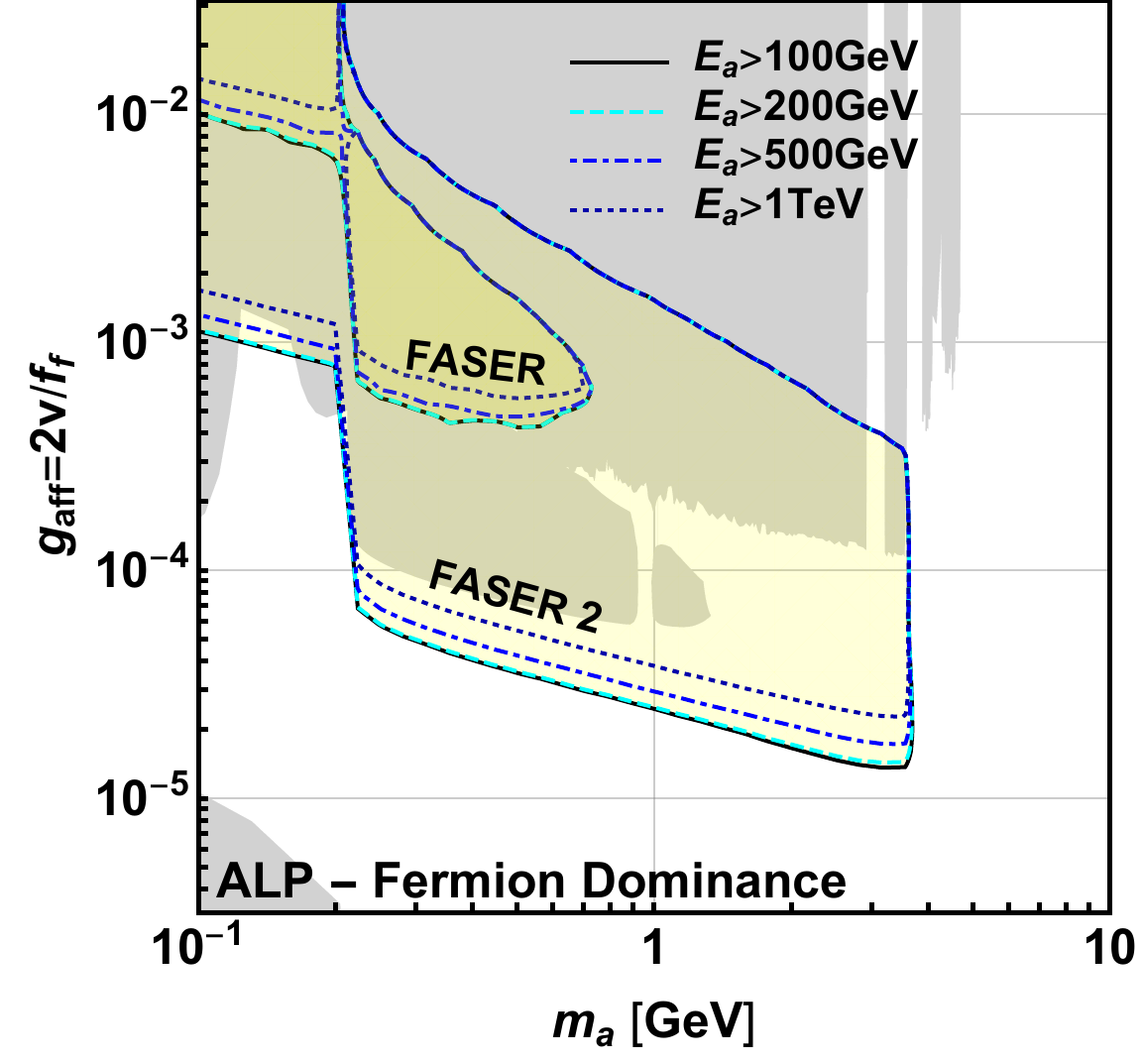}  
\caption{FASER reach for dark photons (left) and ALPs with dominant couplings to fermions (right) for different LLP energy threshold cuts.   
}
\label{fig:eff3}
\end{figure}

Further reducing the energy threshold does not improve the reach for the models considered here. However, as discussed in App.~A of Ref.~\cite{Berlin:2018jbm}, a lower threshold can improve the reach in inelastic dark matter scenarios. 

\subsection{Dependence on Signal Efficiency}
\label{sec:signalefficiency}

Among the other important factors that determine the physics reach of the experiment is the efficiency of the detector response and event reconstruction. A detailed discussion of these effects is beyond the scope of this paper. However, it is useful to note that an initial analysis of these effects was carried out for the FASER Letter of Intent~\cite{Ariga:2018zuc}, focusing on the case of dark photons decay to $e^+ e^-$ pairs.  In particular, it was shown that, even requiring that the $e^+$ and $e^-$ be separated by $\delta=0.3~\mm$ in the first out of several tracking stations, i.e., the one placed right after the fiducial decay volume, does not drastically affect FASER’s reach. In addition, the sensitivity remains basically unaffected if such a strict condition is applied only to the last two tracking stations. Further detailed analyses of the detector efficiency are currently ongoing with the use of Geant4 simulations~\cite{Agostinelli:2002hh} and dedicated software tools under development.

\section{Conclusions}
\label{sec:summary}

The null results of new physics searches in the high-$p_T$ region of $pp$ collisions call for new ideas that could extend the LHC physics reach. The recently approved FASER experiment will extend the LHC's physics program by searching for new light, weakly coupled LLPs in the far forward region of $pp$ collisions, with the potential to discover physics beyond the SM and shed light on dark matter. The detector will be installed in TI12, an existing and unused tunnel 480 m from the ATLAS IP. FASER will run concurrently with the other LHC experiments, requiring no beam modifications and interacting with the accelerator and existing experiments only in requesting luminosity information from ATLAS and bunch crossing timing information from the LHC.

FASER's discovery prospects for the models discussed in this paper are given in \tableref{summary}. A successful installation in LS2 and data taking during Run 3 will assure FASER's sensitivity to new regions of parameter space for dark photons, other light gauge bosons, HNLs with dominantly $\tau$ couplings, and axion-like particles with masses in the 10 MeV to GeV range.  A larger detector, FASER 2, running in the HL-LHC era, will extend this sensitivity to larger masses and will probe currently unconstrained parameter space for all renormalizable portals (dark photons, dark Higgs bosons, and heavy neutral leptons), ALPs with photon, fermion, or gluon couplings, pseudoscalars with Yukawa-like couplings, and many other new particles.  These new physics scenarios discussed here have significant overlap with the benchmark scenarios studied by the CERN Physics Beyond Colliders study group~\cite{Beacham:2019nyx}, and this work provides the details behind the FASER results summarized by that study group. 

Although the LLP models considered here are among the most widely discussed, it is important to note that they do not exhaust the full physics potential of the detectors. In particular, FASER’s discovery potential has already been discussed in other new physics models, including inelastic dark matter~\cite{Berlin:2018jbm}, $R$-parity violating supersymmetry~\cite{Helo:2018qej,Dercks:2018eua}, models with strongly interacting massive particles (SIMPs)~\cite{Hochberg:2018rjs}, and twin Higgs scenarios~\cite{Cheng:2018vaj}. In addition, when more complete models of BSM physics are considered, it is often natural that more than one new light particle can appear, e.g., both a dark photon and a dark Higgs boson, leading to opportunities to simultaneously discovery more than one new particle in FASER and FASER 2. Dedicated analyses of such scenarios, as well as other LLP models, are left for future studies that can be performed employing the detector details described in~\secref{FASERdetails}.

\acknowledgments

We thank our colleagues in the Physics Beyond Colliders study group for useful discussions, particularly David Curtin, Masahiro Ibe, Simon Knapen, Gaia Lanfranchi, and Maxim Pospelov, and we also thank Felix Kahlhoefer for providing us the current constraints for ALPs with dominant couplings to fermions. This work was supported by grants from the Heising-Simons Foundation (Grant Nos.~2018-1135 and 2019-1179) and the Simons Foundation (Grant No.~623683).  J.L.F., F.K., and J.S.~are supported in part by U.S.~National Science Foundation Grant No.~PHY-1620638.  J.L.F.~is supported in part by Simons Investigator Award \#376204. This publication is funded in part by the Gordon and Betty Moore Foundation through Grant GBMF6210 to support the work of F.K.. I.G.~is supported in part by U.S.~Department of Energy Grant DOE-SC0010008. S.T.~is supported in part by the Lancaster-Manchester-Sheffield Consortium for Fundamental Physics under STFC grant ST/L000520/1, by the Polish Ministry of Science and Higher Education under research grant 1309/MOB/IV/2015/0, and by the National Science Centre (NCN) research Grant No.~2015-18-A-ST2-00748.  This work is supported in part by the Swiss National Science Foundation. 

\bibliography{Benchmarks}

\providecommand{\href}[2]{#2}\begingroup\raggedright\begin{thebibliography}{100}

\bibitem{Battaglieri:2017aum}
M.~Battaglieri {\em et al.}, ``{US Cosmic Visions: New Ideas in Dark Matter
  2017: Community Report},''
\href{http://arxiv.org/abs/1707.04591}{{\ttfamily arXiv:1707.04591 [hep-ph]}}.

\bibitem{Boehm:2003hm}
C.~Boehm and P.~Fayet, ``{Scalar dark matter candidates},''
  \href{http://dx.doi.org/10.1016/j.nuclphysb.2004.01.015}{{\em Nucl. Phys.}
  {\bfseries B683} (2004) 219--263},
\href{http://arxiv.org/abs/hep-ph/0305261}{{\ttfamily arXiv:hep-ph/0305261
  [hep-ph]}}.

\bibitem{Feng:2008ya}
J.~L. Feng and J.~Kumar, ``{The WIMPless Miracle: Dark-Matter Particles without
  Weak-Scale Masses or Weak Interactions},''
  \href{http://dx.doi.org/10.1103/PhysRevLett.101.231301}{{\em Phys. Rev.
  Lett.} {\bfseries 101} (2008) 231301},
\href{http://arxiv.org/abs/0803.4196}{{\ttfamily arXiv:0803.4196 [hep-ph]}}.

\bibitem{Bennett:2006fi}
{\bfseries Muon g-2} Collaboration, G.~W. Bennett {\em et al.}, ``{Final Report
  of the Muon E821 Anomalous Magnetic Moment Measurement at BNL},''
  \href{http://dx.doi.org/10.1103/PhysRevD.73.072003}{{\em Phys. Rev.}
  {\bfseries D73} (2006) 072003},
\href{http://arxiv.org/abs/hep-ex/0602035}{{\ttfamily arXiv:hep-ex/0602035
  [hep-ex]}}.

\bibitem{Pohl:2010zza}
R.~Pohl {\em et al.}, ``{The size of the proton},''
\href{http://dx.doi.org/10.1038/nature09250}{{\em Nature} {\bfseries 466}
  (2010) 213--216}.

\bibitem{Krasznahorkay:2015iga}
A.~J. Krasznahorkay {\em et al.}, ``{Observation of Anomalous Internal Pair
  Creation in Be8 : A Possible Indication of a Light, Neutral Boson},''
  \href{http://dx.doi.org/10.1103/PhysRevLett.116.042501}{{\em Phys. Rev.
  Lett.} {\bfseries 116} (2016) 042501},
\href{http://arxiv.org/abs/1504.01527}{{\ttfamily arXiv:1504.01527 [nucl-ex]}}.

\bibitem{Feng:2017uoz}
J.~L. Feng, I.~Galon, F.~Kling, and S.~Trojanowski, ``{ForwArd Search
  ExpeRiment at the LHC},''
  \href{http://dx.doi.org/10.1103/PhysRevD.97.035001}{{\em Phys. Rev.}
  {\bfseries D97} no.~3, (2018) 035001},
\href{http://arxiv.org/abs/1708.09389}{{\ttfamily arXiv:1708.09389 [hep-ph]}}.

\bibitem{Aaboud:2016mmw}
{\bfseries ATLAS} Collaboration, M.~Aaboud {\em et al.}, ``{Measurement of the
  Inelastic Proton-Proton Cross Section at $\sqrt{s}=13$ TeV with the ATLAS
  Detector at the LHC},''
  \href{http://dx.doi.org/10.1103/PhysRevLett.117.182002}{{\em Phys. Rev.
  Lett.} {\bfseries 117} no.~18, (2016) 182002},
\href{http://arxiv.org/abs/1606.02625}{{\ttfamily arXiv:1606.02625 [hep-ex]}}.

\bibitem{VanHaevermaet:2016gnh}
{\bfseries CMS} Collaboration, H.~Van~Haevermaet, ``{Measurement of the
  inelastic proton-proton cross section at $\sqrt{s}$ = 13 TeV},''
  \href{http://dx.doi.org/10.22323/1.265.0198}{{\em PoS} {\bfseries DIS2016}
  (2016) 198},
\href{http://arxiv.org/abs/1607.02033}{{\ttfamily arXiv:1607.02033 [hep-ex]}}.

\bibitem{Ferrari:2005zk}
A.~Ferrari, P.~R. Sala, A.~Fasso, and J.~Ranft, {\em {FLUKA: A Multi-particle
  Transport Code (Program Version 2005)}}.
\newblock CERN Yellow Reports: Monographs. CERN, Geneva, 2005.
\newblock
\url{http://cds.cern.ch/record/898301}.
\newblock

\bibitem{Bohlen:2014buj}
T.~T. B{\"o}hlen, F.~Cerutti, M.~P.~W. Chin, A.~Fasso, A.~Ferrari, P.~G.
  Ortega, A.~Mairani, P.~R. Sala, G.~Smirnov, and V.~Vlachoudis, ``{The FLUKA
  Code: Developments and Challenges for High Energy and Medical
  Applications},''
\href{http://dx.doi.org/10.1016/j.nds.2014.07.049}{{\em Nucl. Data Sheets}
  {\bfseries 120} (2014) 211--214}.

\bibitem{FLUKAstudy}
{CERN Sources, Targets, and Interactions Group}, M.~Sabate-Gilarte, F.~Cerutti,
  A.~Tsinganis, {\em et al.}, ``{Characterization of the radiation field for
  the FASER experiment},''.

\bibitem{Moreno:2013mja}
O.~Moreno, ``{The Heavy Photon Search Experiment at Jefferson Lab},'' in {\em
  {Meeting of the APS Division of Particles and Fields (DPF 2013) Santa Cruz,
  California, USA, August 13-17, 2013}}.
\newblock 2013.
\newblock
\href{http://arxiv.org/abs/1310.2060}{{\ttfamily arXiv:1310.2060
  [physics.ins-det]}}.
\newblock

\bibitem{Dolan:2017osp}
M.~J. Dolan, T.~Ferber, C.~Hearty, F.~Kahlhoefer, and K.~Schmidt-Hoberg,
  ``{Revised constraints and Belle II sensitivity for visible and invisible
  axion-like particles},''
  \href{http://dx.doi.org/10.1007/JHEP12(2017)094}{{\em JHEP} {\bfseries 12}
  (2017) 094},
\href{http://arxiv.org/abs/1709.00009}{{\ttfamily arXiv:1709.00009 [hep-ph]}}.

\bibitem{Ilten:2015hya}
P.~Ilten, J.~Thaler, M.~Williams, and W.~Xue, ``{Dark photons from charm mesons
  at LHCb},'' \href{http://dx.doi.org/10.1103/PhysRevD.92.115017}{{\em Phys.
  Rev.} {\bfseries D92} no.~11, (2015) 115017},
\href{http://arxiv.org/abs/1509.06765}{{\ttfamily arXiv:1509.06765 [hep-ph]}}.

\bibitem{Ilten:2016tkc}
P.~Ilten, Y.~Soreq, J.~Thaler, M.~Williams, and W.~Xue, ``{Proposed Inclusive
  Dark Photon Search at LHCb},''
  \href{http://dx.doi.org/10.1103/PhysRevLett.116.251803}{{\em Phys. Rev.
  Lett.} {\bfseries 116} no.~25, (2016) 251803},
\href{http://arxiv.org/abs/1603.08926}{{\ttfamily arXiv:1603.08926 [hep-ph]}}.

\bibitem{Dobrich:2018ezn}
{\bfseries NA62} Collaboration, B.~D{\"o}brich, ``{Dark Sectors at fixed
  targets: The example of NA62},''
\newblock 2018.
\newblock
\href{http://arxiv.org/abs/1807.10170}{{\ttfamily arXiv:1807.10170 [hep-ex]}}.
\newblock

\bibitem{Gninenko:2018tlp}
S.~N. Gninenko and N.~V. Krasnikov, ``{Probing the muon $g_\mu-2$ anomaly,
  $L_{\mu} - L_{\tau}$ gauge boson and Dark Matter in dark photon
  experiments},'' \href{http://dx.doi.org/10.1016/j.physletb.2018.06.043}{{\em
  Phys. Lett.} {\bfseries B783} (2018) 24--28},
\href{http://arxiv.org/abs/1801.10448}{{\ttfamily arXiv:1801.10448 [hep-ph]}}.

\bibitem{Berlin:2018pwi}
A.~Berlin, S.~Gori, P.~Schuster, and N.~Toro, ``{Dark Sectors at the Fermilab
  SeaQuest Experiment},''
  \href{http://dx.doi.org/10.1103/PhysRevD.98.035011}{{\em Phys. Rev.}
  {\bfseries D98} no.~3, (2018) 035011},
\href{http://arxiv.org/abs/1804.00661}{{\ttfamily arXiv:1804.00661 [hep-ph]}}.

\bibitem{Alekhin:2015byh}
S.~Alekhin {\em et al.}, ``{A facility to Search for Hidden Particles at the
  CERN SPS: the SHiP physics case},''
  \href{http://dx.doi.org/10.1088/0034-4885/79/12/124201}{{\em Rept. Prog.
  Phys.} {\bfseries 79} (2016) 124201},
\href{http://arxiv.org/abs/1504.04855}{{\ttfamily arXiv:1504.04855 [hep-ph]}}.

\bibitem{Evans:2017lvd}
J.~A. Evans, ``{Detecting Hidden Particles with MATHUSLA},''
  \href{http://dx.doi.org/10.1103/PhysRevD.97.055046}{{\em Phys. Rev.}
  {\bfseries D97} no.~5, (2018) 055046},
\href{http://arxiv.org/abs/1708.08503}{{\ttfamily arXiv:1708.08503 [hep-ph]}}.

\bibitem{Curtin:2018mvb}
D.~Curtin {\em et al.}, ``{Long-Lived Particles at the Energy Frontier: The
  MATHUSLA Physics Case},''
\href{http://arxiv.org/abs/1806.07396}{{\ttfamily arXiv:1806.07396 [hep-ph]}}.

\bibitem{Gligorov:2017nwh}
V.~V. Gligorov, S.~Knapen, M.~Papucci, and D.~J. Robinson, ``{Searching for
  Long-lived Particles: A Compact Detector for Exotics at LHCb},''
  \href{http://dx.doi.org/10.1103/PhysRevD.97.015023}{{\em Phys. Rev.}
  {\bfseries D97} no.~1, (2018) 015023},
\href{http://arxiv.org/abs/1708.09395}{{\ttfamily arXiv:1708.09395 [hep-ph]}}.

\bibitem{Gligorov:2018vkc}
V.~V. Gligorov, S.~Knapen, B.~Nachman, M.~Papucci, and D.~J. Robinson,
  ``{Leveraging the ALICE/L3 cavern for long-lived exotics},''
\href{http://arxiv.org/abs/1810.03636}{{\ttfamily arXiv:1810.03636 [hep-ph]}}.

\bibitem{Berlin:2018bsc}
A.~Berlin, N.~Blinov, G.~Krnjaic, P.~Schuster, and N.~Toro, ``{Dark Matter,
  Millicharges, Axion and Scalar Particles, Gauge Bosons, and Other New Physics
  with LDMX},''
\href{http://arxiv.org/abs/1807.01730}{{\ttfamily arXiv:1807.01730 [hep-ph]}}.

\bibitem{Feng:2017vli}
J.~L. Feng, I.~Galon, F.~Kling, and S.~Trojanowski, ``{Dark Higgs Bosons at
  FASER},'' \href{http://dx.doi.org/10.1103/PhysRevD.97.055034}{{\em Phys.
  Rev.} {\bfseries D97} (2018) 055034},
\href{http://arxiv.org/abs/1710.09387}{{\ttfamily arXiv:1710.09387 [hep-ph]}}.

\bibitem{Batell:2017kty}
B.~Batell, A.~Freitas, A.~Ismail, and D.~Mckeen, ``{Flavor-specific scalar
  mediators},'' \href{http://dx.doi.org/10.1103/PhysRevD.98.055026}{{\em Phys.
  Rev.} {\bfseries D98} no.~5, (2018) 055026},
\href{http://arxiv.org/abs/1712.10022}{{\ttfamily arXiv:1712.10022 [hep-ph]}}.

\bibitem{Kling:2018wct}
F.~Kling and S.~Trojanowski, ``{Heavy Neutral Leptons at FASER},''
  \href{http://dx.doi.org/10.1103/PhysRevD.97.095016}{{\em Phys. Rev.}
  {\bfseries D97} no.~9, (2018) 095016},
\href{http://arxiv.org/abs/1801.08947}{{\ttfamily arXiv:1801.08947 [hep-ph]}}.

\bibitem{Helo:2018qej}
J.~C. Helo, M.~Hirsch, and Z.~S. Wang, ``{Heavy neutral fermions at the
  high-luminosity LHC},'' \href{http://dx.doi.org/10.1007/JHEP07(2018)056}{{\em
  JHEP} {\bfseries 07} (2018) 056},
\href{http://arxiv.org/abs/1803.02212}{{\ttfamily arXiv:1803.02212 [hep-ph]}}.

\bibitem{Bauer:2018onh}
M.~Bauer, P.~Foldenauer, and J.~Jaeckel, ``{Hunting All the Hidden Photons},''
  \href{http://dx.doi.org/10.1007/JHEP07(2018)094}{{\em JHEP} {\bfseries 07}
  (2018) 094},
\href{http://arxiv.org/abs/1803.05466}{{\ttfamily arXiv:1803.05466 [hep-ph]}}.

\bibitem{Cheng:2018vaj}
H.-C. Cheng, L.~Li, and R.~Zheng, ``{Coscattering/Coannihilation Dark Matter in
  a Fraternal Twin Higgs Model},''
  \href{http://dx.doi.org/10.1007/JHEP09(2018)098}{{\em JHEP} {\bfseries 09}
  (2018) 098},
\href{http://arxiv.org/abs/1805.12139}{{\ttfamily arXiv:1805.12139 [hep-ph]}}.

\bibitem{Feng:2018noy}
J.~L. Feng, I.~Galon, F.~Kling, and S.~Trojanowski, ``{Axionlike particles at
  FASER: The LHC as a photon beam dump},''
  \href{http://dx.doi.org/10.1103/PhysRevD.98.055021}{{\em Phys. Rev.}
  {\bfseries D98} no.~5, (2018) 055021},
\href{http://arxiv.org/abs/1806.02348}{{\ttfamily arXiv:1806.02348 [hep-ph]}}.

\bibitem{Hochberg:2018rjs}
Y.~Hochberg, E.~Kuflik, R.~Mcgehee, H.~Murayama, and K.~Schutz, ``{Strongly
  interacting massive particles through the axion portal},''
  \href{http://dx.doi.org/10.1103/PhysRevD.98.115031}{{\em Phys. Rev.}
  {\bfseries D98} no.~11, (2018) 115031},
\href{http://arxiv.org/abs/1806.10139}{{\ttfamily arXiv:1806.10139 [hep-ph]}}.

\bibitem{Berlin:2018jbm}
A.~Berlin and F.~Kling, ``{Inelastic Dark Matter at the LHC Lifetime Frontier:
  ATLAS, CMS, LHCb, CODEX-b, FASER, and MATHUSLA},''
  \href{http://dx.doi.org/10.1103/PhysRevD.99.015021}{{\em Phys. Rev.}
  {\bfseries D99} no.~1, (2019) 015021},
\href{http://arxiv.org/abs/1810.01879}{{\ttfamily arXiv:1810.01879 [hep-ph]}}.

\bibitem{Dercks:2018eua}
D.~Dercks, J.~de~Vries, H.~K. Dreiner, and Z.~S. Wang, ``{R-parity Violation
  and Light Neutralinos at CODEX-b, FASER, and MATHUSLA},''
\href{http://arxiv.org/abs/1810.03617}{{\ttfamily arXiv:1810.03617 [hep-ph]}}.

\bibitem{Ariga:2018zuc}
{\bfseries FASER} Collaboration, A.~Ariga {\em et al.}, ``{Letter of Intent for
  FASER: ForwArd Search ExpeRiment at the LHC},''
\href{http://arxiv.org/abs/1811.10243}{{\ttfamily arXiv:1811.10243
  [physics.ins-det]}}.

\bibitem{Beacham:2019nyx}
J.~Beacham {\em et al.}, ``{Physics Beyond Colliders at CERN: Beyond the
  Standard Model Working Group Report},''
\href{http://arxiv.org/abs/1901.09966}{{\ttfamily arXiv:1901.09966 [hep-ex]}}.

\bibitem{Ariga:2018pin}
{\bfseries FASER} Collaboration, A.~Ariga {\em et al.}, ``{Technical Proposal
  for FASER: ForwArd Search ExpeRiment at the LHC},''
\href{http://arxiv.org/abs/1812.09139}{{\ttfamily arXiv:1812.09139
  [physics.ins-det]}}.

\bibitem{Groom:2001kq}
D.~E. Groom, N.~V. Mokhov, and S.~I. Striganov, ``{Muon stopping power and
  range tables 10-MeV to 100-TeV},''
\href{http://dx.doi.org/10.1006/adnd.2001.0861}{{\em Atom. Data Nucl. Data
  Tabl.} {\bfseries 78} (2001) 183--356}.

\bibitem{VanGinneken:1986rf}
A.~Van~Ginneken, ``{Energy Loss and Angular Characteristics of High-Energy
  Electromagnetic Processes},''
\href{http://dx.doi.org/10.1016/0168-9002(86)91146-0}{{\em Nucl. Instrum.
  Meth.} {\bfseries A251} (1986) 21}.

\bibitem{N.Cartiglia:2015gve}
{\bfseries LHC Forward Physics Working Group} Collaboration, K.~Akiba {\em et
  al.}, ``{LHC Forward Physics},''
  \href{http://dx.doi.org/10.1088/0954-3899/43/11/110201}{{\em J. Phys.}
  {\bfseries G43} (2016) 110201},
\href{http://arxiv.org/abs/1611.05079}{{\ttfamily arXiv:1611.05079 [hep-ph]}}.

\bibitem{Pierog:2013ria}
T.~Pierog, I.~Karpenko, J.~M. Katzy, E.~Yatsenko, and K.~Werner, ``{EPOS LHC:
  Test of collective hadronization with data measured at the CERN Large Hadron
  Collider},'' \href{http://dx.doi.org/10.1103/PhysRevC.92.034906}{{\em Phys.
  Rev.} {\bfseries C92} (2015) 034906},
\href{http://arxiv.org/abs/1306.0121}{{\ttfamily arXiv:1306.0121 [hep-ph]}}.

\bibitem{CRMC}
C.~Baus, T.~Pierog, and R.~Ulrich, ``{Cosmic Ray Monte Carlo (CRMC)},''.
  \url{https://web.ikp.kit.edu/rulrich/crmc.html}.

\bibitem{Nadolsky:2008zw}
P.~M. Nadolsky, H.-L. Lai, Q.-H. Cao, J.~Huston, J.~Pumplin, D.~Stump, W.-K.
  Tung, and C.~P. Yuan, ``{Implications of CTEQ global analysis for collider
  observables},'' \href{http://dx.doi.org/10.1103/PhysRevD.78.013004}{{\em
  Phys. Rev.} {\bfseries D78} (2008) 013004},
\href{http://arxiv.org/abs/0802.0007}{{\ttfamily arXiv:0802.0007 [hep-ph]}}.

\bibitem{Cacciari:1998it}
M.~Cacciari, M.~Greco, and P.~Nason, ``{The P(T) spectrum in heavy flavor
  hadroproduction},''
  \href{http://dx.doi.org/10.1088/1126-6708/1998/05/007}{{\em JHEP} {\bfseries
  05} (1998) 007},
\href{http://arxiv.org/abs/hep-ph/9803400}{{\ttfamily arXiv:hep-ph/9803400
  [hep-ph]}}.

\bibitem{Cacciari:2001td}
M.~Cacciari, S.~Frixione, and P.~Nason, ``{The p(T) spectrum in heavy flavor
  photoproduction},''
  \href{http://dx.doi.org/10.1088/1126-6708/2001/03/006}{{\em JHEP} {\bfseries
  03} (2001) 006},
\href{http://arxiv.org/abs/hep-ph/0102134}{{\ttfamily arXiv:hep-ph/0102134
  [hep-ph]}}.

\bibitem{Braaten:1994bz}
E.~Braaten, K.-m. Cheung, S.~Fleming, and T.~C. Yuan, ``{Perturbative QCD
  fragmentation functions as a model for heavy quark fragmentation},''
  \href{http://dx.doi.org/10.1103/PhysRevD.51.4819}{{\em Phys. Rev.} {\bfseries
  D51} (1995) 4819--4829},
\href{http://arxiv.org/abs/hep-ph/9409316}{{\ttfamily arXiv:hep-ph/9409316
  [hep-ph]}}.

\bibitem{Kartvelishvili:1977pi}
V.~G. Kartvelishvili, A.~K. Likhoded, and V.~A. Petrov, ``{On the Fragmentation
  Functions of Heavy Quarks Into Hadrons},''
\href{http://dx.doi.org/10.1016/0370-2693(78)90653-6}{{\em Phys. Lett.}
  {\bfseries 78B} (1978) 615--617}.

\bibitem{Cacciari:2005uk}
M.~Cacciari, P.~Nason, and C.~Oleari, ``{A Study of heavy flavored meson
  fragmentation functions in e+ e- annihilation},''
  \href{http://dx.doi.org/10.1088/1126-6708/2006/04/006}{{\em JHEP} {\bfseries
  04} (2006) 006},
\href{http://arxiv.org/abs/hep-ph/0510032}{{\ttfamily arXiv:hep-ph/0510032
  [hep-ph]}}.

\bibitem{Blumlein:2013cua}
J.~Bl{\"u}mlein and J.~Brunner, ``{New Exclusion Limits on Dark Gauge Forces
  from Proton Bremsstrahlung in Beam-Dump Data},''
  \href{http://dx.doi.org/10.1016/j.physletb.2014.02.029}{{\em Phys. Lett.}
  {\bfseries B731} (2014) 320--326},
\href{http://arxiv.org/abs/1311.3870}{{\ttfamily arXiv:1311.3870 [hep-ph]}}.

\bibitem{deNiverville:2016rqh}
P.~deNiverville, C.-Y. Chen, M.~Pospelov, and A.~Ritz, ``{Light dark matter in
  neutrino beams: production modelling and scattering signatures at MiniBooNE,
  T2K and SHiP},'' \href{http://dx.doi.org/10.1103/PhysRevD.95.035006}{{\em
  Phys. Rev.} {\bfseries D95} no.~3, (2017) 035006},
\href{http://arxiv.org/abs/1609.01770}{{\ttfamily arXiv:1609.01770 [hep-ph]}}.

\bibitem{Liu:2017htz}
Y.-S. Liu and G.~A. Miller, ``{Validity of the Weizs{\"a}cker-Williams
  approximation and the analysis of beam dump experiments: Production of an
  axion, a dark photon, or a new axial-vector boson},''
  \href{http://dx.doi.org/10.1103/PhysRevD.96.016004}{{\em Phys. Rev.}
  {\bfseries D96} no.~1, (2017) 016004},
\href{http://arxiv.org/abs/1705.01633}{{\ttfamily arXiv:1705.01633 [hep-ph]}}.

\bibitem{Holdom:1985ag}
B.~Holdom, ``{Two U(1)'s and Epsilon Charge Shifts},''
\href{http://dx.doi.org/10.1016/0370-2693(86)91377-8}{{\em Phys. Lett.}
  {\bfseries 166B} (1986) 196--198}.

\bibitem{Buschmann:2015awa}
M.~Buschmann, J.~Kopp, J.~Liu, and P.~A.~N. Machado, ``{Lepton Jets from
  Radiating Dark Matter},''
  \href{http://dx.doi.org/10.1007/JHEP07(2015)045}{{\em JHEP} {\bfseries 07}
  (2015) 045},
\href{http://arxiv.org/abs/1505.07459}{{\ttfamily arXiv:1505.07459 [hep-ph]}}.

\bibitem{Moreno:2018xxx}
O.~Moreno, ``{First Results from the Heavy Photon Search Experiment},'' in {\em
  {39th International Conference on High Energy Physics (ICHEP 2018), July 4-11
  2018, Coex, Seoul, Korea}}.
\newblock 2018.

\bibitem{Gninenko:2320630}
{\bfseries NA64} Collaboration, S.~Gninenko {\em et al.}, ``{NA64 Status Report
  2018},'' Tech. Rep. CERN-SPSC-2018-015. SPSC-SR-231, CERN, Geneva, May, 2018.
\newblock \url{https://cds.cern.ch/record/2320630}.

\bibitem{Caldwell:2018atq}
A.~Caldwell {\em et al.}, ``{Particle physics applications of the AWAKE
  acceleration scheme},''
\href{http://arxiv.org/abs/1812.11164}{{\ttfamily arXiv:1812.11164
  [physics.acc-ph]}}.

\bibitem{Ilten:2018crw}
P.~Ilten, Y.~Soreq, M.~Williams, and W.~Xue, ``{Serendipity in dark photon
  searches},'' \href{http://dx.doi.org/10.1007/JHEP06(2018)004}{{\em JHEP}
  {\bfseries 06} (2018) 004},
\href{http://arxiv.org/abs/1801.04847}{{\ttfamily arXiv:1801.04847 [hep-ph]}}.

\bibitem{Gardner:2015wea}
S.~Gardner, R.~J. Holt, and A.~S. Tadepalli, ``{New Prospects in Fixed Target
  Searches for Dark Forces with the SeaQuest Experiment at Fermilab},''
  \href{http://dx.doi.org/10.1103/PhysRevD.93.115015}{{\em Phys. Rev.}
  {\bfseries D93} (2016) 115015},
\href{http://arxiv.org/abs/1509.00050}{{\ttfamily arXiv:1509.00050 [hep-ph]}}.

\bibitem{Dent:2016wcr}
J.~B. Dent, B.~Dutta, S.~Liao, J.~L. Newstead, L.~E. Strigari, and J.~W.
  Walker, ``{Probing light mediators at ultralow threshold energies with
  coherent elastic neutrino-nucleus scattering},''
  \href{http://dx.doi.org/10.1103/PhysRevD.96.095007}{{\em Phys. Rev.}
  {\bfseries D96} no.~9, (2017) 095007},
\href{http://arxiv.org/abs/1612.06350}{{\ttfamily arXiv:1612.06350 [hep-ph]}}.

\bibitem{Ibe:2016dir}
M.~Ibe, W.~Nakano, and M.~Suzuki, ``{Constraints on $L_\mu-L_\tau$ gauge
  interactions from rare kaon decay},''
  \href{http://dx.doi.org/10.1103/PhysRevD.95.055022}{{\em Phys. Rev.}
  {\bfseries D95} no.~5, (2017) 055022},
\href{http://arxiv.org/abs/1611.08460}{{\ttfamily arXiv:1611.08460 [hep-ph]}}.

\bibitem{Bezrukov:2009yw}
F.~Bezrukov and D.~Gorbunov, ``{Light inflaton Hunter's Guide},''
  \href{http://dx.doi.org/10.1007/JHEP05(2010)010}{{\em JHEP} {\bfseries 05}
  (2010) 010},
\href{http://arxiv.org/abs/0912.0390}{{\ttfamily arXiv:0912.0390 [hep-ph]}}.

\bibitem{OConnell:2006rsp}
D.~O'Connell, M.~J. Ramsey-Musolf, and M.~B. Wise, ``{Minimal Extension of the
  Standard Model Scalar Sector},''
  \href{http://dx.doi.org/10.1103/PhysRevD.75.037701}{{\em Phys. Rev.}
  {\bfseries D75} (2007) 037701},
\href{http://arxiv.org/abs/hep-ph/0611014}{{\ttfamily arXiv:hep-ph/0611014
  [hep-ph]}}.

\bibitem{Grinstein:1988yu}
B.~Grinstein, L.~J. Hall, and L.~Randall, ``{Do B meson decays exclude a light
  Higgs?},''
\href{http://dx.doi.org/10.1016/0370-2693(88)90916-1}{{\em Phys. Lett.}
  {\bfseries B211} (1988) 363--369}.

\bibitem{Chivukula:1988gp}
R.~S. Chivukula and A.~V. Manohar, ``{Limits on a Light Higgs Boson},''
  \href{http://dx.doi.org/10.1016/0370-2693(88)90891-X}{{\em Phys. Lett.}
  {\bfseries B207} (1988) 86}.
[Erratum: Phys.~Lett.~B 217, 568 (1989)].

\bibitem{Bezrukov:2013fca}
F.~Bezrukov and D.~Gorbunov, ``{Light inflaton after LHC8 and WMAP9 results},''
  \href{http://dx.doi.org/10.1007/JHEP07(2013)140}{{\em JHEP} {\bfseries 07}
  (2013) 140},
\href{http://arxiv.org/abs/1303.4395}{{\ttfamily arXiv:1303.4395 [hep-ph]}}.

\bibitem{Donoghue:1990xh}
J.~F. Donoghue, J.~Gasser, and H.~Leutwyler, ``{The Decay of a Light Higgs
  Boson},''
\href{http://dx.doi.org/10.1016/0550-3213(90)90474-R}{{\em Nucl. Phys.}
  {\bfseries B343} (1990) 341--368}.

\bibitem{Gunion:1989we}
J.~F. Gunion, H.~E. Haber, G.~L. Kane, and S.~Dawson, ``{The Higgs Hunter's
  Guide},''
{\em Front. Phys.} {\bfseries 80} (2000) 1--404.

\bibitem{McKeen:2008gd}
D.~McKeen, ``{Constraining Light Bosons with Radiative Upsilon(1S) Decays},''
  \href{http://dx.doi.org/10.1103/PhysRevD.79.015007}{{\em Phys. Rev.}
  {\bfseries D79} (2009) 015007},
\href{http://arxiv.org/abs/0809.4787}{{\ttfamily arXiv:0809.4787 [hep-ph]}}.

\bibitem{Winkler:2018qyg}
M.~W. Winkler, ``{Decay and Detection of a Light Scalar Boson Mixing with the
  Higgs},'' \href{http://dx.doi.org/10.1103/PhysRevD.99.015018}{{\em Phys.
  Rev.} {\bfseries D99} no.~1, (2019) 015018},
\href{http://arxiv.org/abs/1809.01876}{{\ttfamily arXiv:1809.01876 [hep-ph]}}.

\bibitem{Ambrosino:2019qvz}
{\bfseries KLEVER} Collaboration, F.~Ambrosino {\em et al.}, ``{KLEVER: An
  experiment to measure BR($K_L\to\pi^0\nu\bar{\nu}$) at the CERN SPS},''
\href{http://arxiv.org/abs/1901.03099}{{\ttfamily arXiv:1901.03099 [hep-ex]}}.

\bibitem{Bird:2004ts}
C.~Bird, P.~Jackson, R.~V. Kowalewski, and M.~Pospelov, ``{Search for dark
  matter in $b \to s$ transitions with missing energy},''
  \href{http://dx.doi.org/10.1103/PhysRevLett.93.201803}{{\em Phys. Rev. Lett.}
  {\bfseries 93} (2004) 201803},
\href{http://arxiv.org/abs/hep-ph/0401195}{{\ttfamily arXiv:hep-ph/0401195
  [hep-ph]}}.

\bibitem{Altmannshofer:2009ma}
W.~Altmannshofer, A.~J. Buras, D.~M. Straub, and M.~Wick, ``{New strategies for
  New Physics search in $B \to K^{*} \nu \bar{\nu}$, $B \to K \nu \bar{\nu}$
  and $B \to X_{s} \nu \bar{\nu}$ decays},''
  \href{http://dx.doi.org/10.1088/1126-6708/2009/04/022}{{\em JHEP} {\bfseries
  04} (2009) 022},
\href{http://arxiv.org/abs/0902.0160}{{\ttfamily arXiv:0902.0160 [hep-ph]}}.

\bibitem{Aihara:2019gcq}
{\bfseries ILC} Collaboration, H.~Aihara {\em et al.}, ``{The International
  Linear Collider. A Global Project},''
\href{http://arxiv.org/abs/1901.09829}{{\ttfamily arXiv:1901.09829 [hep-ex]}}.

\bibitem{Mangano:2018mur}
{\bfseries FCC} Collaboration, M.~Mangano {\em et al.}, ``{Future Circular
  Collider},'' Tech. Rep. CERN-ACC-2018-0056, CERN, Geneva, Dec, 2018.
\newblock \url{http://cds.cern.ch/record/2651294}.

\bibitem{Drewes:2015iva}
M.~Drewes and B.~Garbrecht, ``{Combining experimental and cosmological
  constraints on heavy neutrinos},''
  \href{http://dx.doi.org/10.1016/j.nuclphysb.2017.05.001}{{\em Nucl. Phys.}
  {\bfseries B921} (2017) 250--315},
\href{http://arxiv.org/abs/1502.00477}{{\ttfamily arXiv:1502.00477 [hep-ph]}}.

\bibitem{Shrock:1978ft}
R.~E. Shrock, ``{A Test for the Existence of Effectively Stable Neutral Heavy
  Leptons},''
\href{http://dx.doi.org/10.1103/PhysRevLett.40.1688}{{\em Phys. Rev. Lett.}
  {\bfseries 40} (1978) 1688}.

\bibitem{Gallas:1994xp}
{\bfseries FMMF} Collaboration, E.~Gallas {\em et al.}, ``{Search for neutral
  weakly interacting massive particles in the Fermilab Tevatron wide band
  neutrino beam},''
\href{http://dx.doi.org/10.1103/PhysRevD.52.6}{{\em Phys. Rev.} {\bfseries D52}
  (1995) 6--14}.

\bibitem{Gorbunov:2007ak}
D.~Gorbunov and M.~Shaposhnikov, ``{How to find neutral leptons of the
  $\nu$MSM?},'' \href{http://dx.doi.org/10.1007/JHEP11(2013)101,
  10.1088/1126-6708/2007/10/015}{{\em JHEP} {\bfseries 10} (2007) 015},
  \href{http://arxiv.org/abs/0705.1729}{{\ttfamily arXiv:0705.1729 [hep-ph]}}.
[Erratum: JHEP11,101(2013)].

\bibitem{Drewes:2018irr}
M.~Drewes, J.~Hajer, J.~Klaric, and G.~Lanfranchi, ``{Perspectives to find
  heavy neutrinos with NA62},'' in {\em {53rd Rencontres de Moriond on
  Electroweak Interactions and Unified Theories (Moriond EW 2018) La Thuile,
  Italy, March 10-17, 2018}}.
\newblock 2018.
\newblock
\href{http://arxiv.org/abs/1806.00100}{{\ttfamily arXiv:1806.00100 [hep-ph]}}.
\newblock

\bibitem{Adams:2013qkq}
{\bfseries LBNE} Collaboration, C.~Adams {\em et al.}, ``{The Long-Baseline
  Neutrino Experiment: Exploring Fundamental Symmetries of the Universe},''
\href{http://arxiv.org/abs/1307.7335}{{\ttfamily arXiv:1307.7335 [hep-ex]}}.

\bibitem{Izaguirre:2015pga}
E.~Izaguirre and B.~Shuve, ``{Multilepton and Lepton Jet Probes of
  Sub-Weak-Scale Right-Handed Neutrinos},''
  \href{http://dx.doi.org/10.1103/PhysRevD.91.093010}{{\em Phys. Rev.}
  {\bfseries D91} no.~9, (2015) 093010},
\href{http://arxiv.org/abs/1504.02470}{{\ttfamily arXiv:1504.02470 [hep-ph]}}.

\bibitem{Antusch:2017hhu}
S.~Antusch, E.~Cazzato, and O.~Fischer, ``{Sterile neutrino searches via
  displaced vertices at LHCb},''
  \href{http://dx.doi.org/10.1016/j.physletb.2017.09.057}{{\em Phys. Lett.}
  {\bfseries B774} (2017) 114--118},
\href{http://arxiv.org/abs/1706.05990}{{\ttfamily arXiv:1706.05990 [hep-ph]}}.

\bibitem{Kobach:2014hea}
A.~Kobach and S.~Dobbs, ``{Heavy Neutrinos and the Kinematics of Tau Decays},''
  \href{http://dx.doi.org/10.1103/PhysRevD.91.053006}{{\em Phys. Rev.}
  {\bfseries D91} no.~5, (2015) 053006},
\href{http://arxiv.org/abs/1412.4785}{{\ttfamily arXiv:1412.4785 [hep-ph]}}.

\bibitem{Coloma:2017ppo}
P.~Coloma, P.~A.~N. Machado, I.~Martinez-Soler, and I.~M. Shoemaker,
  ``{Double-Cascade Events from New Physics in Icecube},''
  \href{http://dx.doi.org/10.1103/PhysRevLett.119.201804}{{\em Phys. Rev.
  Lett.} {\bfseries 119} no.~20, (2017) 201804},
\href{http://arxiv.org/abs/1707.08573}{{\ttfamily arXiv:1707.08573 [hep-ph]}}.

\bibitem{Drewes:2018xma}
M.~Drewes, A.~Giammanco, J.~Hajer, M.~Lucente, and O.~Mattelaer, ``{A Heavy
  Metal Path to New Physics},''
\href{http://arxiv.org/abs/1810.09400}{{\ttfamily arXiv:1810.09400 [hep-ph]}}.

\bibitem{Peccei:1977hh}
R.~D. Peccei and H.~R. Quinn, ``{CP Conservation in the Presence of
  Instantons},'' \href{http://dx.doi.org/10.1103/PhysRevLett.38.1440}{{\em
  Phys. Rev. Lett.} {\bfseries 38} (1977) 1440--1443}.
[,328(1977)].

\bibitem{Peccei:1977ur}
R.~D. Peccei and H.~R. Quinn, ``{Constraints Imposed by CP Conservation in the
  Presence of Instantons},''
\href{http://dx.doi.org/10.1103/PhysRevD.16.1791}{{\em Phys. Rev.} {\bfseries
  D16} (1977) 1791--1797}.

\bibitem{Wilczek:1977pj}
F.~Wilczek, ``{Problem of Strong p and t Invariance in the Presence of
  Instantons},''
\href{http://dx.doi.org/10.1103/PhysRevLett.40.279}{{\em Phys. Rev. Lett.}
  {\bfseries 40} (1978) 279--282}.

\bibitem{Weinberg:1977ma}
S.~Weinberg, ``{A New Light Boson?},''
\href{http://dx.doi.org/10.1103/PhysRevLett.40.223}{{\em Phys. Rev. Lett.}
  {\bfseries 40} (1978) 223--226}.

\bibitem{Jaeckel:2010ni}
J.~Jaeckel and A.~Ringwald, ``{The Low-Energy Frontier of Particle Physics},''
  \href{http://dx.doi.org/10.1146/annurev.nucl.012809.104433}{{\em Ann. Rev.
  Nucl. Part. Sci.} {\bfseries 60} (2010) 405--437},
\href{http://arxiv.org/abs/1002.0329}{{\ttfamily arXiv:1002.0329 [hep-ph]}}.

\bibitem{Bauer:2017ris}
M.~Bauer, M.~Neubert, and A.~Thamm, ``{Collider Probes of Axion-Like
  Particles},'' \href{http://dx.doi.org/10.1007/JHEP12(2017)044}{{\em JHEP}
  {\bfseries 12} (2017) 044},
\href{http://arxiv.org/abs/1708.00443}{{\ttfamily arXiv:1708.00443 [hep-ph]}}.

\bibitem{Dobrich:2015jyk}
B.~D{\"o}brich, J.~Jaeckel, F.~Kahlhoefer, A.~Ringwald, and K.~Schmidt-Hoberg,
  ``{ALPtraum: ALP production in proton beam dump experiments},''
  \href{http://dx.doi.org/10.1007/JHEP02(2016)018}{{\em JHEP} {\bfseries 02}
  (2016) 018},
\href{http://arxiv.org/abs/1512.03069}{{\ttfamily arXiv:1512.03069 [hep-ph]}}.

\bibitem{Batell:2009jf}
B.~Batell, M.~Pospelov, and A.~Ritz, ``{Multi-lepton Signatures of a Hidden
  Sector in Rare $B$ Decays},''
  \href{http://dx.doi.org/10.1103/PhysRevD.83.054005}{{\em Phys. Rev.}
  {\bfseries D83} (2011) 054005},
\href{http://arxiv.org/abs/0911.4938}{{\ttfamily arXiv:0911.4938 [hep-ph]}}.

\bibitem{Dolan:2014ska}
M.~J. Dolan, F.~Kahlhoefer, C.~McCabe, and K.~Schmidt-Hoberg, ``{A taste of
  dark matter: Flavour constraints on pseudoscalar mediators},''
  \href{http://dx.doi.org/10.1007/JHEP07(2015)103,
  10.1007/JHEP03(2015)171}{{\em JHEP} {\bfseries 03} (2015) 171},
  \href{http://arxiv.org/abs/1412.5174}{{\ttfamily arXiv:1412.5174 [hep-ph]}}.
[Erratum: JHEP07,103(2015)].

\bibitem{Dobrich:2018jyi}
B.~D{\"o}brich, F.~Ertas, F.~Kahlhoefer, and T.~Spadaro, ``{Model-independent
  bounds on light pseudoscalars from rare $B$-meson decays},''
\href{http://arxiv.org/abs/1810.11336}{{\ttfamily arXiv:1810.11336 [hep-ph]}}.

\bibitem{Domingo:2016yih}
F.~Domingo, ``{Decays of a NMSSM CP-odd Higgs in the low-mass region},''
  \href{http://dx.doi.org/10.1007/JHEP03(2017)052}{{\em JHEP} {\bfseries 03}
  (2017) 052},
\href{http://arxiv.org/abs/1612.06538}{{\ttfamily arXiv:1612.06538 [hep-ph]}}.

\bibitem{Gatto:2016rae}
{\bfseries REDTOP} Collaboration, C.~Gatto, B.~Fabela~Enriquez, and M.~I.
  Pedraza~Morales, ``{The REDTOP project: Rare Eta Decays with a TPC for
  Optical Photons},''
\href{http://dx.doi.org/10.22323/1.282.0812}{{\em PoS} {\bfseries ICHEP2016}
  (2016) 812}.

\bibitem{Aloni:2018vki}
D.~Aloni, Y.~Soreq, and M.~Williams, ``{Coupling QCD-scale axion-like particles
  to gluons},''
\href{http://arxiv.org/abs/1811.03474}{{\ttfamily arXiv:1811.03474 [hep-ph]}}.

\bibitem{Bergsma:1985qz}
{\bfseries CHARM} Collaboration, F.~Bergsma {\em et al.}, ``{Search for Axion
  Like Particle Production in 400-{GeV} Proton - Copper Interactions},''
\href{http://dx.doi.org/10.1016/0370-2693(85)90400-9}{{\em Phys. Lett.}
  {\bfseries 157B} (1985) 458--462}.

\bibitem{Ostapchenko:2010vb}
S.~Ostapchenko, ``{Monte Carlo treatment of hadronic interactions in enhanced
  Pomeron scheme: I. QGSJET-II model},''
  \href{http://dx.doi.org/10.1103/PhysRevD.83.014018}{{\em Phys. Rev.}
  {\bfseries D83} (2011) 014018},
\href{http://arxiv.org/abs/1010.1869}{{\ttfamily arXiv:1010.1869 [hep-ph]}}.

\bibitem{Ahn:2009wx}
E.-J. Ahn, R.~Engel, T.~K. Gaisser, P.~Lipari, and T.~Stanev, ``{Cosmic ray
  interaction event generator SIBYLL 2.1},''
  \href{http://dx.doi.org/10.1103/PhysRevD.80.094003}{{\em Phys. Rev.}
  {\bfseries D80} (2009) 094003},
\href{http://arxiv.org/abs/0906.4113}{{\ttfamily arXiv:0906.4113 [hep-ph]}}.

\bibitem{Riehn:2015oba}
F.~Riehn, R.~Engel, A.~Fedynitch, T.~K. Gaisser, and T.~Stanev, ``{A new
  version of the event generator Sibyll},'' {\em PoS} {\bfseries ICRC2015}
  (2016) 558,
\href{http://arxiv.org/abs/1510.00568}{{\ttfamily arXiv:1510.00568 [hep-ph]}}.

\bibitem{Dulat:2015mca}
S.~Dulat, T.-J. Hou, J.~Gao, M.~Guzzi, J.~Huston, P.~Nadolsky, J.~Pumplin,
  C.~Schmidt, D.~Stump, and C.~P. Yuan, ``{New parton distribution functions
  from a global analysis of quantum chromodynamics},''
  \href{http://dx.doi.org/10.1103/PhysRevD.93.033006}{{\em Phys. Rev.}
  {\bfseries D93} no.~3, (2016) 033006},
\href{http://arxiv.org/abs/1506.07443}{{\ttfamily arXiv:1506.07443 [hep-ph]}}.

\bibitem{Ball:2017nwa}
{\bfseries NNPDF} Collaboration, R.~D. Ball {\em et al.}, ``{Parton
  distributions from high-precision collider data},''
  \href{http://dx.doi.org/10.1140/epjc/s10052-017-5199-5}{{\em Eur. Phys. J.}
  {\bfseries C77} no.~10, (2017) 663},
\href{http://arxiv.org/abs/1706.00428}{{\ttfamily arXiv:1706.00428 [hep-ph]}}.

\bibitem{Agostinelli:2002hh}
{\bfseries GEANT4} Collaboration, S.~Agostinelli {\em et al.}, ``{GEANT4: A
  Simulation toolkit},''
\href{http://dx.doi.org/10.1016/S0168-9002(03)01368-8}{{\em Nucl. Instrum.
  Meth.} {\bfseries A506} (2003) 250--303}.

\end{thebibliography}\endgroup

\end{document}